\documentclass[12pt]{article}
\usepackage{amsmath,bbm,array,amsfonts,float,mathtools,amsthm,mathrsfs}
\usepackage{jheppub}
\usepackage{caption,subcaption}
\usepackage{multirow}
\usepackage[toc,page]{appendix}
\usepackage[utf8]{inputenc}
\usepackage[autostyle=true]{csquotes}

\numberwithin{equation}{section}
\numberwithin{figure}{section}
\numberwithin{table}{section}
\usepackage{todonotes}

\title{Machine Learning Calabi-Yau Hypersurfaces}
\author[a]{David S. Berman,}
\author[b,c,d,e]{Yang-Hui He,}
\author[c,b]{Edward Hirst}

\affiliation[a]{
	Centre for Theoretical Physics, School of Physics and Astronomy, Queen Mary University of London, 327 Mile End Road, London E1 4NS, UK}
\affiliation[b]{
    London Institute for Mathematical Sciences, Royal Institution, London W1S 4BS, UK}
\affiliation[c]{
	Department of Mathematics, City, University of London, EC1V 0HB, UK}
\affiliation[d]{
	Merton College, University of Oxford, OX1 4JD, UK}
\affiliation[e]{
	School of Physics, NanKai University, Tianjin, 300071, P.R. China}

\emailAdd{d.s.berman@qmul.ac.uk}
\emailAdd{hey@maths.ox.ac.uk}
\emailAdd{edward.hirst@city.ac.uk}

\preprint{\begin{flushright}
QMUL-PH-21-55

LIMS-2021-017
\end{flushright}}
\abstract{

We revisit the classic database of weighted-$\mathbb{P}^4$s which admit Calabi-Yau 3-fold hypersurfaces equipped with a diverse set of tools from the machine-learning toolbox.
Unsupervised techniques identify an unanticipated almost linear dependence of the topological data on the weights. This then allows us to identify a previously unnoticed clustering in the Calabi-Yau data. 
Supervised techniques are successful in predicting the topological parameters of the hypersurface from its weights with an accuracy of $R^2 > 95\%$. Supervised learning also allows us to identify weighted-$\mathbb{P}^4$s which admit Calabi-Yau hypersurfaces to $100\%$ accuracy by making use of partitioning supported by the clustering behaviour.
}

\begin{document}
\maketitle

\section{Introduction}
Artificial intelligence has now permeated through all disciplines of human enterprise.
Machine-learning (ML) has become, in this age driven by big data, as indispensable a tool as  calculus was to the Age of Enlightenment \cite{lecun2015deep}.
Perhaps surprisingly, noise-less, pure mathematical data can often be learned at high precision, indicating underlying formulae which have not yet been uncovered from traditional analyses.
Examples of this data driven approach to mathematics may be seen in applications of ML to: the string theory landscape \cite{He:2017aed,He:2017set,Carifio:2017bov,Krefl:2017yox,Ruehle:2017mzq}; abstract algebra \cite{He:2019nzx}; modern number theory \cite{Alessandretti:2019jbs,He:2020eva}; and graph theory \cite{He:2020fdg}. It is hoped that ML might reveal structures in the very nature of mathematics \cite{He:2021oav} and mathematical intuition \cite{davies2021advancing}, deeply embedded in the {\it mathematical data}. Apart from ML itself, the tools that have been developed to enable ML have provided significant new capabilities for old problems. This is exemplified by the use of the autodifferentiation capabilities of \texttt{Tensorflow} to explore the possible vacua of various gauged supergravities, see for example \cite{Comsa:2019rcz,Bobev:2019dik,Krishnan:2020sfg,Bobev:2020ttg,Berman:2021ynm,Berman:2022jqn}.

Amongst its various virtues, string theory pioneered the data-mining of such mathematical data. One should be mindful that this was done shortly after the beginnings of string phenomenology in the late 1980s, long before the dawn of the modern era of ``Big Data'' and modern readily available ultra-fast computing power.
Indeed, when Calabi-Yau manifolds were realized to be \cite{Candelas:1985en} the standard solution to vacuum configurations (see \cite{Bao:2020sqg} for a brief, and \cite{He:2018jtw} a longer, pedagogical review), and hence low-energy particle physics, a programme was introduced by the physics community to compile one of the first databases in algebraic geometry.

These were some of the earliest appearances of ``big data'' in mathematics, beyond compiling digits of $\pi$ or large primes.
The first dataset was the so-called CICYs, which stands for ``complete intersection Calabi-Yau manifolds'' in products of complex projective spaces \cite{Candelas:1987kf,Green:1986ck}; which can be thought of as a generalization of the famous quintic 3-fold in $\mathbb{P}^4$.
Appropriately, one of the first ML experiments in geometry was performed on this dataset \cite{He:2017aed}.
Over the last few years, the initial success has been vastly improved by using more and more sophisticated neural network (NN) architectures and machine learning techniques \cite{Bull:2018uow,Bull:2019cij,Krippendorf:2020gny,He:2020lbz,Douglas:2020hpv,ashmore2021machine,Anderson:2020hux,Erbin:2020tks,Erbin:2020srm,Erbin:2021hmx,Larfors:2021pbb,Bao:2020nbi,Bao:2021auj,Bao:2021olg,Jejjala:2020wcc,Brodie:2021nit,Cole:2021nnt,Halverson:2021aot,gao2021machine,Cole:2019enn,Krippendorf:2021uxu}.

Yet, the CICY dataset has a peculiarity: it is skewed toward negative Euler number.
This would have occurred to Candelas et al.~since they knew about mirror symmetry.
Since the exchange of the two Hodge numbers $(h^{1,1}, h^{2,1})$ would reverse the sign of the Euler number $\chi$; the conjecture that to every Calabi-Yau 3-fold with $(h^{1,1}, h^{2,1})$ there is a mirror with these exchanged would imply the negation of $\chi$.

Therefore, as the second database of geometries in string theory, another generalization of the quintic was undertaken by placing {\it weights} on the ambient $\mathbb{P}^4$ and considering a single, generic Calabi-Yau hypersurface therein \cite{CANDELAS1990383}.
This produced a much more balanced set of Calabi-Yau 3-folds with $\pm$ Euler numbers, and the rough outline of the famous ``mirror plot'' of the distributions of $2(h^{1,1} - h^{2,1})$ vs
$(h^{1,1} + h^{2,1})$ could 
already be seen to emerge.

All these datasets were subsequently subsumed into the dataset created through the extraordinary work of Kreuzer and Skarke \cite{Kreuzer:2000xy,Skarke1996WEIGHTSF}. This set contains the Calabi-Yau hypersurfaces in toric varieties. (Since weighted projective spaces are special types of toric varieties, the set described in \cite{CANDELAS1990383} is a subset of the Kreuzer-Sharke set.)

However, the Kreuzer-Sharke dataset is of astronomical size, containing some half-billion members.
While ML of this set is in progress \cite{Bao:2021ofk}, the much more manageable list of hypersurfaces in weighted $\mathbb{P}^4$, numbering around 8000 (comparable to the CICYs) is natural choice of geometries to study and apply the latest methods from data science.

Thus our motivation is clear.
We shall re-visit the classic database of \cite{CANDELAS1990383} with a modern perspective, continuing the paradigm of machine-learning the string theory landscape and the resulting emergent mathematical structures, using tools from the \texttt{sci-kit learn} library \cite{scikit-learn} implemented in \texttt{python}.
The paper is organized as follows.
In \S\ref{s:CY} we begin with a rapid review of the mathematics of our Calabi-Yau hypersurfaces, emphasizing on the data structure.
In \S\ref{data}, we proceed with analyses of the data, using methods which were unavailable at the time of their creation, such as principle component analysis and topological data analysis.
We then use neural-networks to machine learn the dataset in \S\ref{ml}.
We conclude with a summary and outlook in \S\ref{s:conc}.

All of the data and code are freely available on
GitHub at: 
\url{https://github.com/edhirst/P4CY3ML.git}

\section{Characterising the Calabi-Yau Hypersurfaces}\label{s:CY}
The dataset of focus in this study is that of weighted projective spaces $\mathbb{P}^4_\mathbb{C}(w_i)$, which admit Calabi-Yau (CY) three-fold hypersurfaces within them.

This dataset was constructed in the early 90s alongside other efforts to expand the CY landscape for use in Landau-Ginzburg models and superstring compactification \cite{KS1992,KS1994,CANDELAS1990383,Kreuzer:2000xy,COK1995}.
The dataset is readily available at:  \url{http://hep.itp.tuwien.ac.at/~kreuzer/CY/}, whilst another copy is given with this study's scripts on the corresponding GitHub repository.

A generic weighted projective space generalises the concept of a projective space, defined by taking some $\mathbb{C}^{n+1}$ with coordinates $\{z_1,z_2,...,z_{n+1}\}$ and performing an identification with weights $w_i$ such that
\begin{equation}
    (z_1,z_2,...,z_{n+1}) \sim (\lambda^{w_1}z_1,\lambda^{w_2}z_2,...,\lambda^{w_{n+1}}z_{n+1})\,,
\end{equation}
$\forall \lambda \in \mathbb{C}$, hence defining the projective space $\mathbb{P}^{n}$ with these $n+1$ homogeneous coordinates.
For the projective spaces in consideration $n$ takes the value of 4, and the space is hence defined with 5 weights.
These weights are coprime as a set, such that the projective space definition is free from redundancy from different weight combinations.

Within these weighted projective spaces one can embed hypersurfaces defined by polynomials in these homogeneous coordinates. 
Of particular interest to physicists are those hypersurfaces which are CY in nature.
A defining property of CY manifolds is the vanishing of the first Chern class, and for this to hold within the projective space the hypersurface's polynomial has to have degree $d = \sum_i (w_i)$.

It should be noted that the identifications that are used in constructing the projective space lead to singular sets, which the hypersurfaces can intersect with suitable resolution.
To be consistently defined over these singular sets another property of the polynomial is required: \textit{transversity}.
The transversity property implies that the polynomial equation and its derivative share no common solutions, and this condition translates into a condition on the projective space weights: 
\begin{equation}
    \forall w_i \; \exists w_j \; s.t. \; \frac{\sum_k (w_k) - w_j}{w_i} \in \mathbb{Z}^+.
\end{equation}\label{transverse}
However as exemplified in \cite{CANDELAS1990383}, this condition is necessary but not sufficient for the surface to be CY.
It is therefore of interest to consider the extent to which each of these 5-vector weights properties contribute to determine the very special CY property; and it is this question we look to probe with new tools from data analysis, and machine-learning. 

It has been shown that only a finite number of possible weights permit these CY hypersurfaces. In fact, the dataset of weights consists of just 7555 5-vectors of transverse coprime integers.

Beyond learning the CY property explicitly, we are also interested in exploring if the topological features of the Calabi-Yau can be learnt from the weights. Of specific importance are the non-trivial Hodge numbers, $h^{1,1}$ and $h^{2,1}$, which describe the manifolds cohomology, and the Euler number, $\chi$. These all have a variety of uses in determining physical phenomena.
The formula for Hodge numbers comes from expansion of the Poincaré polynomial $Q(u,v) := \sum_{p,q} h^{p,q}u^pv^q$, the generating function of the Hodge numbers $h^{p,q}$; whilst the formula for Euler number has a direct form \cite{Vafa:1989xc,KS1992,KLRY1998,Batyrev:2020ych}.
Specifically these are
\begin{equation}
\begin{split}
    Q(u,v) & = \frac{1}{uv} \sum_{l=0}^{\sum_i(w_i)} \bigg[ \prod_{\tilde{\theta}_i(l)\in\mathbb{Z}} \frac{(uv)^{q_i}-uv}{1-(uv)^{q_i}} \bigg]_{int} \bigg( v^{size(l)} \bigg(\frac{u}{v}\bigg)^{age(l)}\bigg) \;,\\
    \chi & = \frac{1}{\sum_i(w_i)} \sum_{l,r=0}^{\sum_i(w_i)-1} \bigg[\prod_{i|lq_i \& rq_i \in \mathbb{Z}} \bigg(1-\frac{1}{q_i}\bigg)\bigg]\;,
\end{split}
\end{equation}\label{hodgeeulerformulas}
for weights $w_i$, normalised weights $q_i = w_i/\sum_i(w_i)$, and $u, v$ are the dummy variables of the  Poincaré polynomial. 
For $Q(u,v)$, $\tilde{\theta}_i(l)$ is the canonical representative of $lq_i$ in $(\mathbb{R}/\mathbb{Z})^5$, $age(l) = \sum_{i=0}^4 \tilde{\theta}_i(l)$ and $size(l) = age(l) + age(\sum_i(w_i) - l)$.
Note also for $\chi$, where $\forall \, i \ lq_i \ or \ rq_i \notin \mathbb{Z}$ then the product takes value 1 \cite{Batyrev:2020ych}.

Both formulas require significant computation, involving many non-trivial steps. 
Even if we realize this dataset in the language of the toric geometry of \cite{Kreuzer:2000xy,batyrev2011calabi}, the formulae involve non-trivial sums over faces of various dimension.
It is consequently interesting to examine the performance of machine-learning methods in learning the Euler number/Hodge numbers from the weights, and perhaps predicting the results through the use of possible hidden structures which we hope to uncover in the weights and weight distributions.

\section{Data Analysis}\label{data}
Before we apply the supervised machine-learning methods described in \S\ref{ml}, let us provide some analysis of the fundamentals of the dataset through the use of \textit{principal component analysis} (PCA), \textit{topological data analysis} (TDA), and other unsupervised machine-learning methods. 

\subsection{Datasets}\label{datasets}
In addition to the CY dataset which forms the central focus of this study, we will construct some auxiliary datasets that will help in assessing the learning of the Calabi-Yau property.  These are equivalent datasets of 5-vectors that possess fewer of the necessary properties required to meet the Calabi-Yau property.

The 4 datasets (including the original CY dataset) are composed of: 
\begin{description}
    \item[(a)] 7555 5-vectors of positive random integers,
    \item[(b)] 7555 5-vectors of positive random coprime integers,
    \item[(c)] 7555 transverse 5-vectors of positive random coprime integers,
    \item[(d)] 7555 Calabi-Yau 5-vectors.
\end{description}
These datasets were specifically constructed so as not to form a filtration, therefore at each stage the dataset generated was ensured to not include data which satisfies the additional conditions at the next level.
To clarify, each 5-vector in set (a) had weights which shared a common factor, in set (b) all 5-vectors did not satisfy condition \ref{transverse}, and those in set (c) where not in the CY list of (d). 

\begin{figure}[h!]
	\centering
	\begin{subfigure}{0.45\textwidth}
		\centering
		\includegraphics[width=\textwidth]{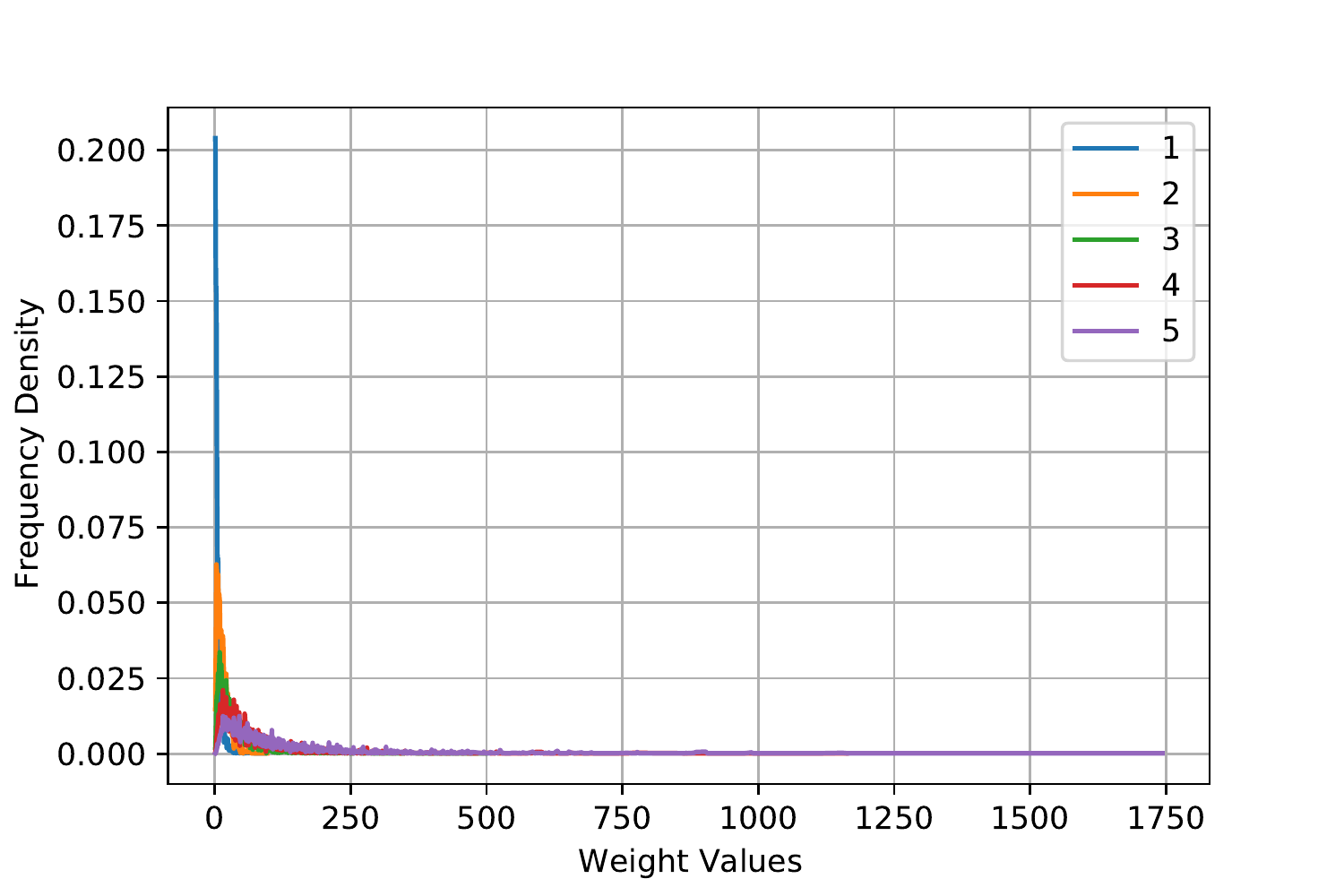}
		\caption{}\label{CYWeightFreq_full}
	\end{subfigure} 
    \begin{subfigure}{0.45\textwidth}
    	\centering
    	\includegraphics[width=\textwidth]{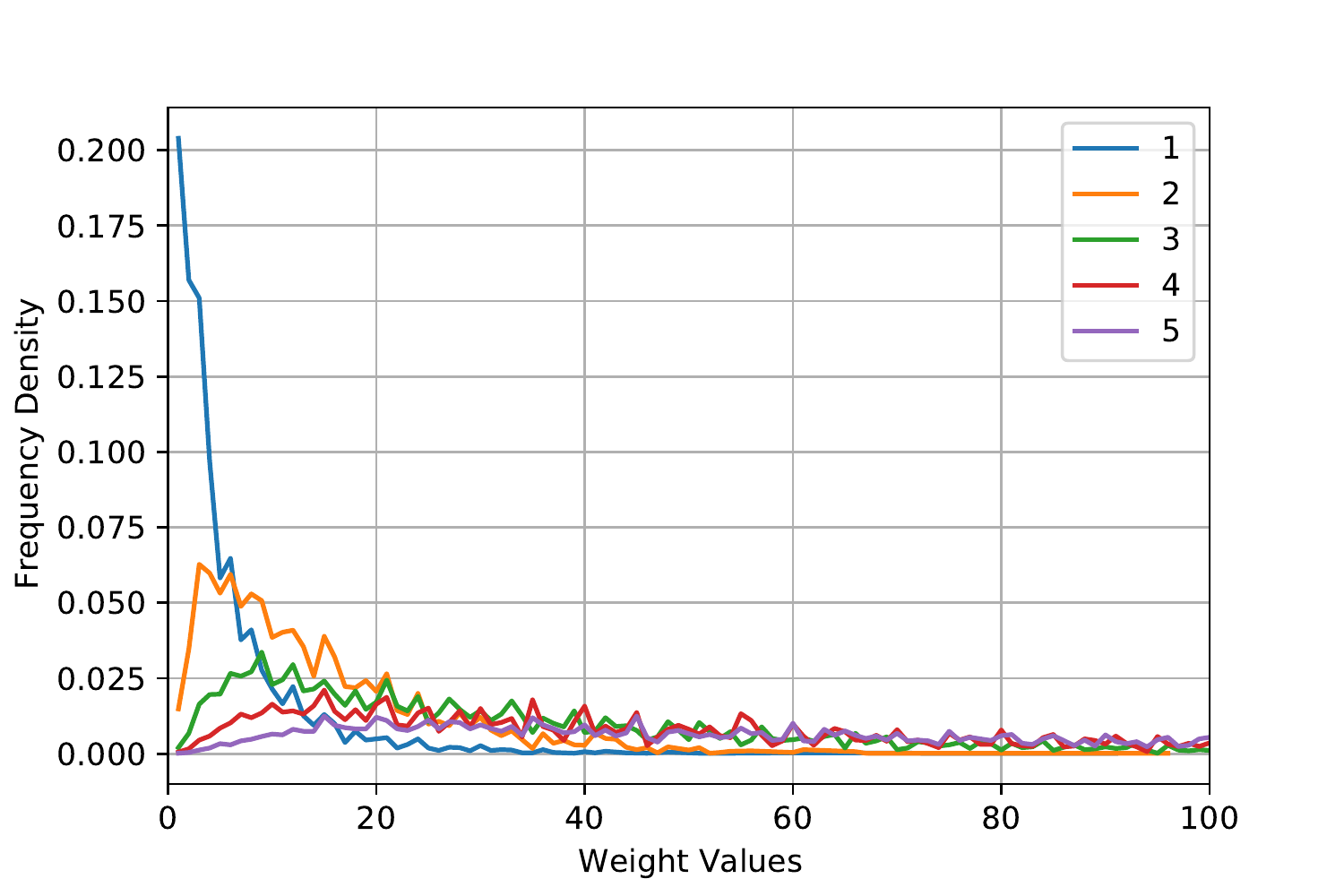}
    	\caption{}\label{CYWeightFreq_zoomed}
    \end{subfigure}
\caption{Frequency distribution of each of the CY 5-vector weights, $w_i$ (labelled by $i: 1-5$). Figure (b) shows the same data as (a), but restricted to lower entries so as to highlight the low value behaviour, due to the entry sorting.}\label{WeightFreqDists}
\end{figure}

\begin{figure}[h!]
    \centering
    \includegraphics[width=0.5\textwidth]{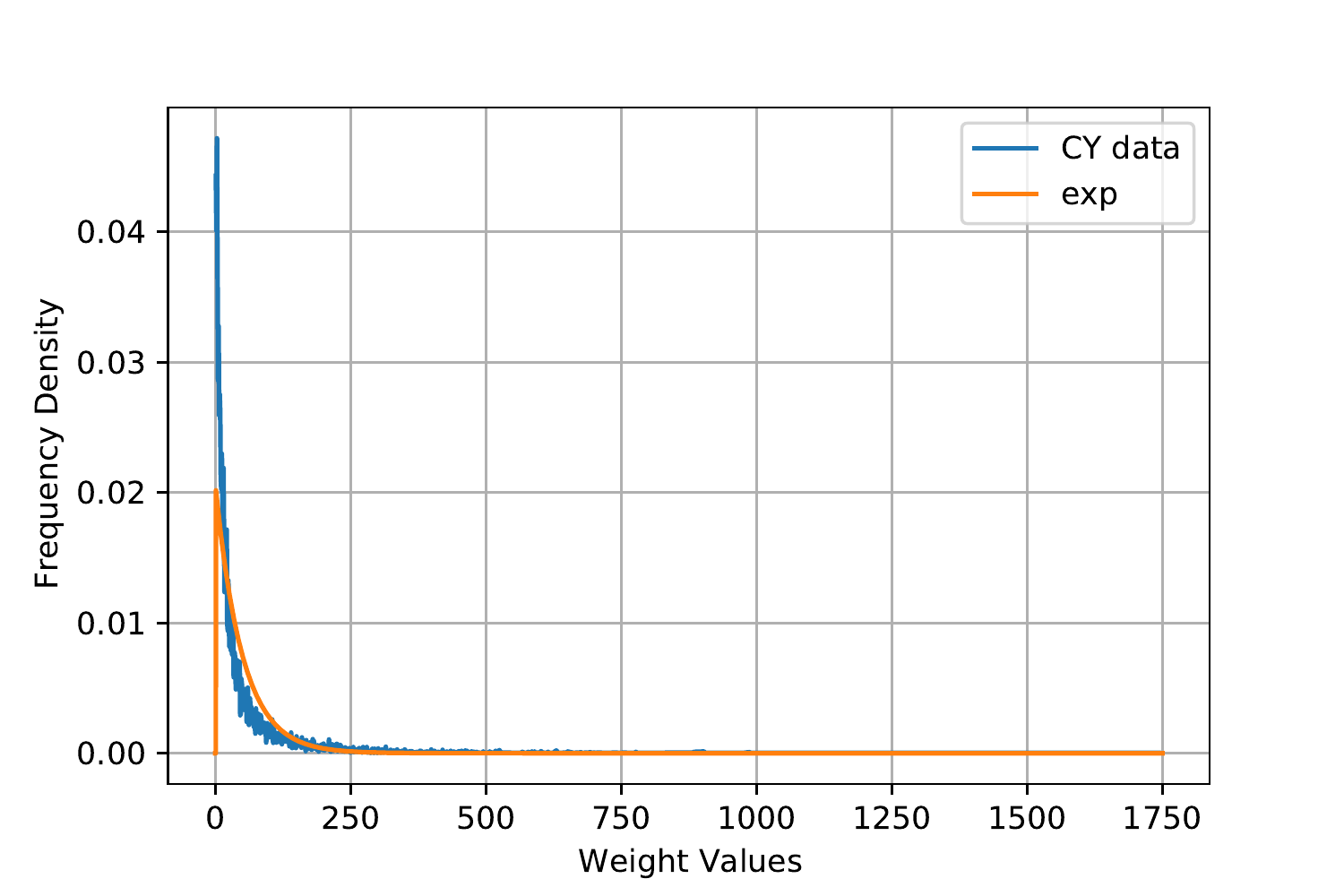}
    \caption{Frequency distribution for \textit{all} weights occurring across all 5-vectors in the CY dataset. Plot also shows the fitted exponential distribution, with scale parameter 49.536 (to 3 decimal places).}
    \label{CYallweights_exp}
\end{figure}

To introduce a consistency across the datasets, all the 5-vectors entries are sorted in increasing order.
Initially the weights for each of the datasets (a-c) were sampled using a discretised uniform distribution, $U(1,2000)$, bound above by 2000 to mimic the highest value in the CY dataset of 1743.
However as shown in figure \ref{CYWeightFreq_full} the weights follow a distribution far from that of a uniform distribution. 
Therefore to make the generated data more representative, an exponential distribution was fitted to the histogram of all weights in the CY dataset, as shown in figure \ref{CYallweights_exp}. 
Fitting was performed using the \texttt{scipy} library.

\begin{figure}[h!]
	\centering
	\begin{subfigure}{0.45\textwidth}
		\centering
		\includegraphics[width=\textwidth]{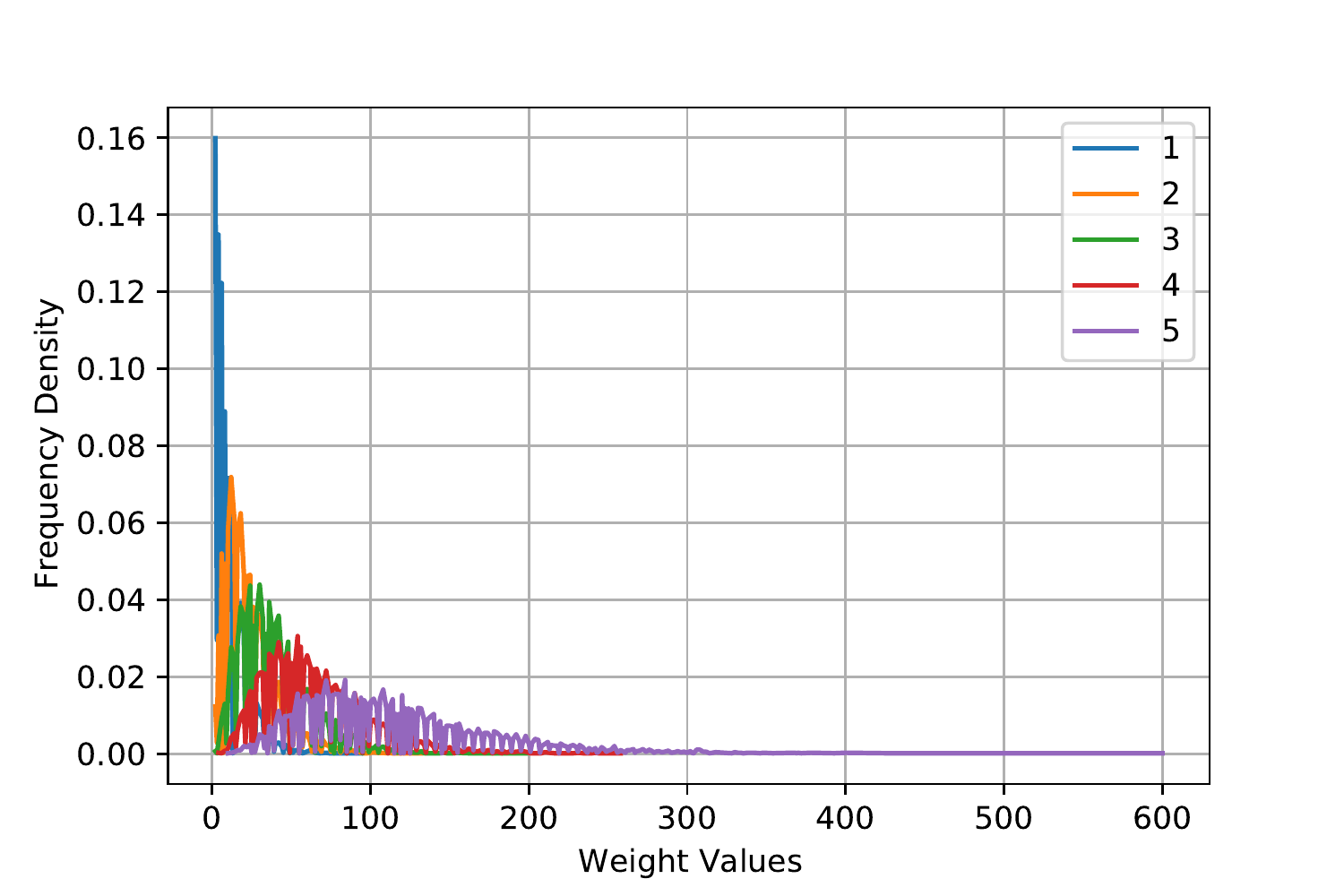}
		\caption{Random Integers}\label{RandomWeightFreq}
	\end{subfigure} 
    \begin{subfigure}{0.45\textwidth}
    	\centering
    	\includegraphics[width=\textwidth]{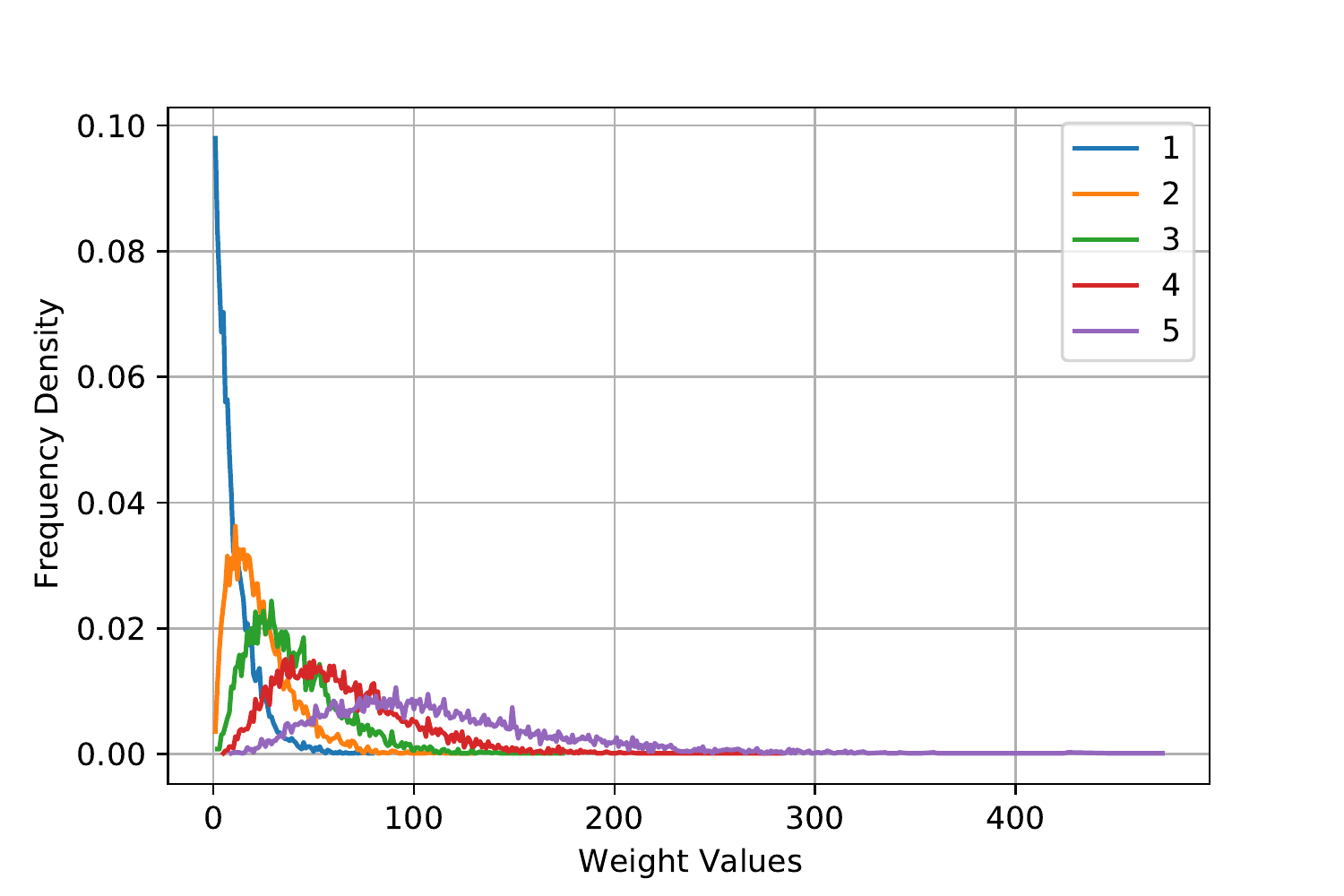}
    	\caption{Random Coprime Integers}\label{CoprimeWeightFreq}
    \end{subfigure} \\ 
	\begin{subfigure}{0.45\textwidth}
		\centering
		\includegraphics[width=\textwidth]{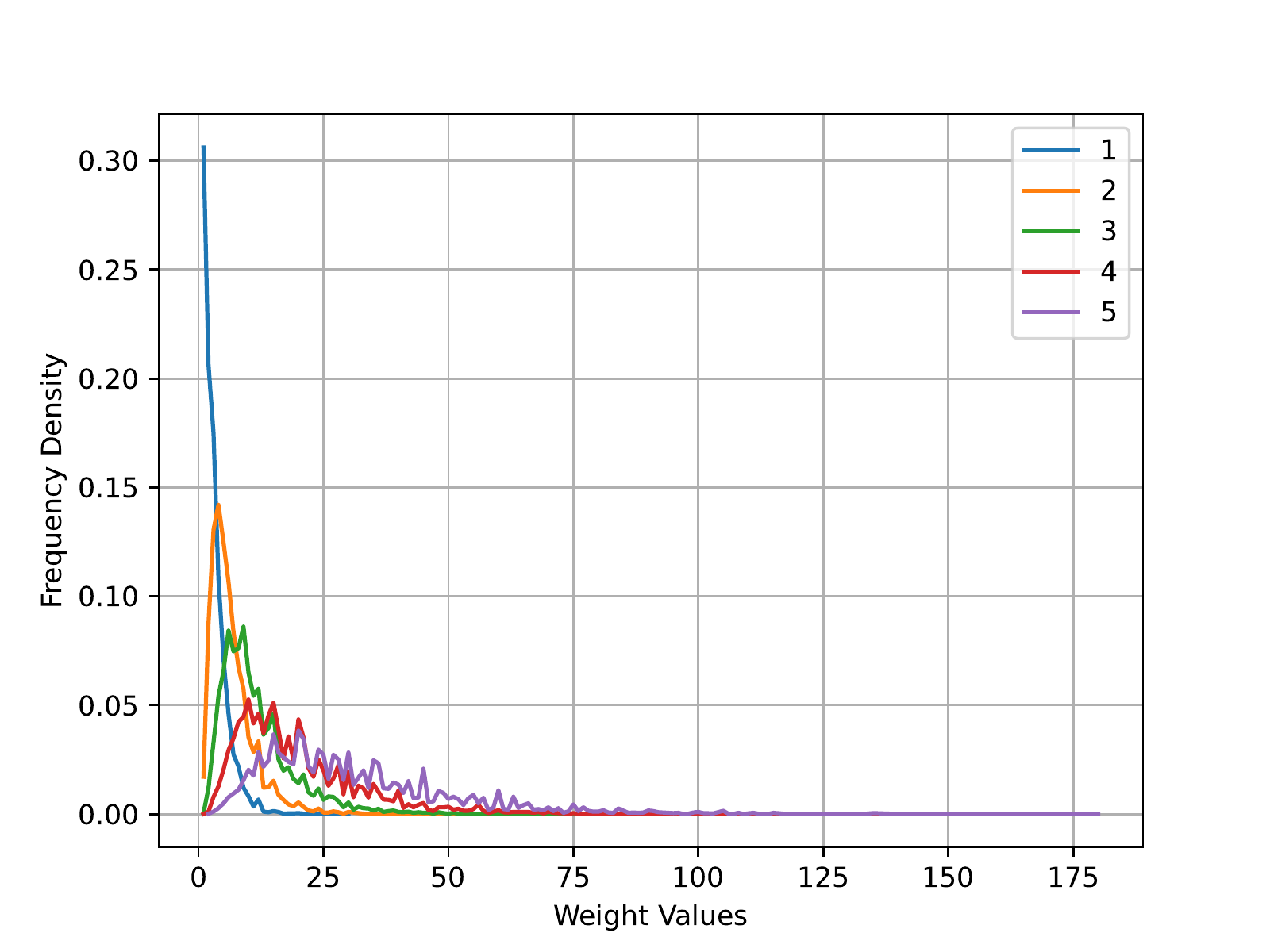}
		\caption{Random Transverse Coprime Integers}\label{TransverseWeightFreq}
	\end{subfigure} 
\caption{Frequency distributions for 5-vector weights, $w_i$ (labelled by $i: 1-5$), for each of the generated datasets. The weights were generated using an exponential distribution fitted to the CY weight data; and hence distributions show similar behaviour across all datasets, and to the CY dataset.}\label{FakeDataWeightFreqDists}
\end{figure}

This exponential distribution was instead then used to sample weights, and as can be seen in figures \ref{FakeDataWeightFreqDists}, the frequency distributions of the weights for each of the datasets align much closer to that of the CY data.
For reference the weight histograms are shown for the uniform distribution sampling in appendix \ref{uniformweightdists}.

\paragraph{Aside: Coprimality}
It is interesting to note that the probability of $k$ randomly chosen integers being coprime is: $1/\zeta(k)$; via the Riemann zeta function.
Hence the probability of a random 5-vector of integers being coprime is $ \sim 0.964$, and therefore the dataset (b) is relatively more common than the dataset (a).
Effectively it is easy to randomly produce weighted projective spaces.

\subsection{Principal Component Analysis}\label{pca}
Datasets of vectors can be analysed for hidden  structures through the use of principal component analysis (PCA).
This method, generally considered to be the simplest form of unsupervised machine-learning, diagonalises the data's covariance matrix and sorts the respective eigenvalues and eigenvectors. 

The covariance matrix of a dataset, computes the covariance between each pairing of the constituent vector elements, defined as:
\begin{equation}
    K_{ij} = E(w_i - E(w_i)) \cdot E(w_j - E(w_j))\;.
\end{equation}
Since our weight entries are within the field of integers the covariance matrix is symmetric. Diagonalising this matrix finds the orthogonal linear combinations of the vector entries which dictate the directions of largest variance.
Therefore the result of this diagonalisation is to identify the data's principal components, which are then sorted in decreasing order according to their respective eigenvalues.
The first component then gives the direction where the data has the most alignment and hence the highest variance, with successive decreasing variance until the final entry gives the direction along which the data has the lowest variance.

The structure of the dataset can then be most easily observed through consideration of these principal components. In this study PCA was applied to each of the datasets under consideration independently. 

In each case the variance eigenvalues were at least 5 times larger for the first principal component compared to the others. In particular for the CY dataset the first principal component was 2 orders of magnitude larger than the others.
This indicates that much of the variation, and hence data structure, is dominated by a single dimension.

Usually a scaling is applied to the data prior to the PCA. 
The 'scaling' process both centres and scales the data such that each entry (i.e. weight) has mean value 0 and standard deviation 1 across the dataset; hence replacing each weight by its respective standardised score.
However for this analysis scaling was not used since the data's general structure is based on the relative sizes between the  weights (which are sorted according to their size).
These relative sizes between weights across each vector are lost through the scaling process, which scales each weight independent of the rest of the vector.

As the data is not scaled one may think that the latter weights of each vector would dominate the behaviour (since the weights are ordered). This would lead the covariance matrix to be near-diagonal, and the principal components would align closely to the original vector entries.
However, as shown by the covariance matrix for the CY dataset in equation \ref{CY_pcainfo}, the matrix is not diagonal and the eigenvectors have significant contribution from multiple components.

\begin{equation}\label{CY_pcainfo}
\scriptsize{
K_{CY} = \begin{pmatrix}
41 & 43 & 109 & 250 & 404 \\
43 & 119 & 278 & 642 & 1017 \\
109 & 278 & 1795 & 3626 & 5562 \\
250 & 642 & 3626 & 8588 & 12941 \\
404 & 1017 & 5562 & 12941 & 20018 
\end{pmatrix},\;
\varepsilon_{CY} = \begin{pmatrix}
0.016 & 0.041 & 0.229 & 0.531 & 0.815 \\
0.021 & 0.036 & -0.973 & 0.100 & 0.205 \\
0.120 & 0.206 & 0.034 & -0.823 & 0.514 \\
0.417 & 0.875 & 0.023 & 0.173 & -0.172 \\
0.900 & -0.435 & 0.003 & 0.018 & -0.008 
\end{pmatrix},\; \lambda_{CY} = \begin{pmatrix}
30071 \\
233 \\
161 \\
74 \\
21
\end{pmatrix},}
\end{equation}
for eigenvectors as rows of $\varepsilon_{CY}$ with respective eigenvalues in $\lambda_{CY}$; where covariance and eigenvalue entries are given to the nearest integer, and eigenvector entries to 3 decimal places.
This implies that the PCA structure is more subtle than a trivial projection.
The covariance matrices, eigenvectors and eigenvalues for the other datasets are provided for comparison in appendix \ref{pca_appendix}.

\begin{figure}[h!]
	\centering
	\begin{subfigure}{0.45\textwidth}
		\centering
		\includegraphics[width=\textwidth]{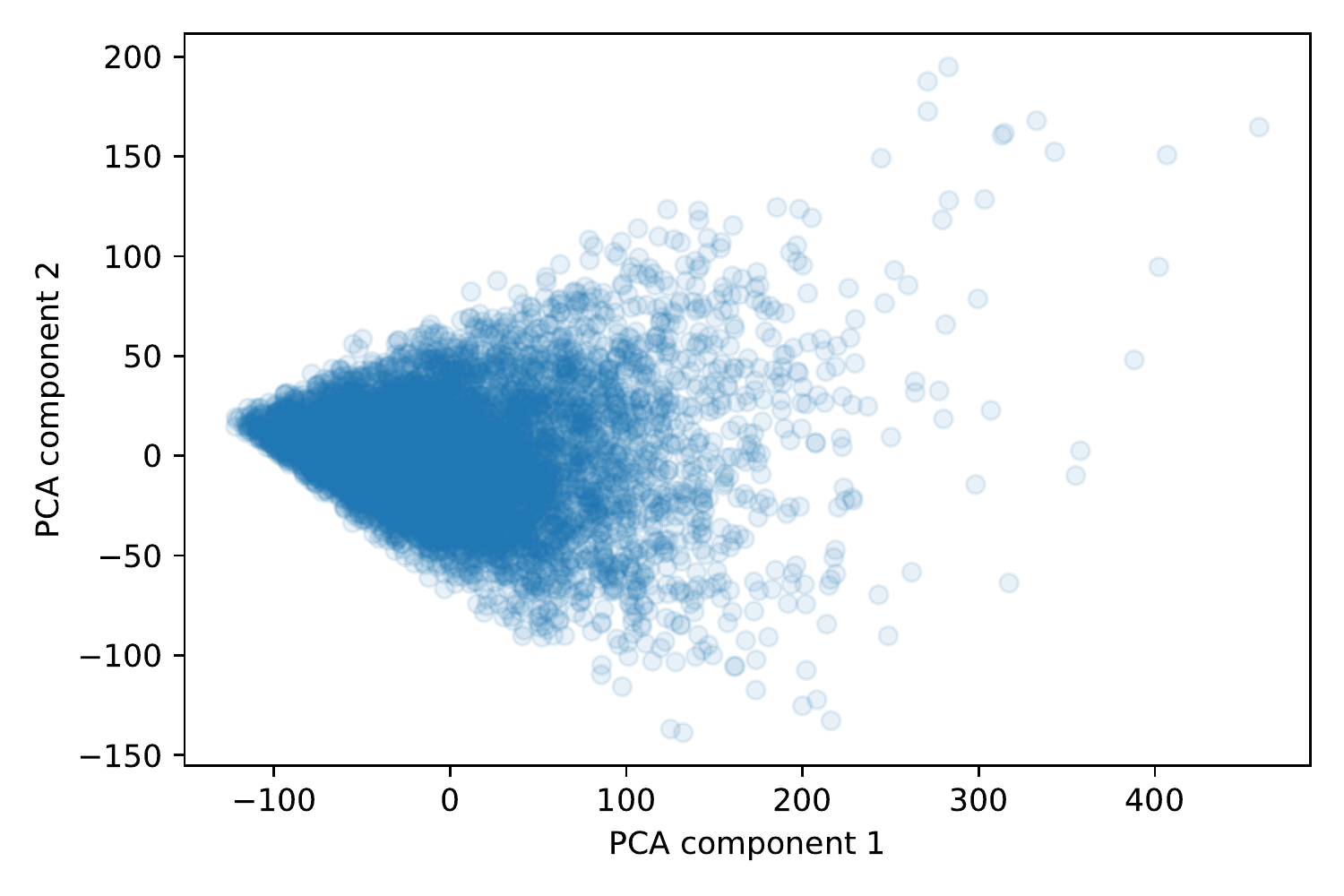}
		\caption{Random Integers}\label{RandomPCA}
	\end{subfigure} 
    \begin{subfigure}{0.45\textwidth}
    	\centering
    	\includegraphics[width=\textwidth]{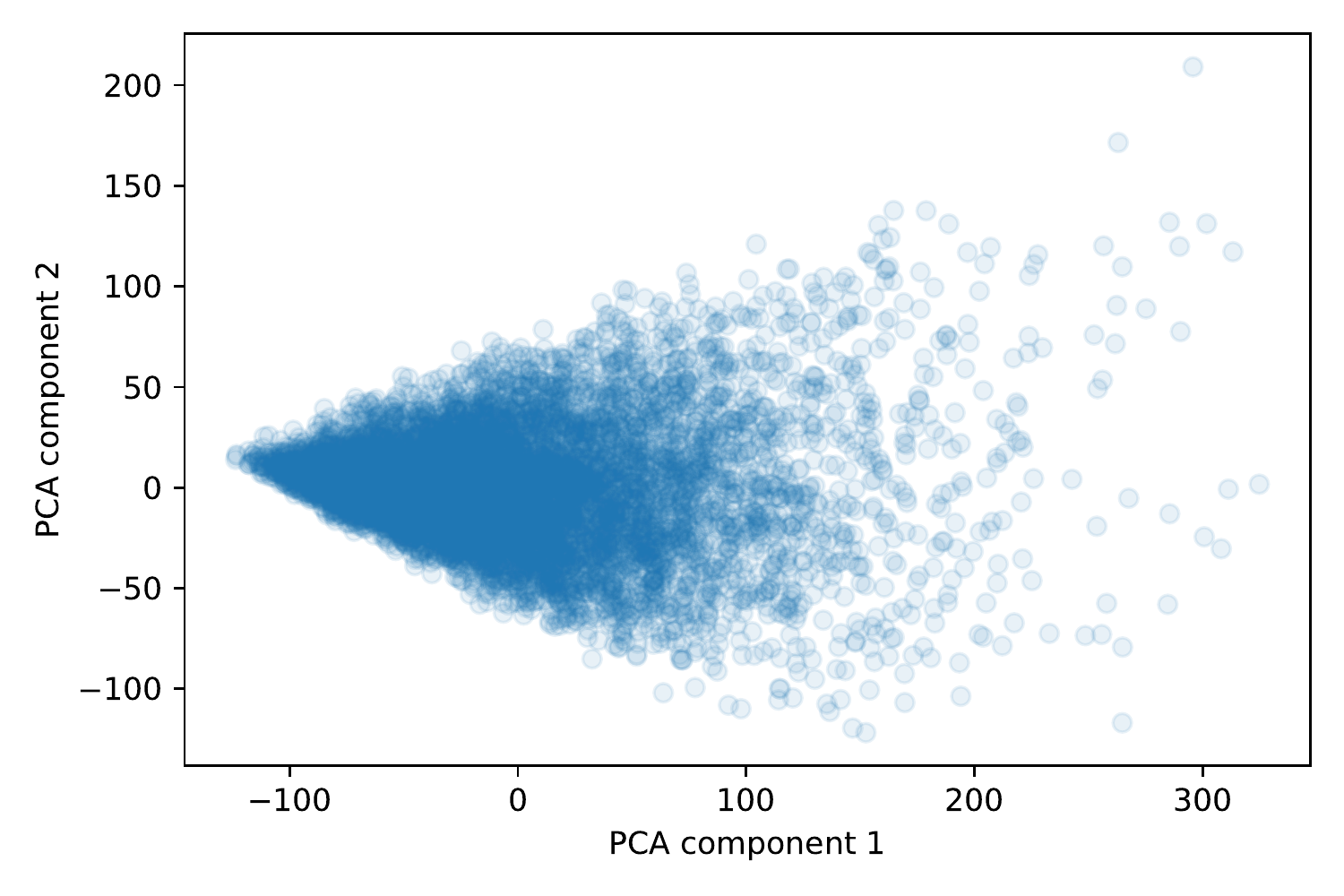}
    	\caption{Random Coprime Integers}\label{CoprimePCA}
    \end{subfigure} \\ 
	\begin{subfigure}{0.45\textwidth}
		\centering
		\includegraphics[width=\textwidth]{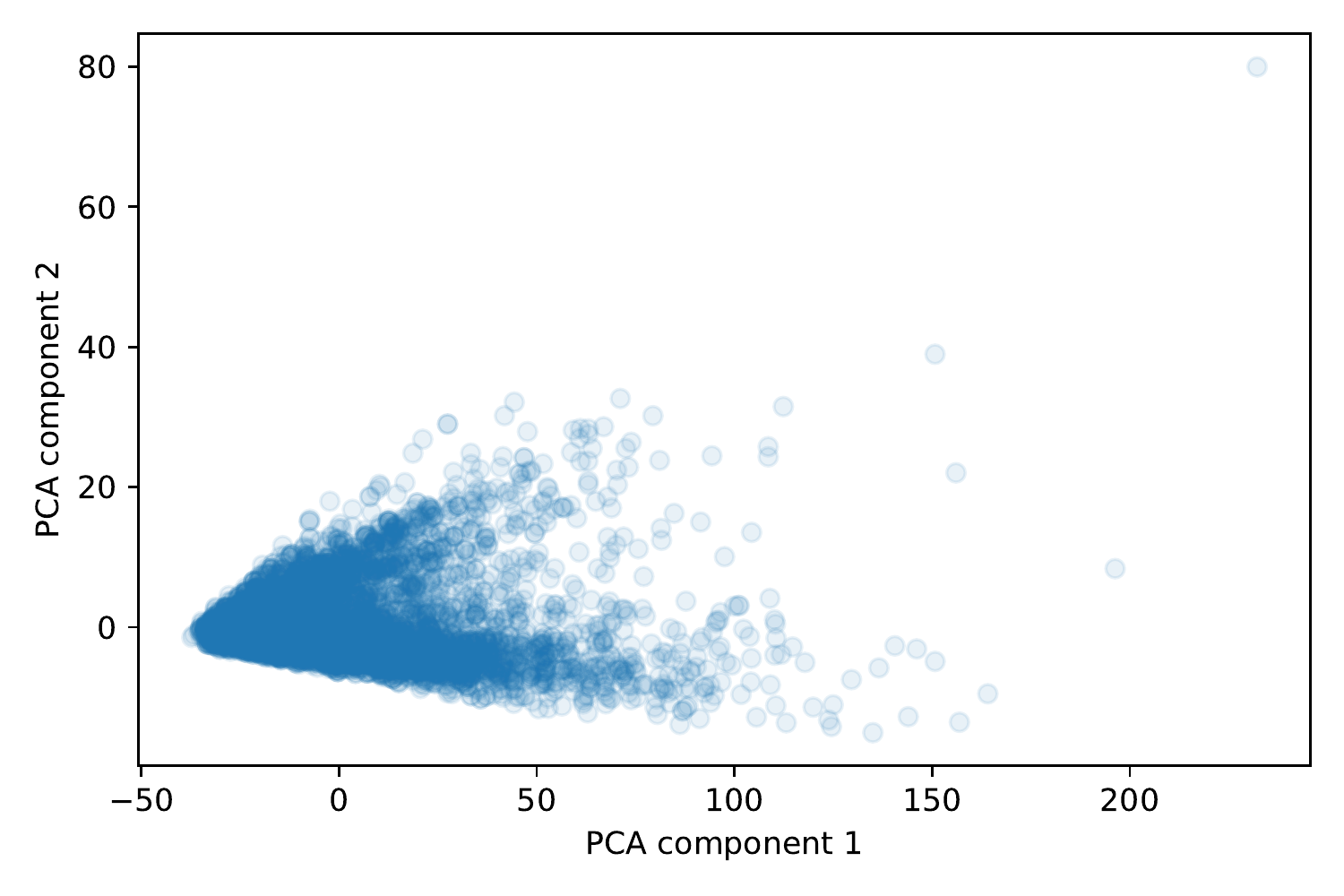}
		\caption{Random Transverse Coprime Integers}\label{TransversePCA}
	\end{subfigure} 
	\begin{subfigure}{0.45\textwidth}
		\centering
		\includegraphics[width=\textwidth]{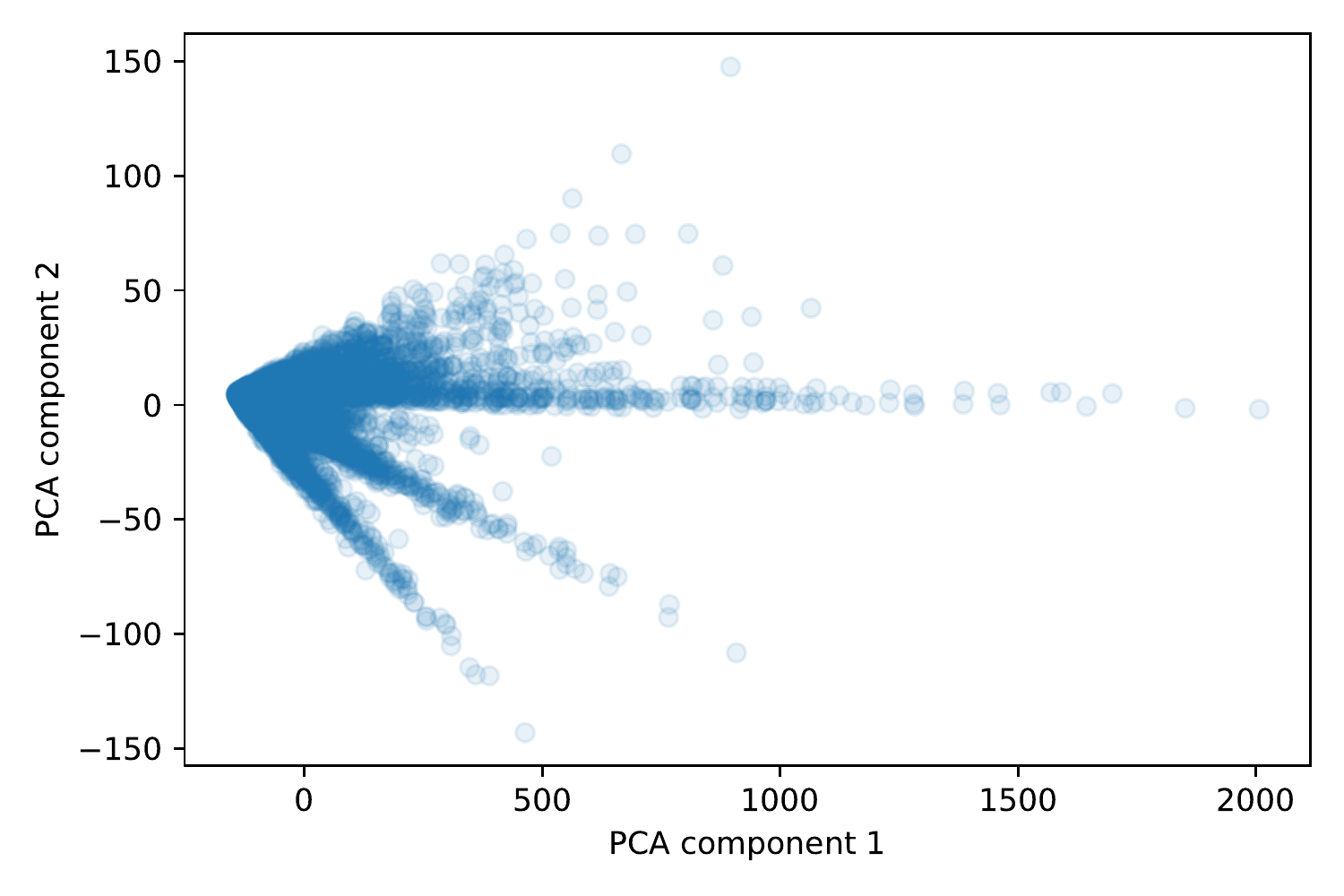}
		\caption{CY Weights}\label{CYPCA}
	\end{subfigure} 
\caption{2d PCA plots for the 4 considered datasets. As more of the conditions are added, more structure appears, in particular there is some form of distinct class separation for the CY weights.}\label{PCADatasets}
\end{figure}

To relatively compare the datasets' PCAs, the normalised vectors of eigenvalues are given in equation \ref{pca_evlaue_norms}, for the random 'R', coprime 'C', transverse 'T', and Calabi-Yau 'CY' datasets respectively. 
They show that the first component significantly dominates, and hence lower dimensional representation of the data through PCA will usefully depict the data's underlying linear structure.
\begin{equation}\label{pca_evlaue_norms}
\footnotesize{
\lambda_{R} = \begin{pmatrix}
0.75534 \\
0.16297 \\
0.05274 \\
0.02059 \\
0.00837
\end{pmatrix},\quad \lambda_{C} = \begin{pmatrix}
0.74845 \\
0.16856 \\
0.05417 \\
0.01997 \\
0.00885
\end{pmatrix},\quad \lambda_{T} = \begin{pmatrix}
0.91388 \\
0.04211 \\
0.02578 \\
0.01334 \\
0.00489
\end{pmatrix},\quad \lambda_{CY} = \begin{pmatrix}
0.98399 \\
0.00764 \\
0.00525 \\
0.00242 \\
0.00070
\end{pmatrix}.}
\end{equation}
Hence for the sake of visualisation, the first 2 components of each datapoint's principal component projection are plotted as a 2-dimensional scatter diagram for each dataset.
These components show the directions with the most variation, and hence display the underlying structure most clearly.
The 2d PCA plots are given in figure \ref{PCADatasets}, for each of the 4 datasets considered.

The cone-like bounding structure of all plots shows the effects of the weight ordering. This is simply that as the first component's value increases (most correlated to the largest, and hence last, weight in the 5-vector) the range of values the second component (roughly correlated to the second-largest / second-last weight) can take increases.
Or put more simply, the second-last weight takes values up to the size of the last weight and so this places cone-like bounds on the plots.
All plots also show higher densities at lower values of the principal components which is also related to this effect.

The PCA plots show that as more of the necessary conditions are added to the datasets, more structure is apparent in the projected outputs.
 First note, the coprime condition causes a negligible change to the distribution of weights.
The transverse condition however has a significant effect. The second components become much more limited and the data begins to separate into approximately two forks.
Most exciting, is the jump to the full Calabi-Yau data. Now the PCA shows a clear clustering of the 5-vectors at higher values of the first principal component.
This distinct separation into clear lines of datapoints shows a rich structure to the weights of Calabi-Yau projective spaces, which is \textbf{not} present for spaces with just the transverse condition.
The reasons for this separation are unclear, however we make conjectural statements about a potential relation to the spaces' Hodge numbers due to a similar structural separation in section \S\ref{hodgeplots}.

A final note is that the PCA used here was explicitly linear, and hence probes the simplest kind of implicit structure. 
More technically involved methods of principal component analysis involve kernel methods, called 'kernel-PCA'. 
Kernel methods were also used to analyse these datasets, for a variety of traditional kernels (including Gaussian, sigmoid, and an array of polynomial kernels), and functionality for this is provided in the respective code scripts.
However, none of these methods produced as distinct a clustering separation as that for the linear kernel.
Indicating, that surprisingly, the most prominent implicit structure of the Calabi-Yau weights takes a linear form.

\subsection{Topological Data Analysis}
Principal Component Analysis allows for a 2d visualisation of the 5d CY data.
Through the PCA with linear kernel, a linear clustering structure was uncovered in the data. To visualise the extension of this behaviour to the full 5d space we turn to tools from topological data analysis; specifically \textit{persistent homology}.

The persistent homology of a dataset is constructed through a filtration of Vietoris-Rips complexes.
The full CY dataset is first plotted in $\mathbb{R}^5$ with each weight a coordinate, such that each weighted-$\mathbb{P}^4$ is now represented by a point (0-simplex) in the $\mathbb{R}^5$ space due to its respective 5-vector.

5-sphere's of radius $d$ are then drawn around each point, and the range of $d$ values are taken from $0 \longmapsto \infty$.
Initially all the spheres will be independent with no overlap, but as $d$ increases the spheres will begin to overlap more frequently.
The complex is then constructed by drawing an $n$-simplex between $n$ points where all their spheres overlap.

Therefore as $d$ increases more simplices are added to the complex, and at each $d$-value where there is a change to the complex we have a stage in the complex's filtration.
The complex hence grows up until a point where all possible simplices lie in the complex. This is where the filtration terminates (no further changes as $d \longmapsto \infty$).

The role of persistent homology in the analysis of this filtration is to examine how long cycles of $n$-simplices last throughout the filtration before they become filled by the $(n+1)$-simplices they bound.
Specifically $H_n$ examines how long cycles of $n$-simplices exist until becoming filled by $(n+1)$-simplices.

\begin{figure}[h!]
    \centering
    \includegraphics[width=0.5\textwidth]{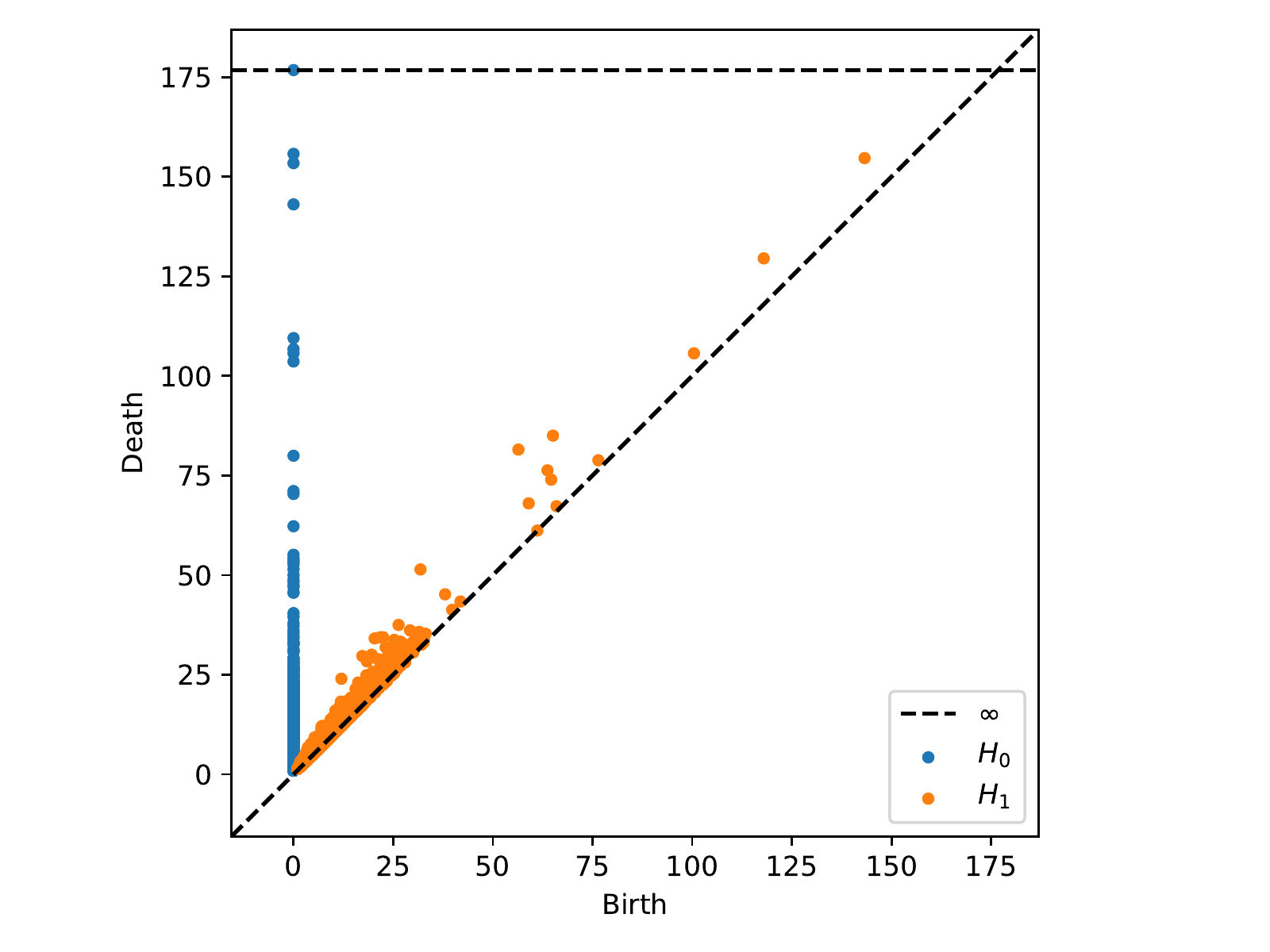}
    \caption{Persistent diagram for the $H_0$ and $H_1$ homology groups of the CY data's Vietoris-Rips complex filtration.}
    \label{CYpersistenthomology}
\end{figure}

This persistent homology for the CY data was computed for $H_0$ and $H_1$ (higher $H_n$ up to $n=4$ can be computed in 5d space but are incredibly computationally expensive in terms of memory for $n \geq 2$).
The persistence diagram for this analysis is shown in figure \ref{CYpersistenthomology}, where the diagram plots all members of $H_0$ and $H_1$ as points with their respective $d$ values of birth (cycle creation) and death (cycle filling).
For specific computation of the persistent homology the python package 'ripser' was used \cite{ctralie2018ripser}; whilst to review previous application of these techniques to the string landscape please see \cite{Cirafici:2015pky,Cole:2018emh}.

As can be seen from the diagram all the members of $H_0$ are blue points born at $d=0$, these are each of the 0-cycles (i.e. 0-simplices / datapoints) that exist until they are connected by an edge (1-simplex) to any other datapoint.
The behaviour shows that there are some datapoints that are significantly far away from the rest of the data and hence join/die much later in the filtration.
These points are those with large weight values such that they are far from the origin in the $\mathbb{R}^5$ embedding.

Conversely all members of $H_1$ are points in orange, and as expected all these 1-cycles (i.e. cycles of 1-simplices/edges which are not boundaries of 2-simplices/triangles) lie close to the diagonal line in the persistence diagram.
This behaviour indicates a short life of each cycle, a behaviour typical of noise in the dataset.
Since traditionally it is only points far from the diagonal that indicate significant persistent structure, there is hence not higher dimensional structure formation or non-trivial topology in the data which would deter from the linear clustering behaviour seen through the PCA.

\subsection{Analysis of CY Topological Properties}\label{hodgeplots}
In addition to the weights used to represent these Calabi-Yau weighted projective spaces, the non-trivial Hodge numbers, $\{h^{1,1},h^{2,1}\}$, for the surfaces are also provided with the KS databases \cite{Kreuzer:2000xy}, and repeated with this study's GitHub.

\begin{figure}[h!]
	\centering
	\begin{subfigure}{0.45\textwidth}
		\centering
		\includegraphics[width=\textwidth]{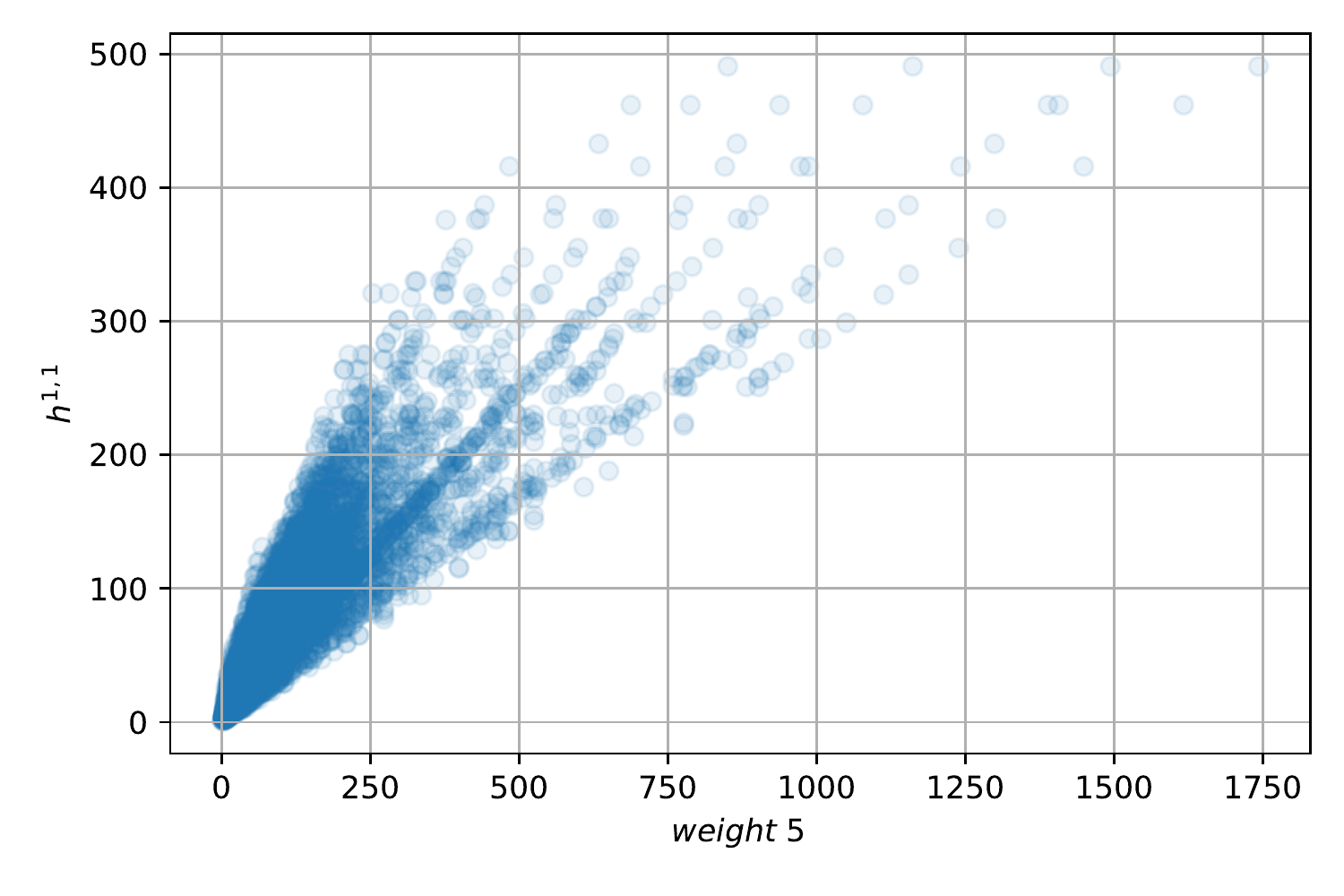}
		\caption{}\label{w5_h11}
	\end{subfigure} 
    \begin{subfigure}{0.45\textwidth}
    	\centering
    	\includegraphics[width=\textwidth]{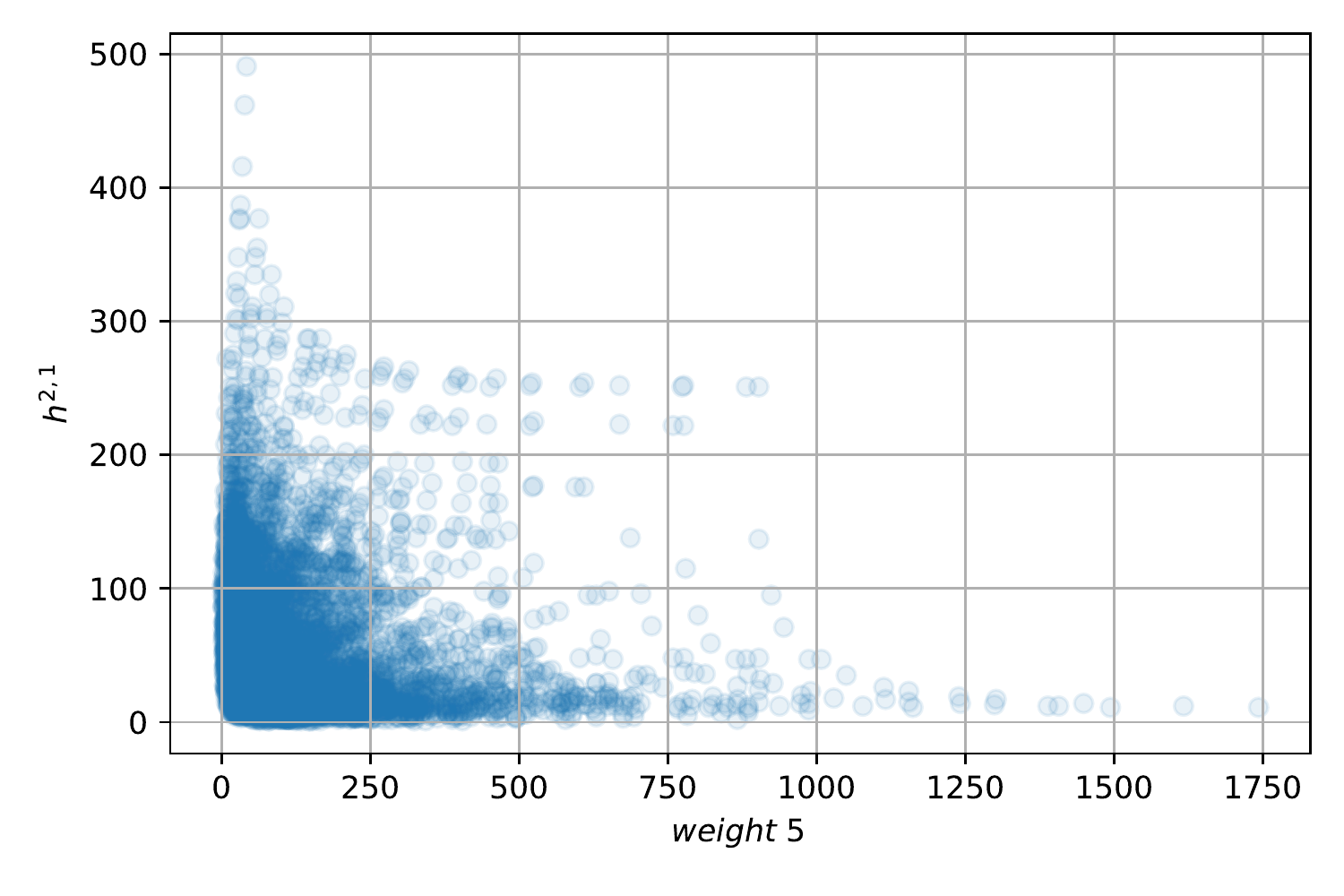}
    	\caption{}\label{w5_h21}
    \end{subfigure}
\caption{Distribution of Calabi-Yau weighted projective spaces, according to their final (and largest) weight and (a) $h^{1,1}$ or (b) $h^{2,1}$ respectively.}\label{w5_hodges}
\end{figure}

This provides more information for analysis the spectrum of CY weights. 
Simple plotting of these weights produces an astonishingly familiar structure, one which is exemplified best when the CY's Hodge numbers are plotted against the final (and hence largest) weight, as shown in figure \ref{w5_hodges}.

The behaviour in figure \ref{w5_h11} shows a similar form of fork-like splitting of the datapoints as in the PCA of figure \ref{CYPCA}, even with a central fork particularly more dominant than the others.
This seemingly linear behaviour between final weight and $h^{1,1}$ is quite surprising, and here again the CY hypersurfaces appear to be separating themselves into classes, according to the ratio of $h^{1,1}$ to the final weight, $w_5$.
On the contrary, the behaviour in figure \ref{w5_h21}, follows the familiar mirror symmetry plot \cite{CANDELAS1990383}, complimenting the linear behaviour with $h^{1,1}$ such that their combination will preserve this structure.

Similar behaviour also occurs for the other weights in the 5-vectors, despite less obvious clustering. 
Plots of these relations are given in appendix \ref{ExtraHodgeWeightPlots}.

\begin{figure}[h!]
    \centering
    \includegraphics[width=0.5\textwidth]{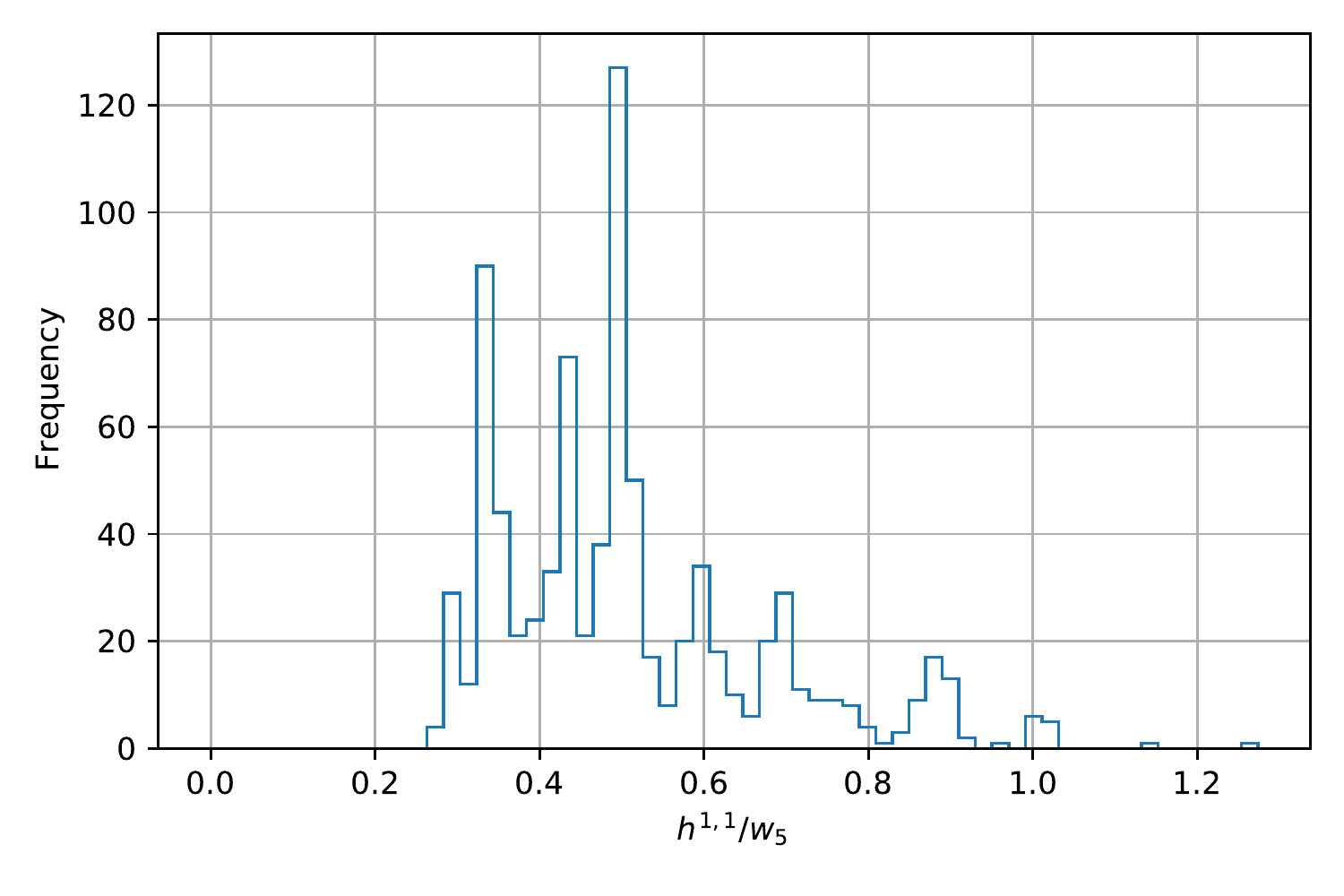}
    \caption{Frequency of the ratio between $h^{1,1}$ and the largest weight, $w_5$, for the CY data with $w_5>250$ (where structure more prominent). Peaks indicate a natural clustering.}
    \label{h11_w5_ratio_histogram}
\end{figure}

To further examine this clustering phenomena we plot a histogram of the ratio $h^{1,1}/w_5$ in figure \ref{h11_w5_ratio_histogram}.
Note for this plot only datapoints with $w_5 > 250$ were used since this was where the class separation was more prominent such that the cluster identification would be improved.
As can be seen from the peaks in the figure, there is a clear clustering behaviour. 
Therefore we reexamine this data of ratios with the use of K-Means clustering.

\subsubsection{Clustering for $h^{1,1}$ Classes}
As motivated by the formation of a set of linear relationships between $w_5$ and $h^{1,1}$ shown in figure \ref{w5_h11}, and the peak occurrence in the histogram of ratios in figure \ref{h11_w5_ratio_histogram}, unsupervised clustering methods were used to examine this behaviour.

The 'outer' ratio data used to produce the histogram plot, where clustering was more prominent, provides a very suitable database for 1-dimensional clustering.
The method used was K-Means clustering, which takes an input predefined number of clusters, initialises mean values for each cluster, and iteratively updates these means such that the final sum of squared distances from each datapoint to its nearest cluster's mean is minimised.
This measure is known as the K-Means inertia,
\begin{equation}
    \mathscr{I} = \sum_\mathscr{C}\sum_{i \in \mathscr{C}}(\mu_\mathscr{C}-r_i)^2\;
\end{equation}
for clusters, $\mathscr{C}$, with respective means, $\mu_\mathscr{C}$, and all datapoints, $i$, exclusively in their nearest cluster with ratios, $r_i$.

\begin{figure}[h!]
    \centering
    \includegraphics[width=0.5\textwidth]{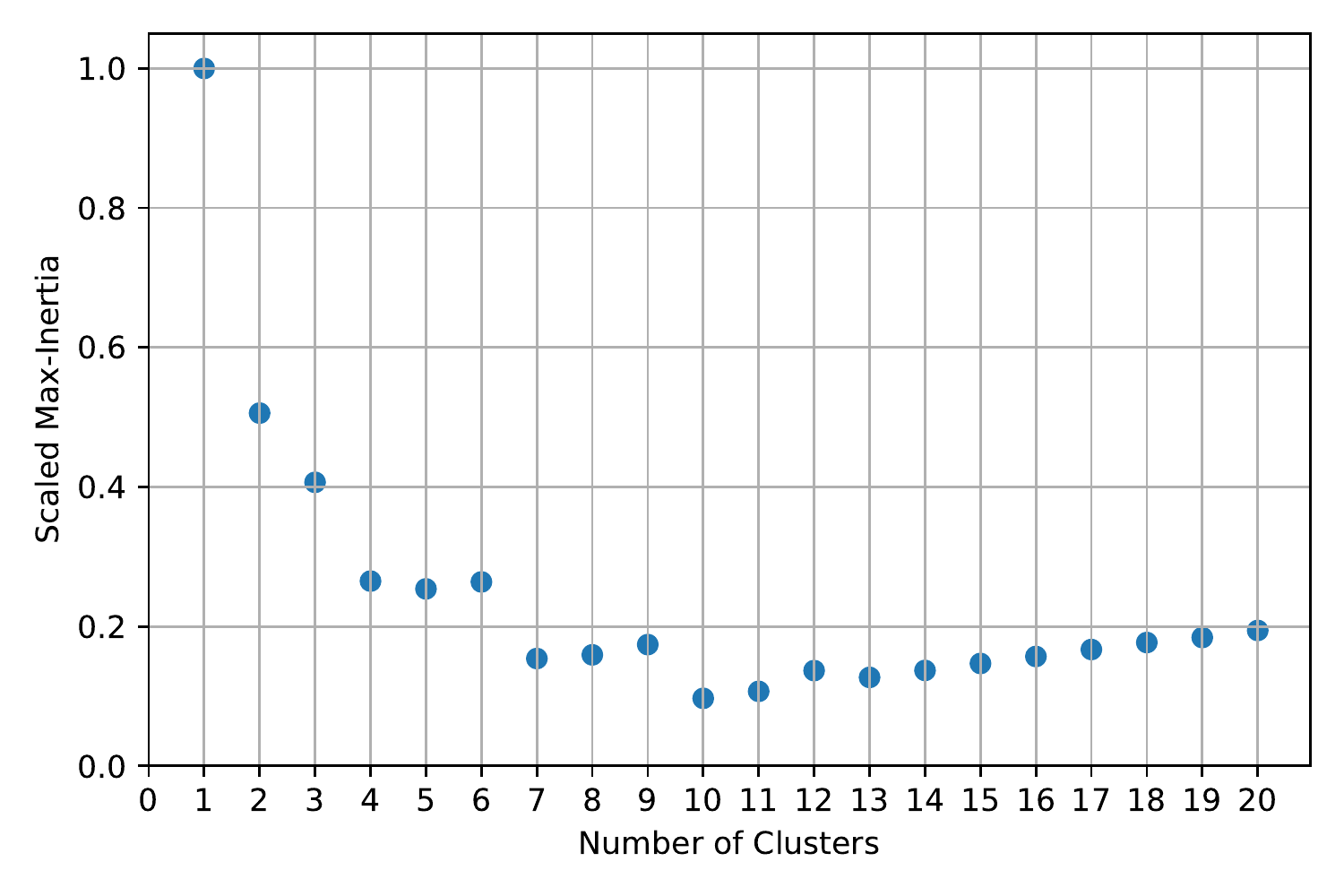}
    \caption{Plot of Scaled Max-Inertia as the number of clusters used for K-Means clustering varies. The minimum identifies an optimum number of clusters: 10.}
    \label{scaledmaxinertia}
\end{figure}

Determining the optimal number of clusters to use is a standard problem in K-Means clustering, to motivate this choice we use a novel measure we call 'scaled-max-inertia'.
This measure identifies the maximum squared-distance any point is from its closest cluster centre, normalises it according to that maximum squared-distance from using only one cluster, and adds a weight factor to penalise using an excessive number of clusters.
We define this to be:
\begin{equation}
    \mathscr{I}_{max} = \frac{\text{Max}_i(\mu_\mathscr{C}-r_i)^2}{\text{Max}_i(\mu_1-r_i)^2} + \frac{(k-1)}{100}\;,
\end{equation}
where $\text{Max}_i$ determines the maximum over all ratios, $r_i$, examining the squared distance to either the closest cluster's mean, $\mu_\mathscr{C}$ or the single cluster's mean, $\mu_1$; then weighting by the number of clusters, $k$.
A plot of scaled-max-inertia against number of clusters identifies an optimum of 10 clusters, as shown in figure \ref{scaledmaxinertia}.

\begin{figure}[h!]
    \centering
    \includegraphics[width=0.5\textwidth]{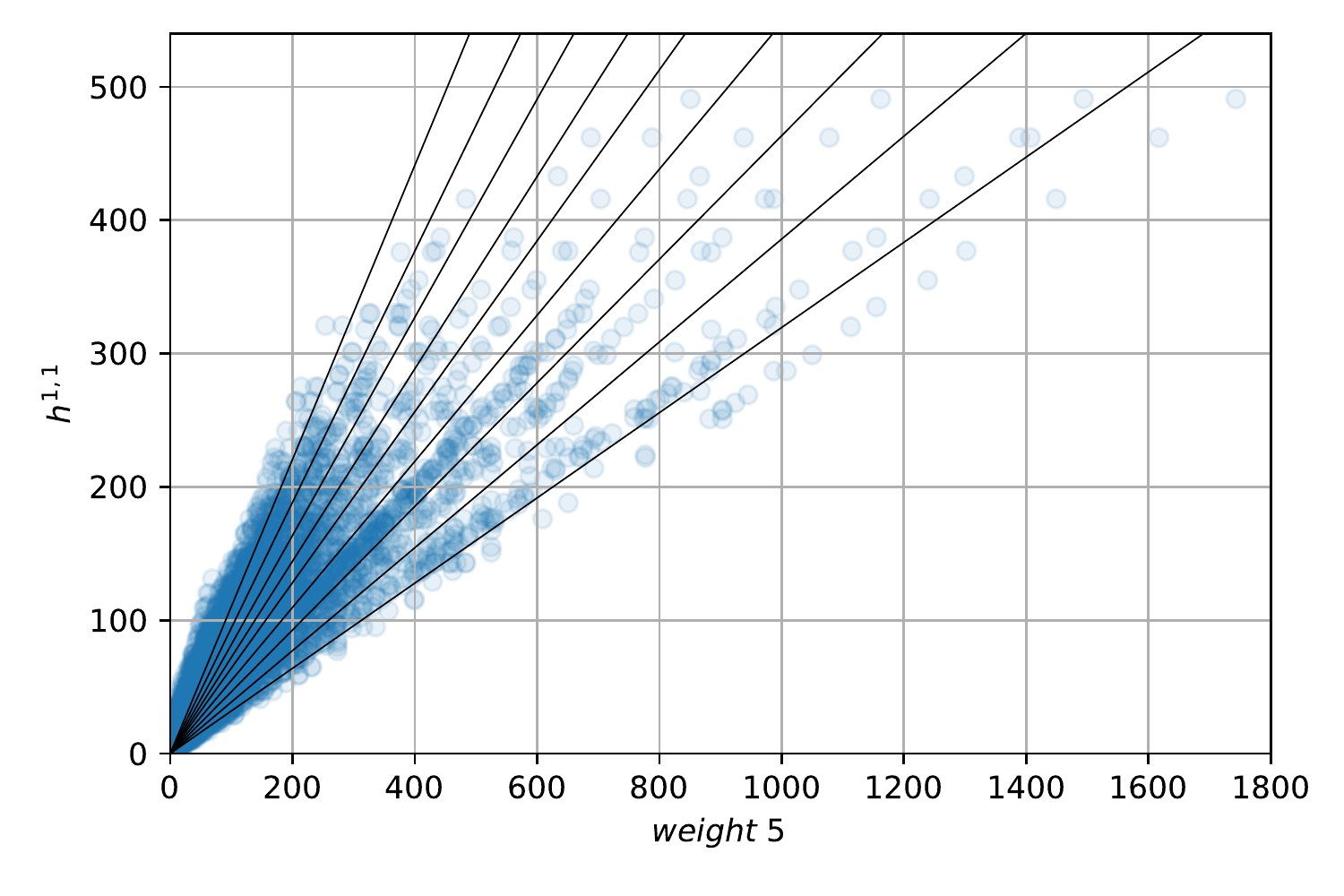}
    \caption{Plot of the bounds of the 10 clusters produced on the outer data ($w_5>250)$ via K-Means clustering.}
    \label{clustbounds}
\end{figure}

Using the optimal number of 10 clusters, the separation matches up exceptionally for the outer data, as shown by plots of the cluster bounds in figure \ref{clustbounds}.
The clusters sizes for the clusters moving anticlockwise about the plot, for increasing ratio, are: $[103,354,454,734,626,623,643,895,1419,1704]$, highlighting that there is a greater density of points at low $w_5$ as expected, since this was why 'outer' data was focused on for clustering.

To measure the clustering performance we use the standard Inertia measure over the \textit{full} dataset, however normalised by the number of ratios across the dataset, $\hat{\mathscr{I}}$, and an equivalent measure also normalised by the range of the ratios:
\begin{equation}
    \hat{\mathscr{I}} = 0.0266\,,\quad \frac{\hat{\mathscr{I}}}{max(r_i)-min(r_i)} = 0.00084\,,
\end{equation}
These values show that clustering performed exceptionally well, as each ratio in the full CY dataset was less than 0.1\% of the ratio-range away from its nearest cluster.
Therefore confirming the distinct \textit{linear} behaviour observed, as well as the class separation.
The distinct classes of CY 5-vectors are also provided in the GitHub.

\section{Machine Learning}\label{ml}
After use of unsupervised ML methods in section \S\ref{data}, we now turn to use of supervised ML methods for learning of the topological properties, as well as the CY property.

\subsection{Architectures}\label{architectures}
The problems addressed by supervised ML in this study fit into both of the field's typical styles: regression, and classification.

The first set of problems learnt in section \S\ref{mltopo} learn the topological Hodge numbers (and related Euler number) from the CY 5-vectors of weights.
Since the output Hodge numbers can take a large range of integer values the problem was formulated as a regression problem.
For this a Multi-Layer Perceptron Regressor was used to learn each output from the input weights. 
This regressor is a type of Neural Network, and the one used specifically had layer sizes of [32,64,32], with ReLU activation, and used the \textit{Adam} \cite{kingma2017adam} optimiser method to minimise a mean-squared error loss.
The fitting used a batchsize of 200, and ran up to 200 epochs until the tolerance of 0.0001 was reached for loss updating.

The second set of problems considered in section \S\ref{mlCY} sort to determine which dataset a 5-vector belonged to, either by binary classification between each dataset and the CY dataset, or a multiclassification among all 4 datasets.
Since these were classification problems an array of different classifiers were used to perform the learning.

The first classifier was a Logistic Regressor, as perhaps the simplest form of classifier.
This Logistic Regressor had a tolerance of 1 for learning the weight behaviour, a C-value of 100 such that there was a low amount of regularisation, and used Newtons method for solving, such that multiclassification could also be performed.
The second classifier was a Support Vector Machine with a simple linear kernel, and here a higher regularisation due to a C-value 1.
The third and final classifier used was a Neural Network Classifier (Multi-Layer Perceptron also), this time with the same hyperparameters as the Regressor except now with a cross-entropy loss function.

\subsubsection{Measures}
To assess learning performance consistent measures are required. 
For this, dependent on the problem being a regression or classification, different measures were selected as follows.

\paragraph{Regressors:} The most standard regressor measure is Mean-Squared Error, \textit{MSE}, which was used for the regressor loss function.
However MSE should be considered in relation to the square of the range of output values to be useful, hence a preferable measure also used was Mean-Absolute-Percentage Error, \textit{MAPE}.
Although it should be noted MAPE has its own drawbacks where it is incalculable when $y_{true}=0$ for any of the date inputs. 
Both these measures are unbounded above and take optimal value of 0 which indicates perfect prediction.

The final regressor measure used was $R^2$, this evaluates how well a regressor is performing by comparing the proximity of the predicted output to the proximity of the mean (which would be the prediction for a null model regressor).
For this measure 1 is optimal, 0 means that prediction is not better than just predicting the true mean each time, and $<0$ means worse than just predicting the mean.
The equations and output bounds for these measures are given in equation \ref{regressormeasures}.
\begin{equation}\label{regressormeasures}
\begin{split}
    MSE & = \frac{1}{n}\sum(y_{pred}-y_{true})^2 \qquad\ \in [0,\infty)\;,\\
    MAPE & = \frac{1}{n}\sum \bigg|\frac{y_{pred}-y_{true}}{y_{true}}\bigg| \qquad\ \in [0,\infty)\;,\\
    R^2 & = 1-\frac{\sum (y_{true}-y_{pred})^2}{\sum (y_{true}-y_{true mean})^2} \in (-\infty,1]\;,
\end{split}
\end{equation}
summing over all predicted, $y_{pred}$, and true, $y_{true}$, outputs in the test data. In addition for $R^2$ the mean of the true values over the test data outputs, $y_{truemean}$, was also used.

\paragraph{Classifiers:} Trained classifiers predict on input test data by assigning them to classes, this leads to a natural sorting of true (row) vs predicted (column) class frequencies over all the test data, arranged into a confusion matrix, \textit{CM}.
From the confusion matrix the normalised sum over the diagonal gives the \textit{Accuracy} which is the proportion of test data correctly classified.
However simple accuracy has problems associated to bias data, therefore a better measure of learning is Matthew's Correlation Coefficient, \textit{MCC}.
Both these measures have optimum learning with values of 1, where all test data inputs are allocated to their true class.
Equations for these two measures used are given in equation \ref{classifiermeasures}.
\begin{equation}\label{classifiermeasures}
\begin{split}
    CM & = \begin{pmatrix} TP & FN \\ FP & TN \end{pmatrix}\,,\\
    Accuracy & = \frac{TP + TN}{TP + TN + FP + FN} \in [0,1]\,,\\
    MCC & = \frac{TP\cdot TN - FP \cdot FN}{\sqrt{(TP + FP)\cdot (TP + FN) \cdot (TN + FP) \cdot (TN + FN)}} \in [-1,1]\,,
\end{split}
\end{equation}
for the binary classification case, where generalisations exist for the multiclassification case.

For all problems 5-fold cross-validation was used, whereby 5 independent versions of each architecture were trained and tested on 5 different train:test partitions of the data, and the learning measures then averaged and standard error computed.

\subsection{ML Topological Parameters}\label{mltopo}
Topological parameters provide key information about a Calabi-Yau manifold which are essential in the computation of physical phenomena when these manifolds are used for superstring compactifications.

This CY subset from weighted $\mathbb{P}^4$s provides a simple scenario whereby the Hodge numbers (and hence Euler number) can be computed directly from the weights of the toric space that the Calabi-Yau is a hypersurface of.
Although it should be noted these formulas are quite non-trivial, as discussed in section \S\ref{s:CY}.

Both of these formulas, given in equation \ref{hodgeeulerformulas}, require greatest common divisor computations throughout their evaluation. Machine-learning methods famously perform badly when approximating these styles of equations and so one would expect the simple Neural Network architecture used here to not be particularly successful.

The results for the machine-learning of the non-trivial Hodge numbers, and the Euler number are given in table \ref{weights_to_hodgeeuler}.
The Hodge number data, provided by \cite{Kreuzer:2000xy}, is also made available on the GitHub with the Calabi-Yau weight data, and from here the Euler numbers can be calculated using $\chi = 2(h^{1,1}-h^{2,1})$.

\begin{table}[h!]
\centering
\begin{tabular}{|c|cccc|}
\hline
\multirow{2}{*}{Measure} & \multicolumn{4}{c|}{Property} \\ \cline{2-5} & \multicolumn{1}{c|}{$h^{1,1}$} & \multicolumn{1}{c|}{$h^{2,1}$} & \multicolumn{1}{c|}{$[h^{1,1},h^{2,1}]$} & $\chi$ \\ \hline
$R^2$                       & \multicolumn{1}{c|}{\begin{tabular}[c]{@{}c@{}}0.9630\\ $\pm$ 0.0015\end{tabular}} & \multicolumn{1}{c|}{\begin{tabular}[c]{@{}c@{}}0.9450\\ $\pm$ 0.0133\end{tabular}} & \multicolumn{1}{c|}{\begin{tabular}[c]{@{}c@{}}0.9470\\ $\pm$ 0.0041\end{tabular}} & \begin{tabular}[c]{@{}c@{}}0.9510\\ $\pm$ 0.0023\end{tabular} \\ \hline
MAPE                     & \multicolumn{1}{c|}{\begin{tabular}[c]{@{}c@{}}0.1493\\ $\pm$ 0.0027\end{tabular}} & \multicolumn{1}{c|}{\begin{tabular}[c]{@{}c@{}}0.2519\\ $\pm$ 0.0152\end{tabular}} & \multicolumn{1}{c|}{\begin{tabular}[c]{@{}c@{}}0.2375\\ $\pm$ 0.018\end{tabular}}  & -                                                          \\ \hline
MSE                      & \multicolumn{1}{c|}{\begin{tabular}[c]{@{}c@{}}166.9\\ $\pm$ 10.0\end{tabular}}    & \multicolumn{1}{c|}{\begin{tabular}[c]{@{}c@{}}147.0\\ $\pm$ 35.6\end{tabular}}    & \multicolumn{1}{c|}{\begin{tabular}[c]{@{}c@{}}186.9\\ $\pm$13.9\end{tabular}}     & \begin{tabular}[c]{@{}c@{}}1746.1\\ $\pm$ 82.4\end{tabular}     \\ \hline
\end{tabular}
\caption{Learning each of the topological parameters from the Calabi-Yau 5-vectors of weights. Note the final column is Euler number $\chi = 2(h^{1,1}-h^{2,1})$, and since it can evaluate to 0 its MAPE value is not defined. Measurement of learning performance uses 5-fold cross-validation to provide an average and standard error on each measure's value.}
\label{weights_to_hodgeeuler}
\end{table}

The results show a surprisingly successful predictive ability for the Hodge numbers and Euler number, particularly with $R^2$ values exceeding 0.9.
The MAPE values show the Hodge numbers are consistently predicted to be only around 20\% off from their true values, whilst the MSE values provide a less physical measure of learning but are included for reference since they were used as the regressor loss.

Considering the complexity of the equation forms in equation \ref{hodgeeulerformulas}, it is impressive the Neural Network can learn any correlating behaviour for computation of Hodge numbers or Euler number from the weights alone.
In addition, the relatively better performance in learning $h^{1,1}$ may be due to the apparent linear relationship to the weights as exemplified in section \S\ref{hodgeplots}.

\subsection{ML CY Property}\label{mlCY}
The conditions for a 5-vector of weights to represent a weighted projective space which can admit a Calabi-Yau hypersurface are highly non-trivial.
As discussed in section \S\ref{datasets}, the necessary conditions of coprimality and transversity are probed through generation of equivalent datasets, with which the CY dataset can be compared.

Due to the exponential generation techniques making these weights more representative, differentiating which dataset a 5-vector belongs to is not possible by eye.
Therefore it is natural to wish to consider the effectiveness of machine-learning to this classification problem: learning the Calabi-Yau nature.

Introduced in section \S\ref{architectures}, three architectures were used to learn to differentiate the Calabi-Yau weights from each of the other datasets: random integers, coprime random integers, and transverse coprime random integers in binary classification problems.
Furthermore they were also used to differentiate all 4 datasets in a multiclassification problem.

Results for this learning are given in table \ref{mlcyresults}.
Measures show that Neural Networks can well differentiate the Calabi-Yau weights from each of the other datasets.  
As expected there is minimal difference due to introduction of coprimality, since this is a common behaviour for 5-vectors as mentioned in section \S\ref{datasets}.
Once transversity was included into the dataset, the binary classification performance dropped. 
However performance was still surprisingly good. 

A further surprise was the equally good performance of the Logistic Regressor and Support Vector Machine.
These simple architectures could accurately classify approximately three-quarters of the data even without using transversity (where this condition was in both CY and compared dataset). 

\begin{table}[h!]
\centering
\begin{tabular}{|c|c|cccc|}
\hline
\multirow{2}{*}{Architecture} & \multirow{2}{*}{Measure} & \multicolumn{4}{c|}{Dataset} \\ \cline{3-6} & & \multicolumn{1}{c|}{Random} & \multicolumn{1}{c|}{Coprime} & \multicolumn{1}{c|}{Transverse} & All \\ \hline
\multirow{2}{*}{\begin{tabular}[c]{@{}c@{}}Logistic\\ Regressor\end{tabular}}       & Accuracy                 & \multicolumn{1}{c|}{\begin{tabular}[c]{@{}c@{}}0.7152\\ $\pm$ 0.0035\end{tabular}} & \multicolumn{1}{c|}{\begin{tabular}[c]{@{}c@{}}0.7199\\ $\pm$ 0.0037\end{tabular}} & \multicolumn{1}{c|}{\begin{tabular}[c]{@{}c@{}}0.7430\\ $\pm$ 0.0065\end{tabular}} & \begin{tabular}[c]{@{}c@{}}0.4825\\ $\pm$ 0.0035\end{tabular} \\ \cline{2-6} & MCC                      & \multicolumn{1}{c|}{\begin{tabular}[c]{@{}c@{}}0.4352\\ $\pm$ 0.0065\end{tabular}} & \multicolumn{1}{c|}{\begin{tabular}[c]{@{}c@{}}0.4467\\ $\pm$ 0.0073\end{tabular}} & \multicolumn{1}{c|}{\begin{tabular}[c]{@{}c@{}}0.5003\\ $\pm$ 0.0121\end{tabular}} & \begin{tabular}[c]{@{}c@{}}0.3141\\ $\pm$ 0.0043\end{tabular} \\ \hline
\multirow{2}{*}{\begin{tabular}[c]{@{}c@{}}Support\\ Vector\\ Machine\end{tabular}} & Accuracy                 & \multicolumn{1}{c|}{\begin{tabular}[c]{@{}c@{}}0.7253\\ $\pm$ 0.0029\end{tabular}} & \multicolumn{1}{c|}{\begin{tabular}[c]{@{}c@{}}0.7116\\ $\pm$ 0.0029\end{tabular}}         & \multicolumn{1}{c|}{\begin{tabular}[c]{@{}c@{}}0.7464\\ $\pm$ 0.0014\end{tabular}}         & \begin{tabular}[c]{@{}c@{}}0.4732\\ $\pm$ 0.0070\end{tabular}         \\ \cline{2-6} & MCC                      & \multicolumn{1}{c|}{\begin{tabular}[c]{@{}c@{}}0.4605\\ $\pm$ 0.0054\end{tabular}} & \multicolumn{1}{c|}{\begin{tabular}[c]{@{}c@{}}0.4374\\ $\pm$ 0.0054\end{tabular}}         & \multicolumn{1}{c|}{\begin{tabular}[c]{@{}c@{}}0.5174\\ $\pm$ 0.0029\end{tabular}}         & \begin{tabular}[c]{@{}c@{}}0.3060\\ $\pm$ 0.0078\end{tabular}         \\ \hline
\multirow{2}{*}{\begin{tabular}[c]{@{}c@{}}Neural\\ Network\end{tabular}} & Accuracy                 & \multicolumn{1}{c|}{\begin{tabular}[c]{@{}c@{}}0.9189\\ $\pm$ 0.0037\end{tabular}} & \multicolumn{1}{c|}{\begin{tabular}[c]{@{}c@{}}0.9178\\ $\pm$ 0.0030\end{tabular}} & \multicolumn{1}{c|}{\begin{tabular}[c]{@{}c@{}}0.7575\\ $\pm$ 0.0024\end{tabular}} & \begin{tabular}[c]{@{}c@{}}0.5881\\ $\pm$ 0.0048\end{tabular} \\ \cline{2-6} & MCC                      & \multicolumn{1}{c|}{\begin{tabular}[c]{@{}c@{}}0.8380\\ $\pm$ 0.0073\end{tabular}} & \multicolumn{1}{c|}{\begin{tabular}[c]{@{}c@{}}0.8377\\ $\pm$ 0.0056\end{tabular}} & \multicolumn{1}{c|}{\begin{tabular}[c]{@{}c@{}}0.5306\\ $\pm$ 0.0059\end{tabular}} & \begin{tabular}[c]{@{}c@{}}0.4615\\ $\pm$ 0.0072\end{tabular} \\ \hline
\end{tabular}
\caption{Machine-learning results for three different architectures performing binary classification between the CY data and each specified dataset; and in addition multiclassification across all 4 datasets (labelled 'All'). Learning is measured using Accuracy and MCC with 5-fold cross-validation to provide an average and standard error on each measure's value.}
\label{mlcyresults}
\end{table}

Multiclassification of all datasets was not as strong. However within these measures the identification of the Calabi-Yau data was considerably better, with most of the performance reduction due to misclassifying between random, coprimality, and transversity.
To exemplify this we give a sample confusion matrix for the multiclassification with the Logistic Regressor:
\begin{equation}
\footnotesize{
    CM_{LR} = \begin{pmatrix}
    0.116 & 0.013 & 0.029 & 0.091  \\
    0.076 & 0.083 & 0.074 & 0.020 \\
    0.074 & 0.078 & 0.062 & 0.019 \\
    0.026 & 0.004 & 0.008 & 0.228
    \end{pmatrix}}\;,
\end{equation}
where row indicates true class and column predicted class for each of: random, coprime, transverse, CY respectively.
The final entry shows nearly all the Calabi-Yau data is correctly classified (0.25 indicates the full quarter of the accumulated datasets).
Therefore measures will indicate lower performance where the other conditions cannot be differentiated, and it is likely that these conditions are not the most prominent conditions to indicate the Calabi-Yau property.

To further examine the learning performance we next look explicitly at the misclassifications of the Calabi-Yau data, using again links to the Hodge numbers to identify areas of difficulty. 

\subsubsection{Misclassification Analysis with Hodge Numbers}\label{misclassifications}
Since the Logistic Regressor performed comparably to the other architectures, and is a significantly simpler architecture than the neural network, its use for misclassification analysis seemed the most appropriate.

Due to the simple structure, only 50 5-vectors in each non-CY dataset were used to train the regressor with another 50 CY 5-vectors. 
The regressor was then used to predict the class of all the CY data, producing accuracies of: 78\%, 81\%, 61\% when trained with each of the random, coprime and transverse datasets respectively.

\begin{figure}[h!]
	\centering
	\begin{subfigure}{0.45\textwidth}
		\centering
		\includegraphics[width=\textwidth]{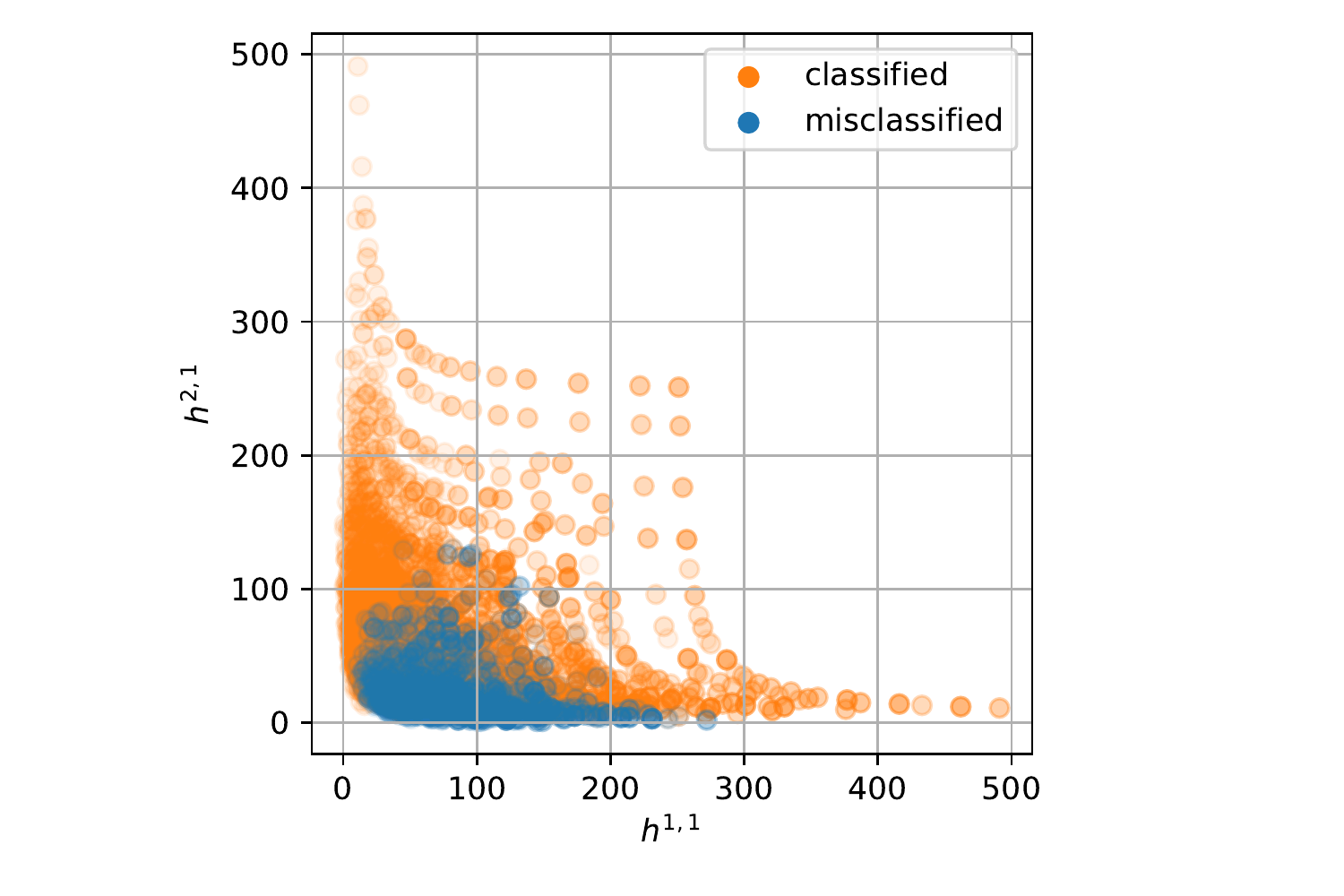}
		\vspace{-0.8cm}
		\caption{Random Integers \\ (1899 misclassified)}\label{RandomMisclassify}
	\end{subfigure} 
    \begin{subfigure}{0.45\textwidth}
    	\centering
    	\includegraphics[width=\textwidth]{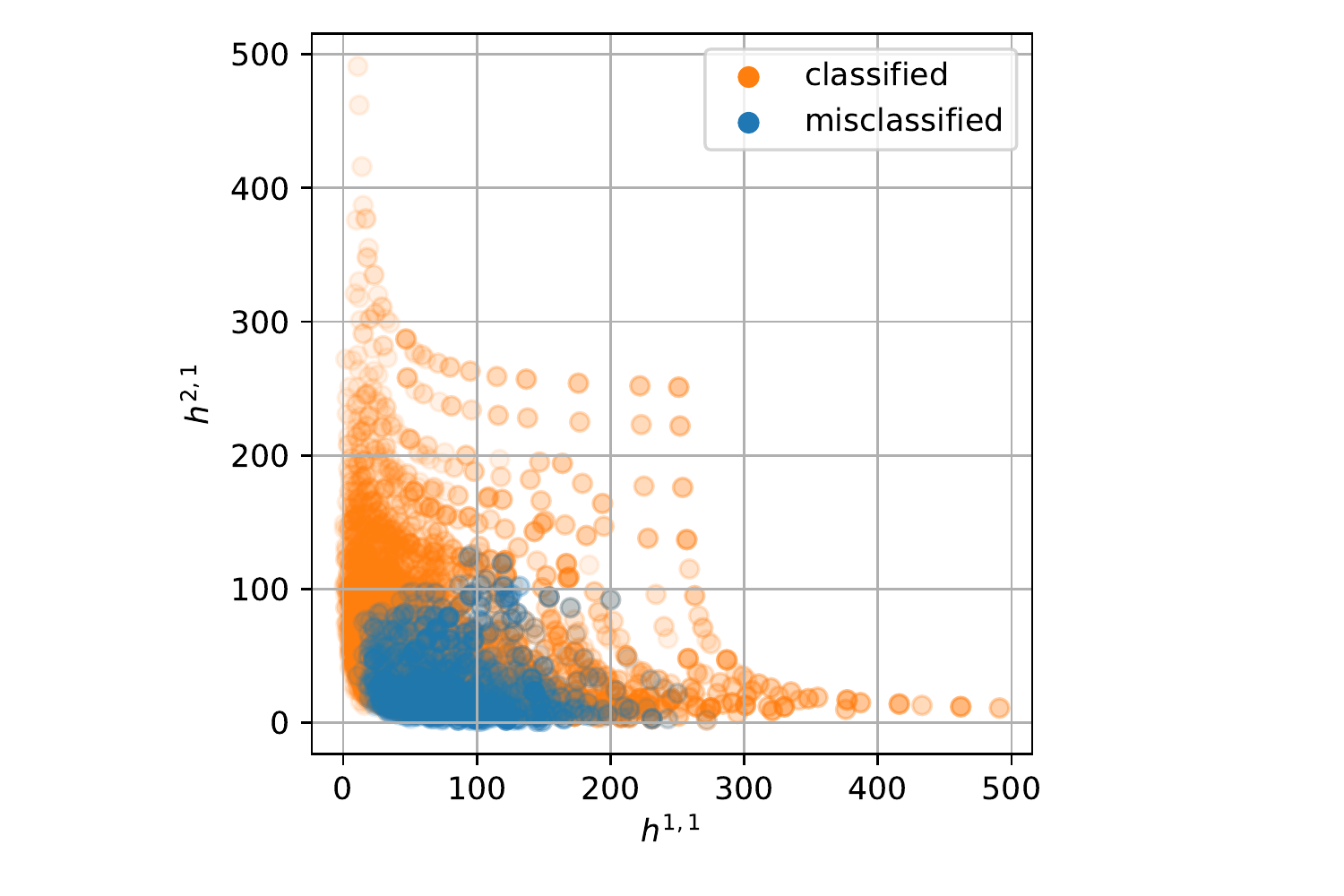}
    	\vspace{-0.8cm}
    	\caption{Random Coprime Integers \\ (1847 misclassified)}\label{CoprimeMisclassify}
    \end{subfigure} \\ 
	\begin{subfigure}{0.45\textwidth}
		\centering
		\includegraphics[width=\textwidth]{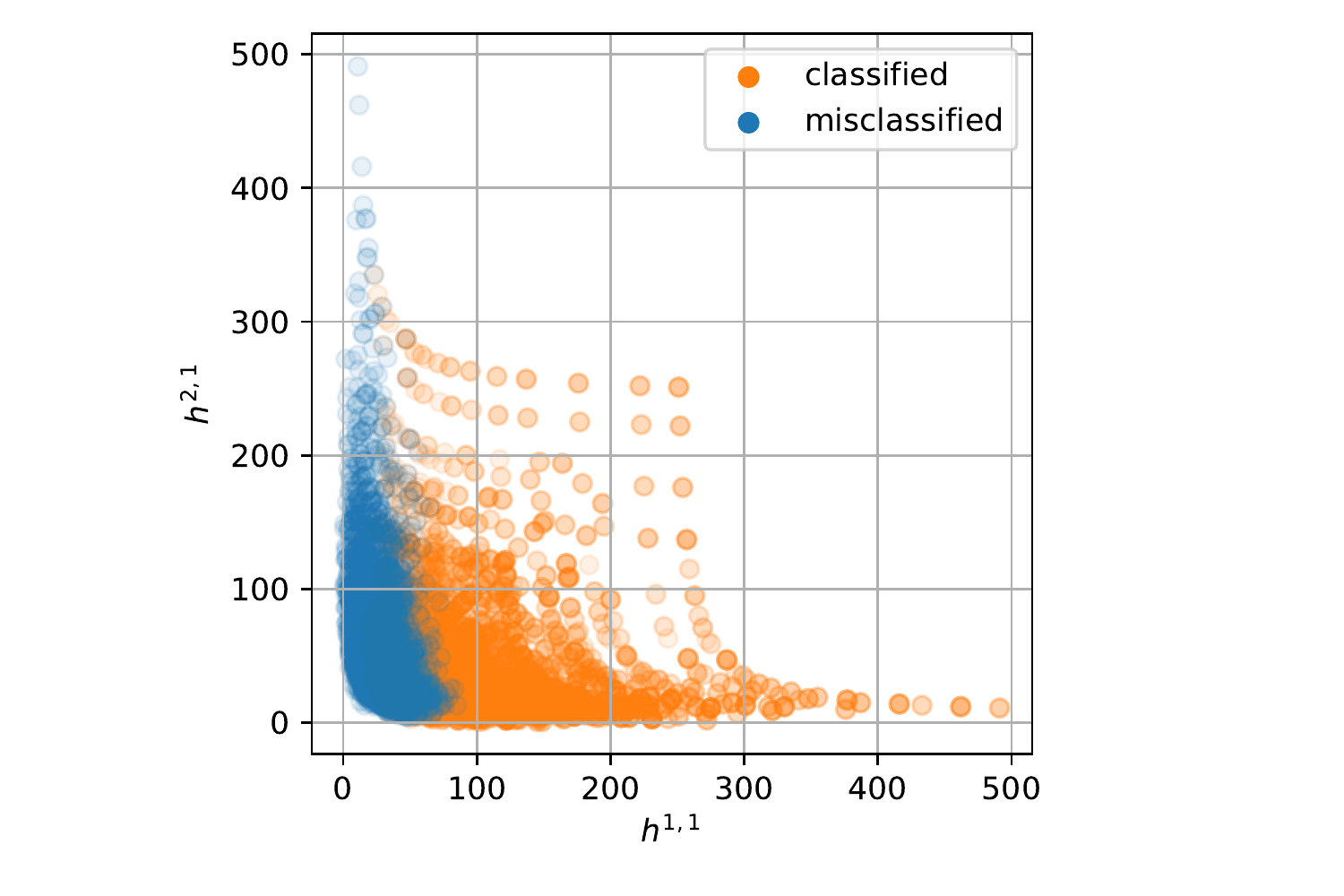}
		\vspace{-0.8cm}
		\caption{Random Transverse Coprime Integers \\ (2739 misclassified)}\label{TransverseMisclassify}
	\end{subfigure} 
\caption{A Logistic Regressor, trained on $50$ CY 5-vectors and $50$ non-CY 5-vectors, predicts whether all of the CY 5-vectors are CY or not. The plot shows the distribution of the CY surfaces according to their Hodge numbers. Those in blue are misclassified as non-CY, those in orange are correctly classified to be CY. The non-CY vectors come from datasets of Random, Coprime, or Transverse 5-vectors respectively.}\label{HodgeMisclassification}
\end{figure}

Perhaps more curious is the distribution of these CY misclassifications with respect to their Hodge numbers, plotted in figure \ref{HodgeMisclassification}.
Training Random and Coprime datasets in both cases leads to perfect classification of CY spaces with high $h^{2,1}$, whereas training with Transverse data leads to perfect classification with high $h^{1,1}$.

For reference both other architectures had similar performance with respect to Hodge numbers, as documented in appendix \ref{additionalmisclass}.

To examine further this relationship, we bin the CY data according to each of the Hodge numbers and only train and test on 5-vectors in each bin's range. This is detailed in section \S\ref{partitioning}.

\subsubsection{Hodge Partitioning}\label{partitioning}
To investigate the dependence of the learning performance on the Hodge numbers, the CY dataset was binned in two independent ways.
The first was according to $h^{2,1}$, and the second according to $h^{1,1}$.
The bin bounds were optimised such that an approximately consistent number of CYs had Hodge numbers within each bin's bounds, with a preset number of 50 bins used (selected to have a suitable bin size $>100$).
Plots of these bin frequencies are given in figures \ref{h21part_binfreq} and \ref{h11part_binfreq}.

This produced a CY dataset associated to each bin, with which a non-CY 5-vector dataset was randomly sampled.
For the $h^{2,1}$ partition the Random dataset was used to sample as many non-CY 5-vectors for each bin, such that the datasets were balanced.
As training-behaviour for the Random and Coprime datasets was so similar, only the Random dataset was used in this investigation.
Conversely, for the $h^{1,1}$ partition the Transverse dataset was used.
These choices of non-CY datasets used for training were selected such that they aligned with the predicted behaviour of section \S\ref{misclassifications}, where Random-training improves high-$h^{2,1}$ performance, and Transverse-training improves high-$h^{1,1}$ performance.

For each bin's now balanced dataset an independent Logistic Regressor (with architecture as before) was initialised, trained and tested.
A random 80\% sample of the data was used for training, with testing on the remaining 20\% complement.
For each bin, the initialisation, training, and testing was repeated 20 times, such that variances on the measures could be calculated.
Accuracies were recorded for each bin regressor, as well as the final 5 weights used to define the LR.

Accurracies across the bins for both partitions are given in figures \ref{h21part_acc} and \ref{h11part_acc}, with their respective accuracy variances in \ref{h21part_accvar} and \ref{h11part_accvar}.
There are near perfect predictions at the upper ends of these partitions, with relatively very small variances.
Determination of the CY property is hence considerably easier for surfaces whose Hodge numbers take extreme values, and pre-training against data with or without the transverse condition can significantly aid learning depending on what values the Hodge numbers take.

Finally, the 5 averaged LR weights are plotted for each bin (with respective variances surrounding them) in figures \ref{h21part_weights} and \ref{h11part_weights}.
As can be seen by comparing the relative weight sizes, in both cases at the higher ends of the partitions the first two weights particularly dominate the regression.
Since each LR weight aligns with the projective space weight, this indicates at these extremes where learning is particular strong, only the first two (i.e. lowest) weights are needed to identify whether the weighted projective space admits a Calabi-Yau hypersurface. 
Where only the CY dataset has the transversity property (i.e. training against Random) the first weight is the most significant, whilst where transversity is in both datasets (i.e. training against Transverse) the second weight is the most significant.

\begin{figure}[H]
	\centering
	\begin{subfigure}{0.48\textwidth}
		\centering
		\includegraphics[width=\textwidth]{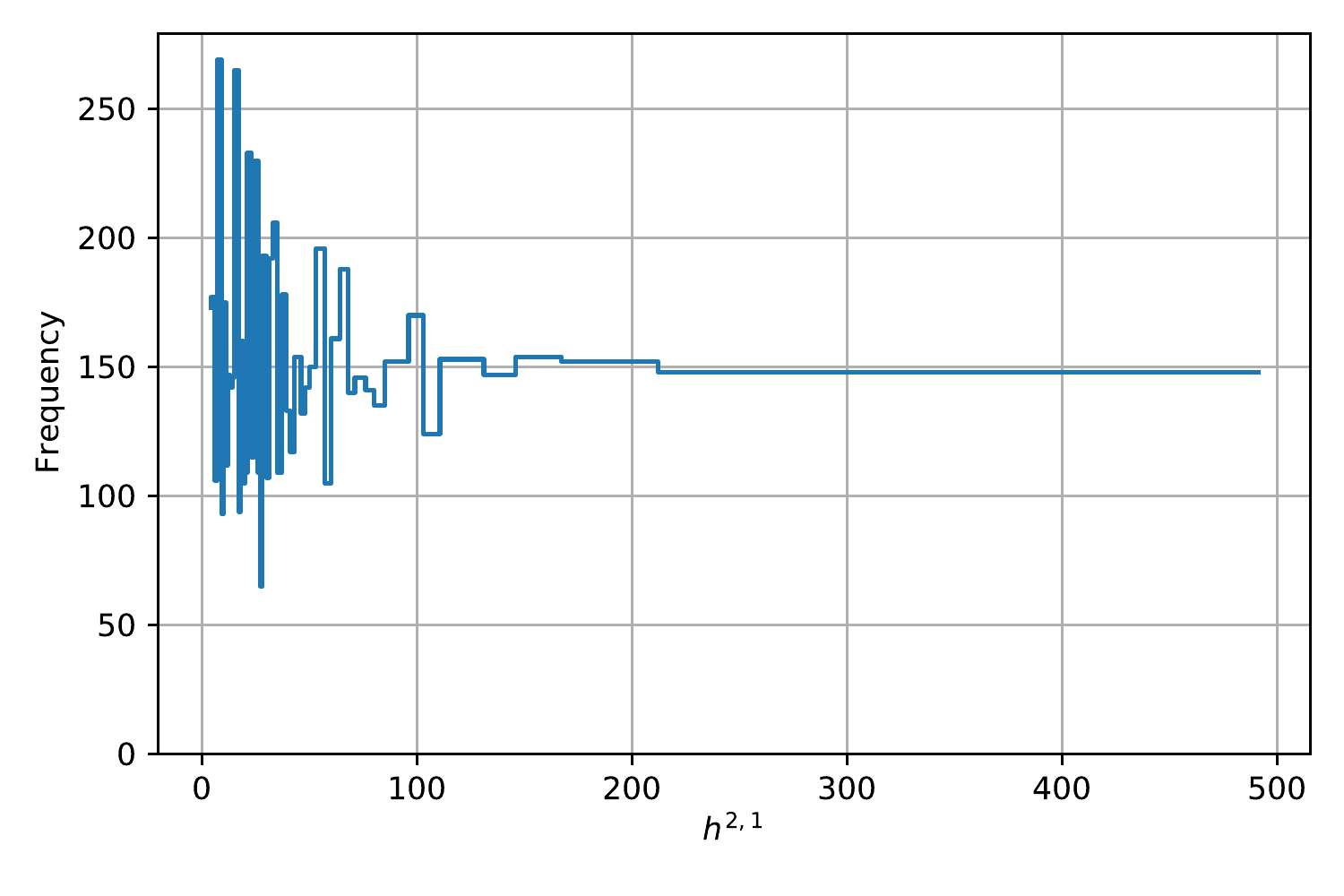}
		\caption{Bin frequencies for $h^{2,1}$ partition}\label{h21part_binfreq}
	\end{subfigure} 
    \begin{subfigure}{0.48\textwidth}
    	\centering
    	\includegraphics[width=\textwidth]{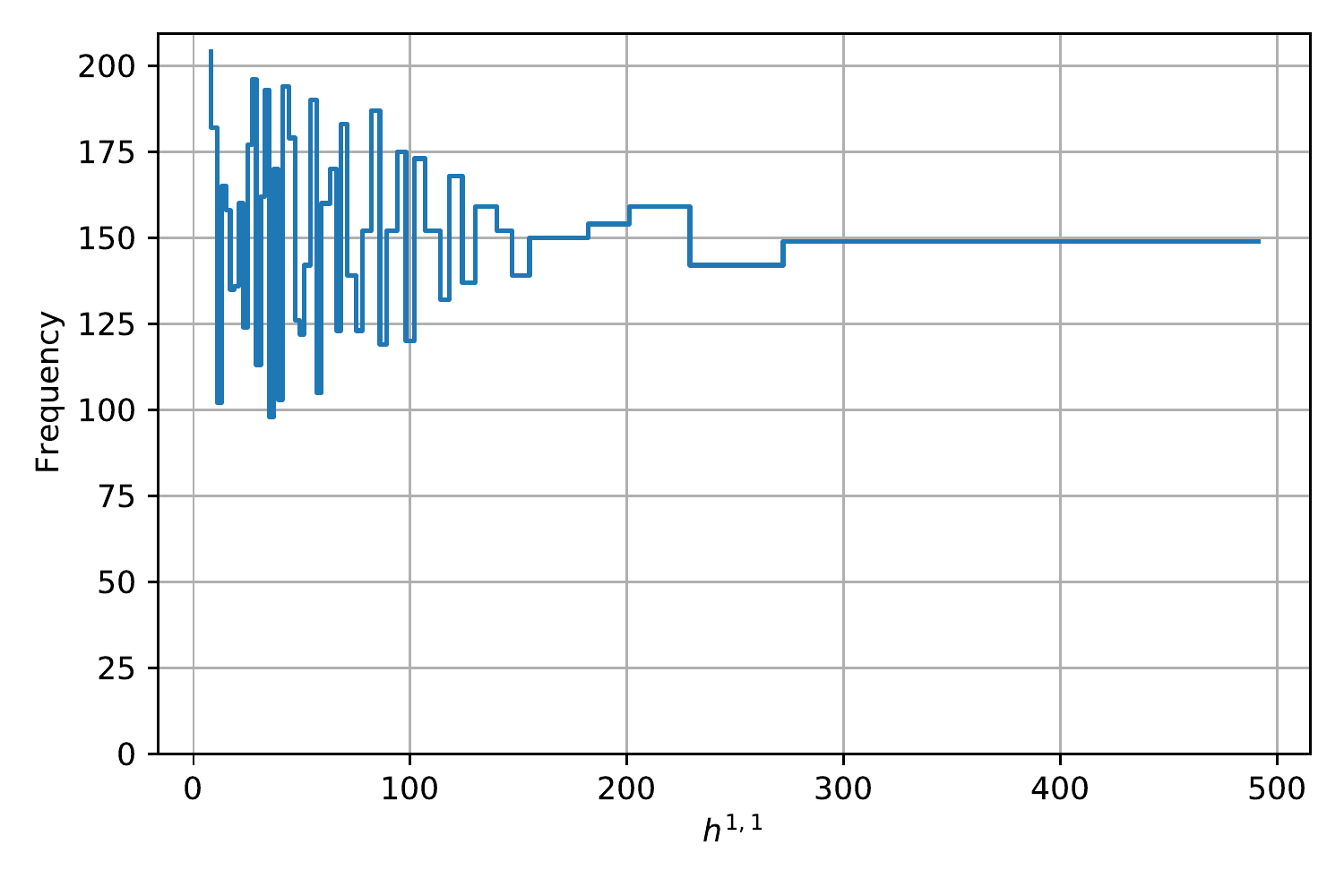}
    	\caption{Bin frequencies for $h^{1,1}$ partition}\label{h11part_binfreq}
    \end{subfigure} 
\end{figure}
\begin{figure}[H]\ContinuedFloat
	\begin{subfigure}{0.48\textwidth}
		\centering
		\includegraphics[width=\textwidth]{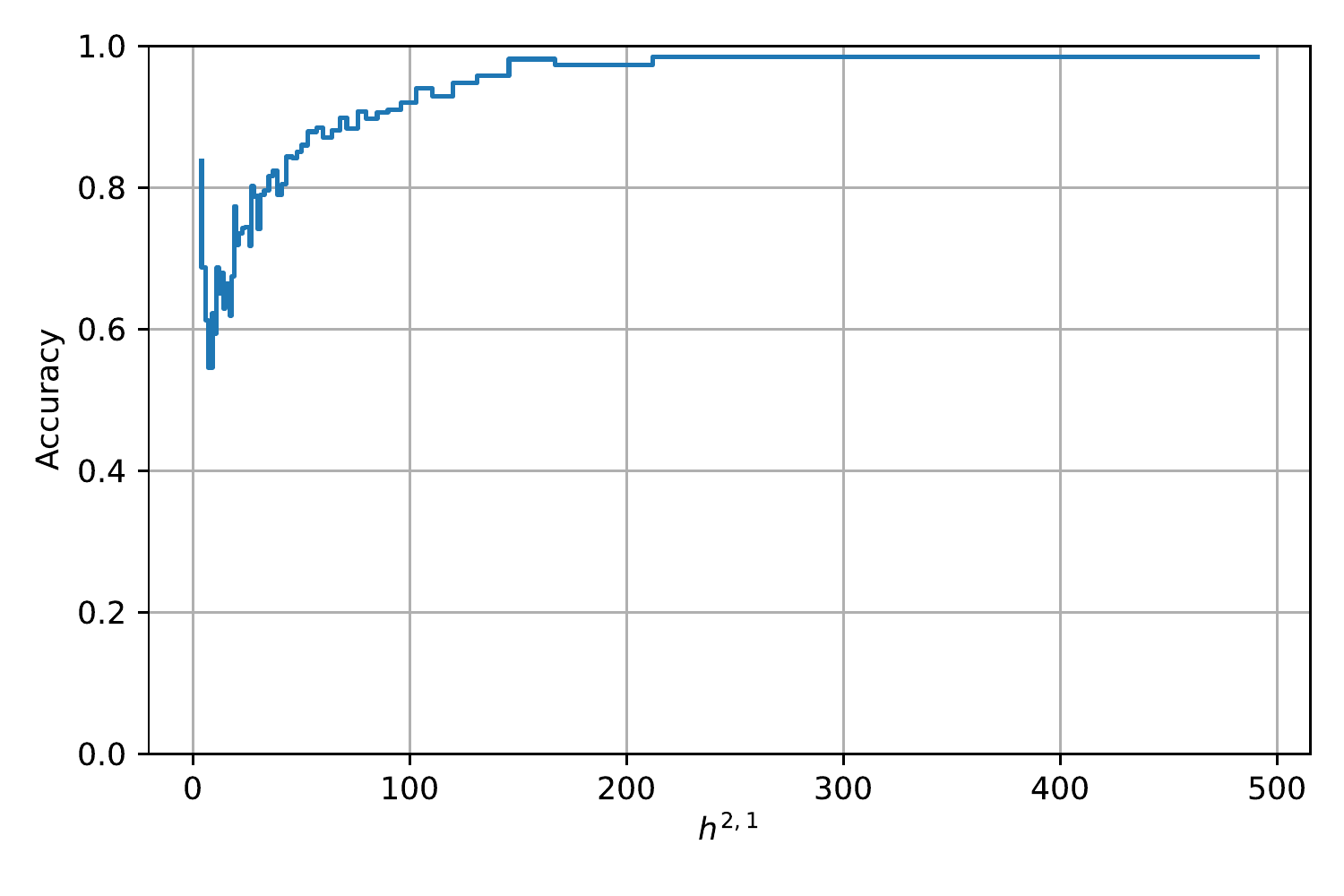}
		\caption{LR (Random-trained) Accuracies for $h^{2,1}$ partition}\label{h21part_acc}
	\end{subfigure} 
    \begin{subfigure}{0.48\textwidth}
    	\centering
    	\includegraphics[width=\textwidth]{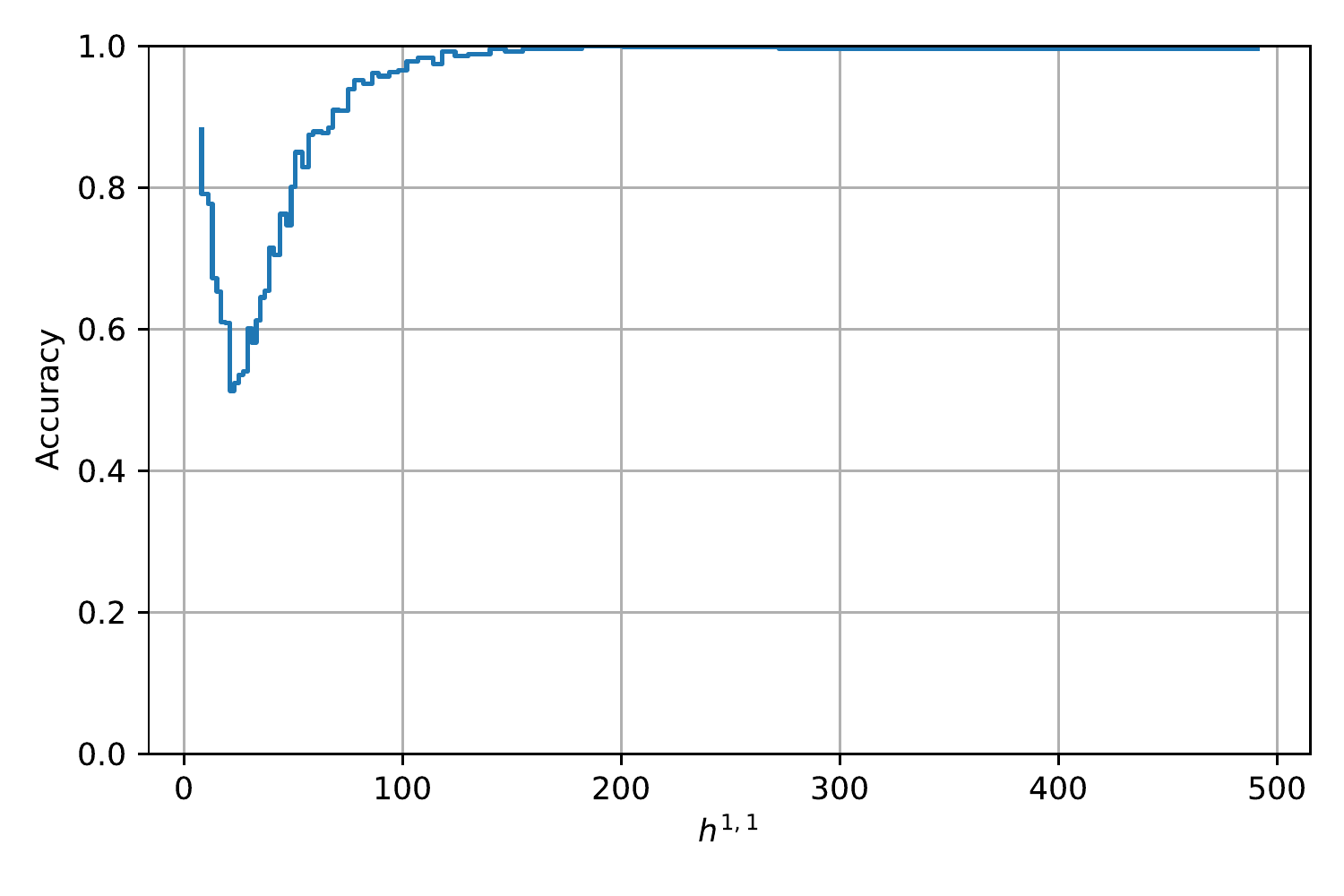}
    	\caption{LR (Transverse-trained) Accuracies for $h^{1,1}$ partition}\label{h11part_acc}
    \end{subfigure} \\
	\begin{subfigure}{0.48\textwidth}
		\centering
		\includegraphics[width=\textwidth]{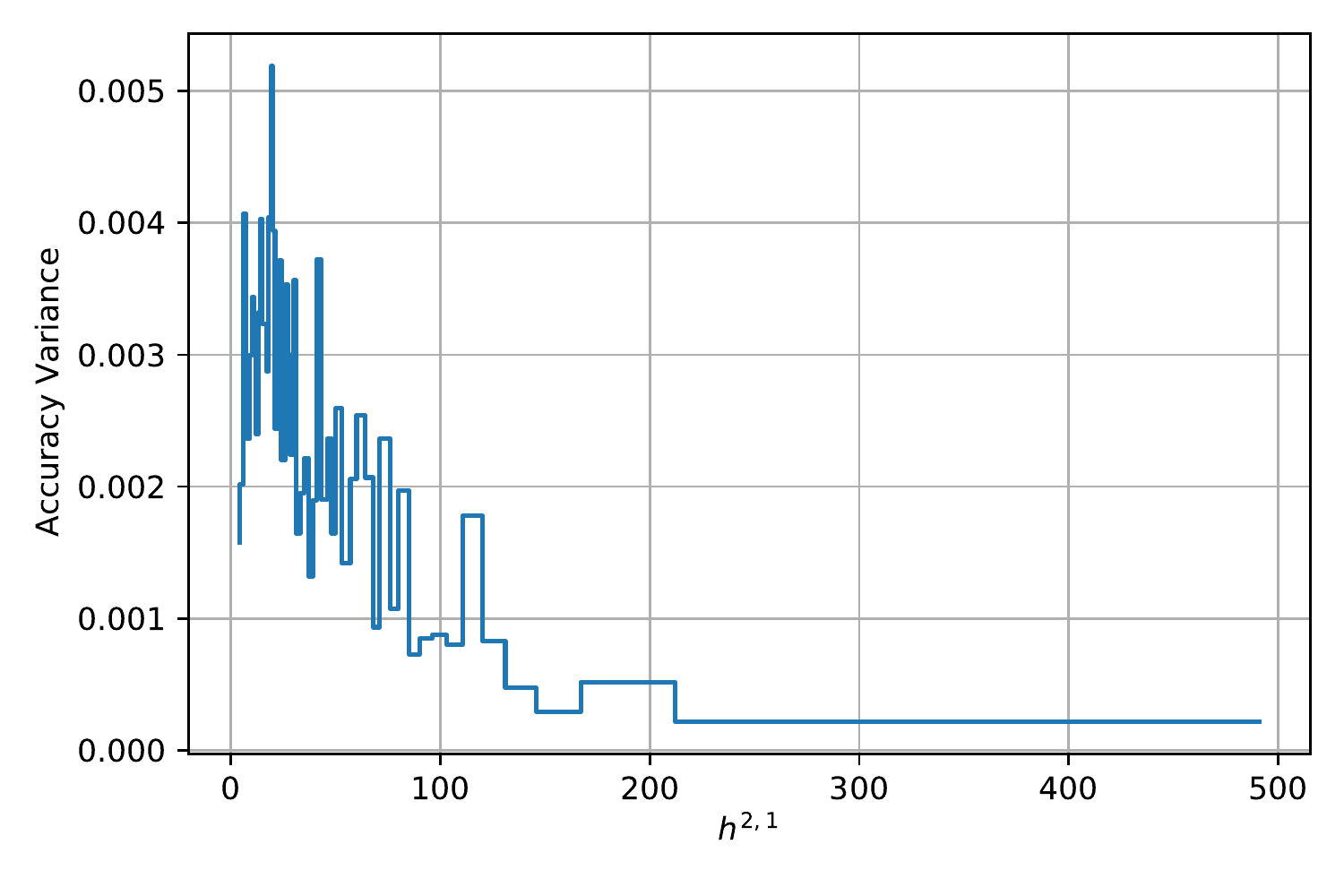}
		\caption{Variances of the LR (Random-trained) accuracies for $h^{2,1}$ partition}\label{h21part_accvar}
	\end{subfigure} 
    \begin{subfigure}{0.48\textwidth}
    	\centering
    	\includegraphics[width=\textwidth]{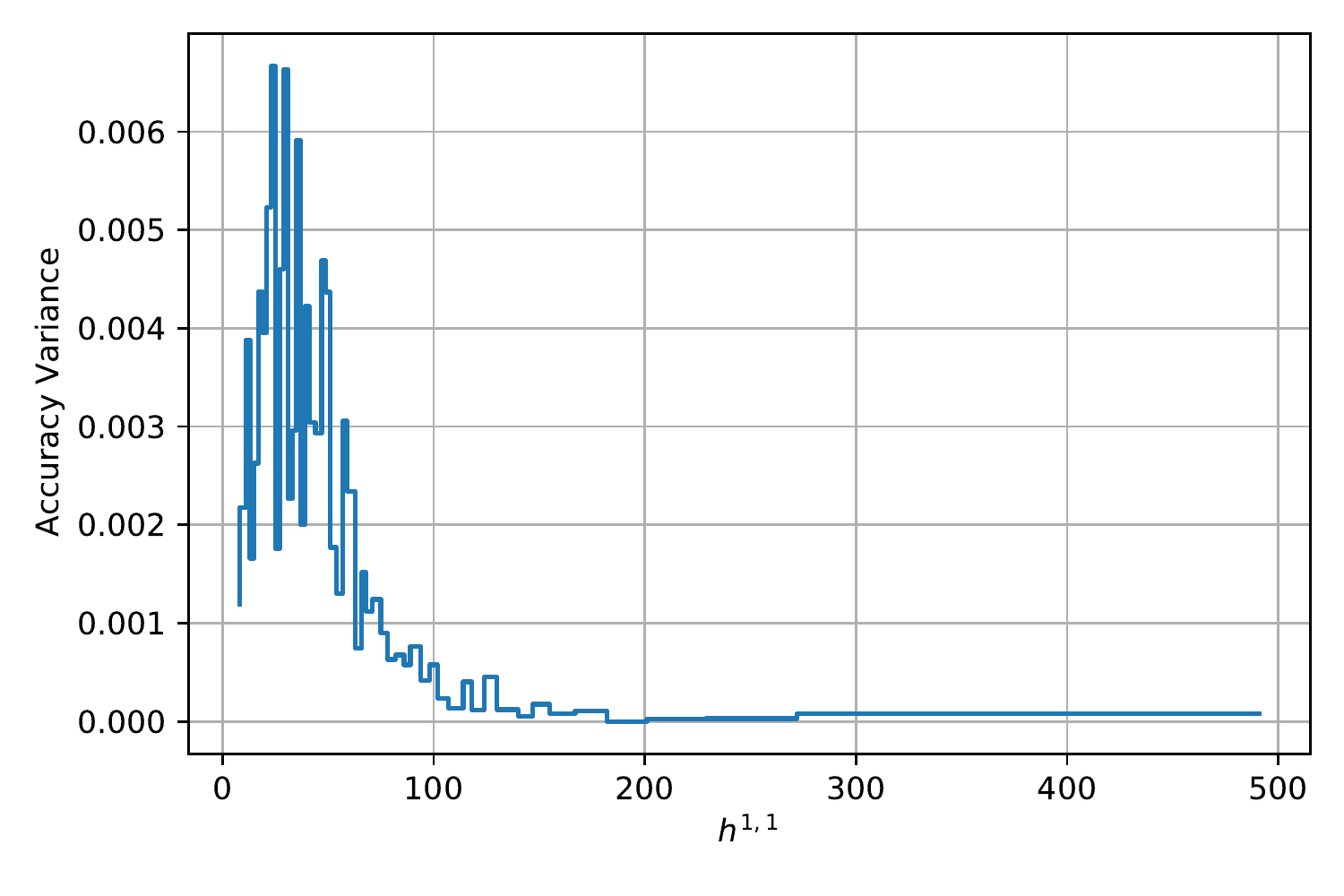}
    	\caption{Variances of the LR (Transverse-trained) accuracies for $h^{1,1}$ partition}\label{h11part_accvar}
    \end{subfigure}\\
	\begin{subfigure}{0.48\textwidth}
		\centering
		\includegraphics[width=\textwidth]{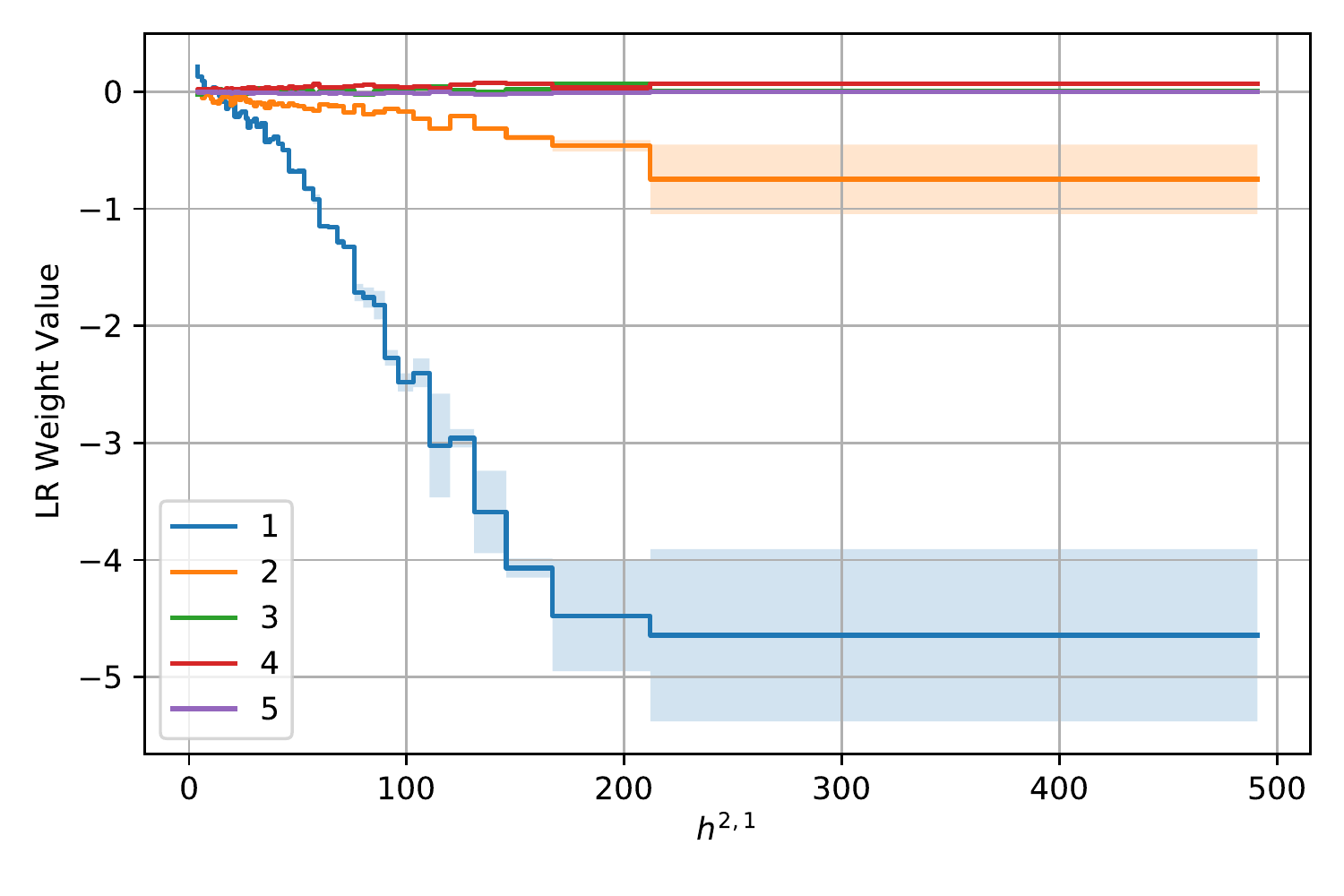}
		\caption{LR (Random-trained) weights for $h^{2,1}$ partition, plotted with variance bars}\label{h21part_weights}
	\end{subfigure} 
    \begin{subfigure}{0.48\textwidth}
    	\centering
    	\includegraphics[width=\textwidth]{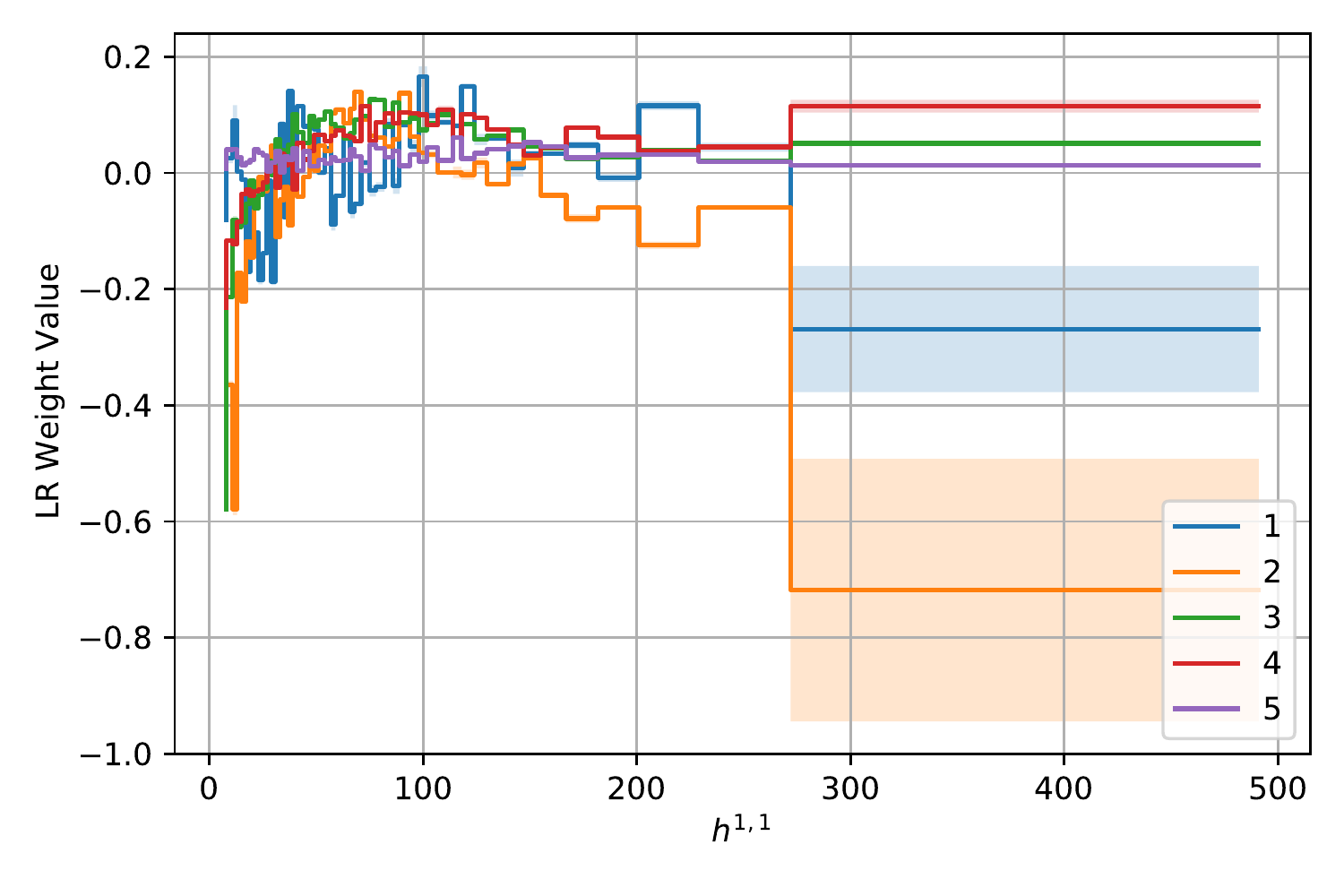}
    	\caption{LR (Transverse-trained) weights for $h^{1,1}$ partition, plotted with variance bars}\label{h11part_weights}
    \end{subfigure}
\caption{Relevant plots for Logistic Regressor learning of the 5-vectors being CY or non-CY. Where the non-CY data was the Random data then binning was according to $h^{2,1}$, where it was Transverse data then according to $h^{1,1}$. The CY data was binned according to either $h^{1,1}$ or $h^{2,1}$ Figures (a) \& (b) show the number of CYs in each Hodge partition bin (half the dataset used in each case as non-CYs cannot be plotted without known Hodge numbers). Figures (c) \& (d) show the average accuracies for the LR learning in each case, with (e) \& (f) the respective variances (very small comparatively). Finally, figures (g) \& (h) show the averaged trained LR weights, plotted with their variances as bands about the average values.}\label{HodgePartitioning}
\end{figure}

\section{Summary \& Outlook}\label{s:conc}
Through the use of unsupervised machine-learning methods we were able to identify a linear clustering structure of the weighted projective spaces that admit Calabi-Yau hypersurfaces.
This structure was first observed through PCA, corroborated with TDA, and then observed again due to relations with the hypersurface's Hodge numbers.

Supervised machine-learning methods then learnt to predict Hodge numbers from the weights directly to a surprisingly exceptional accuracy, perhaps making use of this simple structure.
In addition, simple classifier architecture could detect whether a generic weighted-$\mathbb{P}^4$ admitted a Calabi-Yau hypersurface from the weights alone, and with specific pre-training could reach perfect performance at certain extremes of Hodge numbers.

Further analysis into this Calabi-Yau clustering behaviour for weighted-$\mathbb{P}^4$s would hope to uncover its source, simultaneously explaining the success of machine-learning techniques on this dataset.

\section*{Acknowledgement}
The authors would like to thank Prof. V. Batyrev for clarifying discussion.
DSB is partially supported by Pierre Andurand.
YHH would like to thank STFC for grant ST/J00037X/1.
EH would like to thank STFC for a PhD studentship.

\appendix
\section{Appendix}
\subsection{Uniformly Sampled Weight Distributions}\label{uniformweightdists}
\begin{figure}[h!]
	\centering
	\begin{subfigure}{0.45\textwidth}
		\centering
		\includegraphics[width=\textwidth]{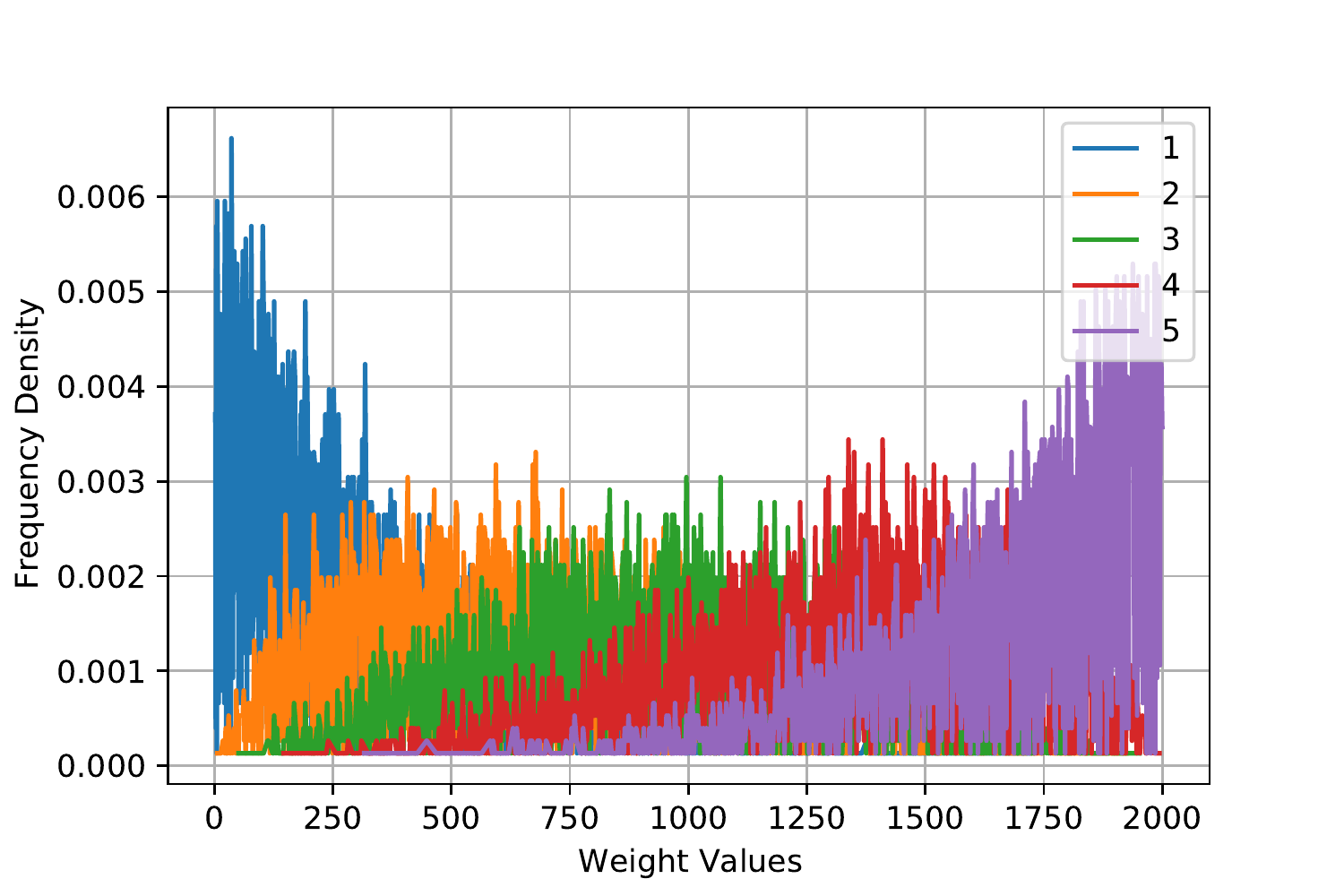}
		\caption{Random Integers}\label{RandomWeightFreqUniform}
	\end{subfigure} 
    \begin{subfigure}{0.45\textwidth}
    	\centering
    	\includegraphics[width=\textwidth]{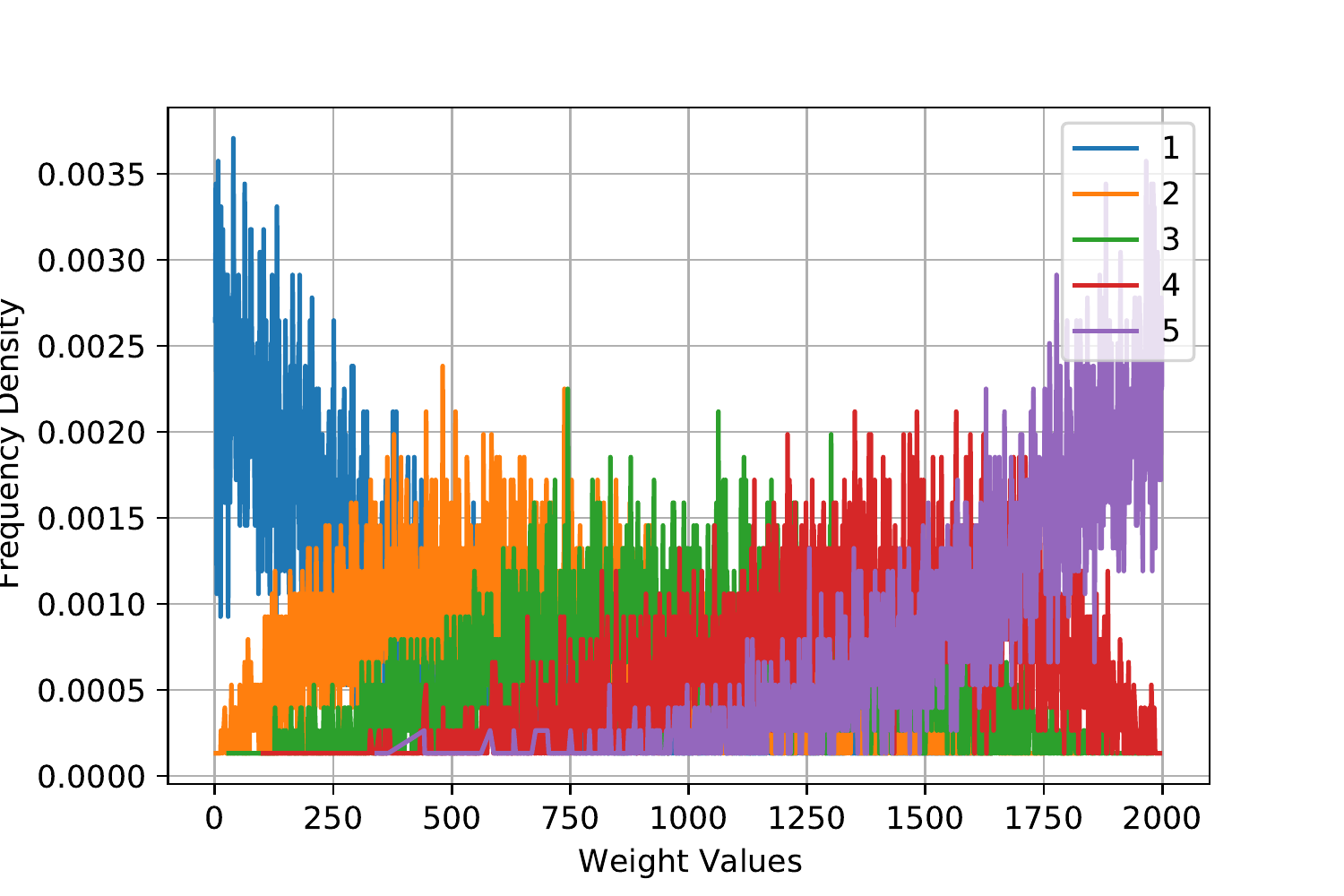}
    	\caption{Random Coprime Integers}\label{CoprimeWeightFreqUniform}
    \end{subfigure}
\caption{Frequency distributions for 5-vector weights, $w_i$ (labelled by $i: 1-5$), for the generated datasets of random integers and random coprime integers. Weights were generated using a discretised uniform distribution, $U(1,2000)$. Distributions show a spread across the range (accounting for the sorting), and hence do not well mimic the CY dataset.}\label{FakeDataWeightFreqDistsUniform}
\end{figure}

For reference, the weight frequency distributions for two of the three generated datasets: (a) Random integers, (b) Random coprime integers; as discussed in section \S\ref{datasets}, are shown below \ref{FakeDataWeightFreqDistsUniform}, where the weights were sampled uniformly using a discretisation of $U(1,2000)$.

Note the dataset of transverse random coprime integers could not be generated using a uniform distribution. 
Since the probability of five random integers in this range each dividing another weight negated from the sum is so improbable, no examples were generated running the code for a multiple days on a supercomputing cluster.
However generation with an exponential distribution took the order of minutes. 
Hence the transverse property most likely is has a significant contribution to the exponential weight distribution behaviour of the CY data.

\subsection{Additional PCA Information}\label{pca_appendix}
Further to the PCA information provided for the CY dataset in section \S\ref{pca}, the covariance matrices, eigenvectors, and eigenvalues are given for the other three datasets here.
They are respectively labelled 'R' for the random dataset, 'C' for coprime dataset, and 'T' for transverse dataset.
The covariance matrices, $K$, and eigenvalues, $\lambda$, are given to the nearest integer, whilst eigenvectors, rows of $\varepsilon$, are given to 3 decimal places.

\begin{equation}\label{R_pcainfo}
\scriptsize{
K_{R} = \begin{pmatrix}
97 & 98 & 98 & 96 & 107 \\
98 & 251 & 250 & 245 & 255 \\
98 & 250 & 530 & 514 & 542 \\
96 & 245 & 514 & 1122 & 1157 \\
107 & 255 & 542 & 1157 & 3614 
\end{pmatrix},\;
\varepsilon_{R} = \begin{pmatrix}
0.039 & 0.094 & 0.191 & 0.375 & 0.902 \\
-0.121 & -0.298 & -0.519 & -0.669 & 0.424 \\
-0.253 & -0.517 & -0.520 & 0.626 & -0.085 \\
-0.469 & -0.591 & 0.640 & -0.145 & 0.006 \\
-0.837 & 0.535 & -0.117 & 0.006 & 0.003 
\end{pmatrix},\; \lambda_{R} = \begin{pmatrix}
4241 \\
915 \\
296 \\
116 \\
47
\end{pmatrix},}
\end{equation}

\begin{equation}\label{C_pcainfo}
\scriptsize{
K_{C} = \begin{pmatrix}
100 & 100 & 101 & 91 & 89 \\
100 & 254 & 255 & 254 & 249 \\
101 & 255 & 527 & 534 & 527 \\
91 & 254 & 534 & 1166 & 1163 \\
89 & 249 & 527 & 1163 & 3418 
\end{pmatrix},\;
\varepsilon_{C} = \begin{pmatrix}
0.036 & 0.098 & 0.199 & 0.400 & 0.889 \\
-0.124 & -0.297 & -0.514 & -0.657 & 0.448 \\
-0.284 & -0.532 & -0.497 & 0.617 & -0.095 \\
-0.457 & -0.570 & 0.662 & -0.168 & 0.009 \\
-0.833 & 0.543 & -0.109 & -0.003 & 0.000 
\end{pmatrix},\; \lambda_{C} = \begin{pmatrix}
4091 \\
921 \\
296 \\
109 \\
48
\end{pmatrix},}
\end{equation}

\begin{equation}\label{T_pcainfo}
\scriptsize{
K_{T} = \begin{pmatrix}
6 & 7 & 8 & 12 & 19 \\
7 & 20 & 25 & 35 & 55 \\
8 & 25 & 62 & 85 & 125 \\
12 & 35 & 85 & 173 & 246 \\
19 & 55 & 125 & 246 & 417 
\end{pmatrix},\;
\varepsilon_{T} = \begin{pmatrix}
0.040 & 0.114 & 0.264 & 0.507 & 0.812 \\
0.102 & 0.332 & 0.712 & 0.349 & -0.501 \\
0.198 & 0.467 & 0.321 & -0.746 & 0.286 \\
-0.428 & -0.660 & 0.556 & -0.253 & 0.091 \\
-0.875 & 0.473 & -0.105 & 0.018 & -0.001 
\end{pmatrix},\; \lambda_{T} = \begin{pmatrix}
620 \\
29 \\
17 \\
9 \\
3
\end{pmatrix}.}
\end{equation}

\subsection{Additional Hodge Plots} \label{ExtraHodgeWeightPlots}
Further to the plots of the two non-trivial Hodge numbers of the CY surfaces, $\{h^{1,1},h^{2,1}\}$, against the final 5-vector weights in section \S\ref{hodgeplots}, additional plots of these Hodge numbers against the other weights are given here in figure \ref{ExtraHodgePlots} for reference.
\begin{figure}[H]
	\centering
	\begin{subfigure}{0.48\textwidth}
		\centering
		\includegraphics[width=\textwidth]{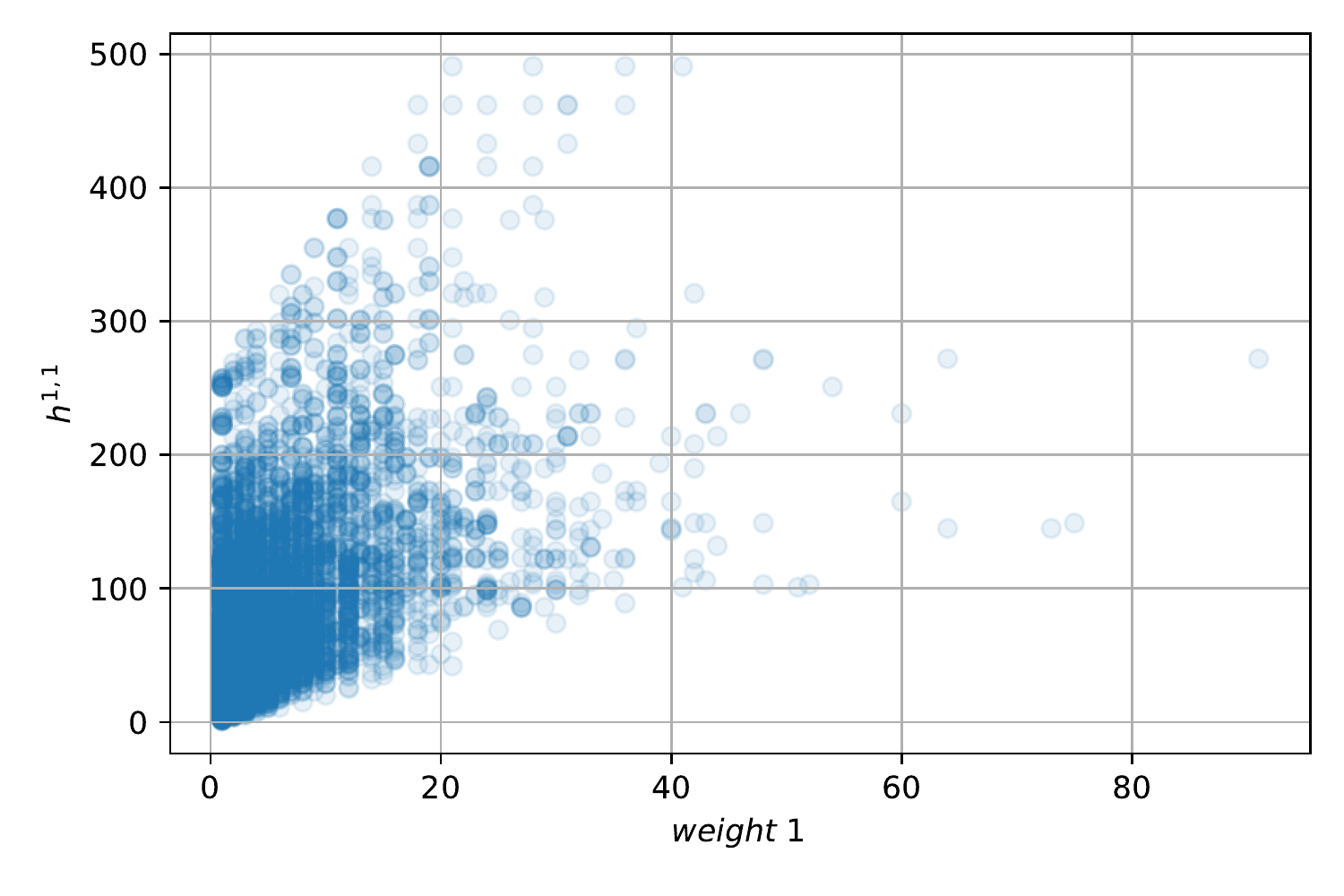}
		\caption{}\label{h11w1}
	\end{subfigure} 
    \begin{subfigure}{0.48\textwidth}
    	\centering
    	\includegraphics[width=\textwidth]{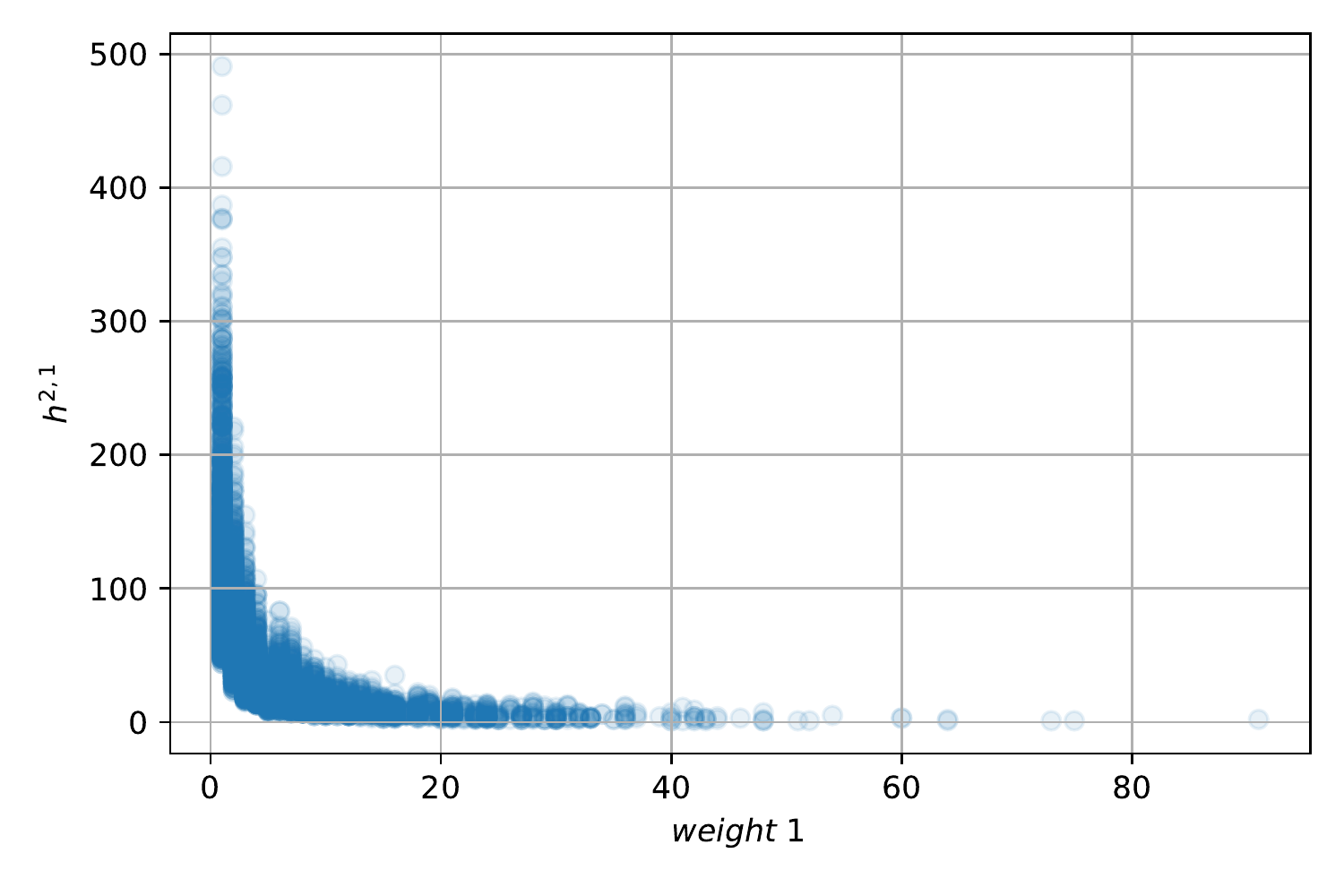}
    	\caption{}\label{h21w1}
    \end{subfigure} \\
	\begin{subfigure}{0.48\textwidth}
		\centering
		\includegraphics[width=\textwidth]{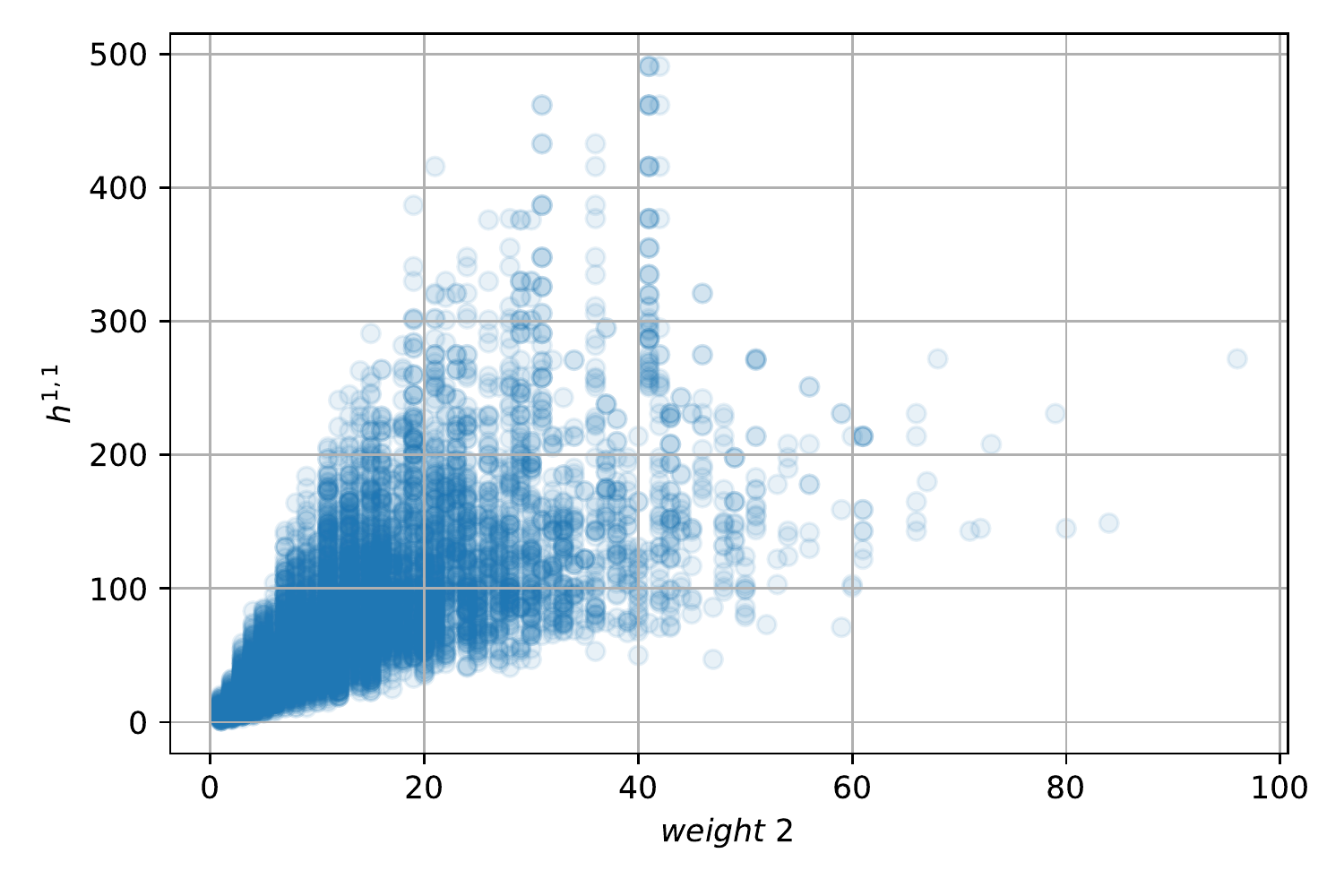}
		\caption{}\label{h11w2}
	\end{subfigure} 
    \begin{subfigure}{0.48\textwidth}
    	\centering
    	\includegraphics[width=\textwidth]{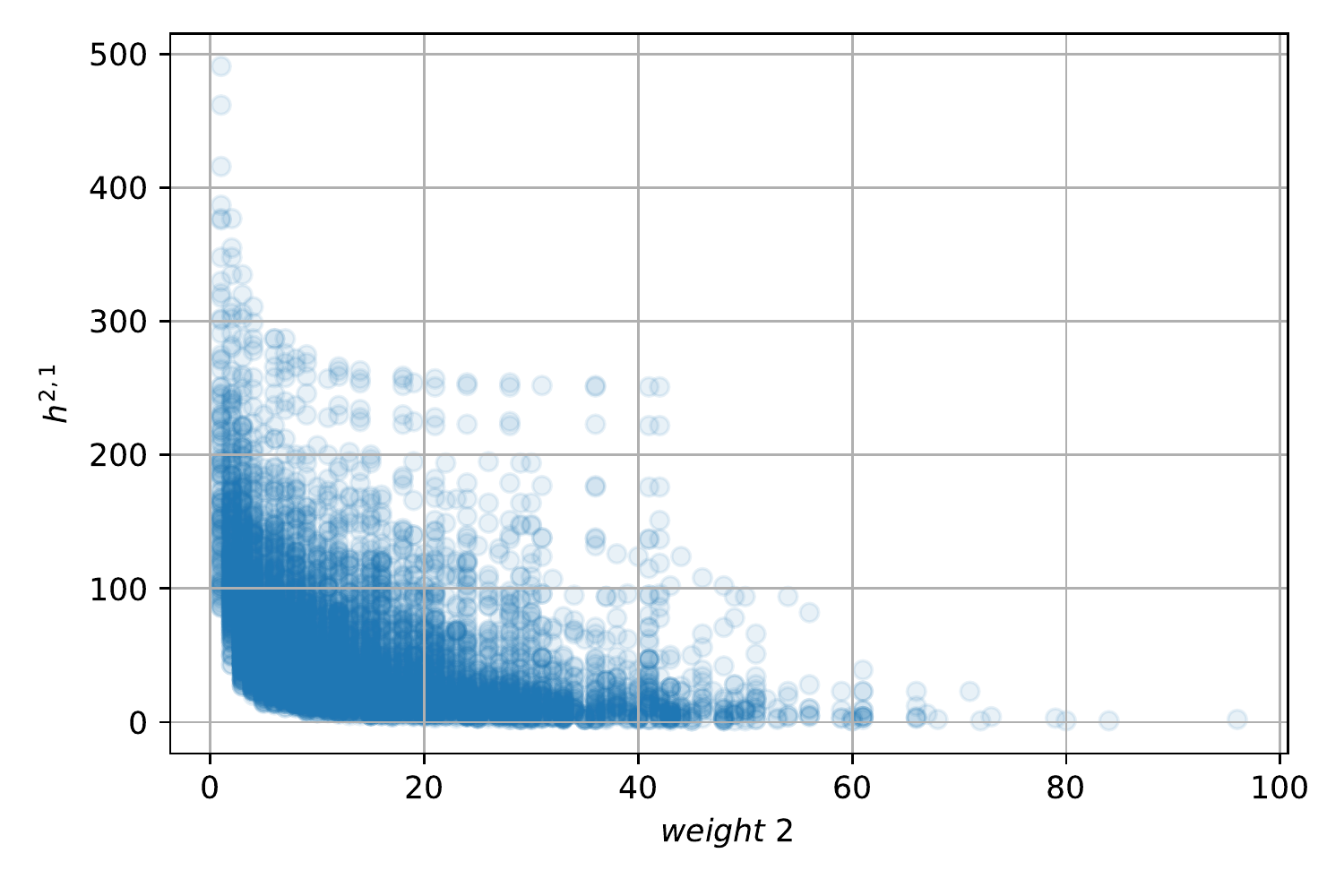}
    	\caption{}\label{h21w2}
    \end{subfigure} \\
	\begin{subfigure}{0.48\textwidth}
		\centering
		\includegraphics[width=\textwidth]{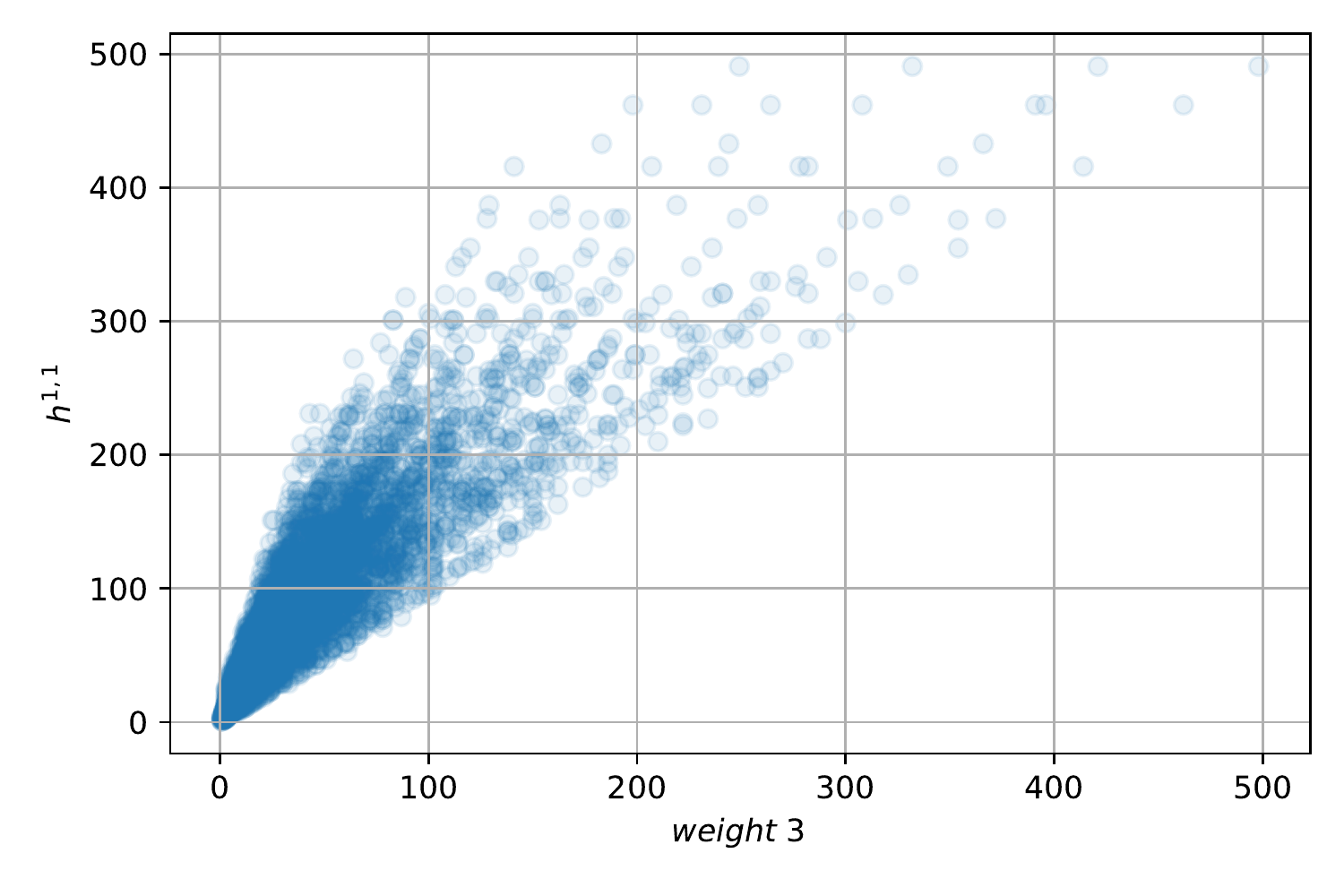}
		\caption{}\label{h11w3}
	\end{subfigure} 
    \begin{subfigure}{0.48\textwidth}
    	\centering
    	\includegraphics[width=\textwidth]{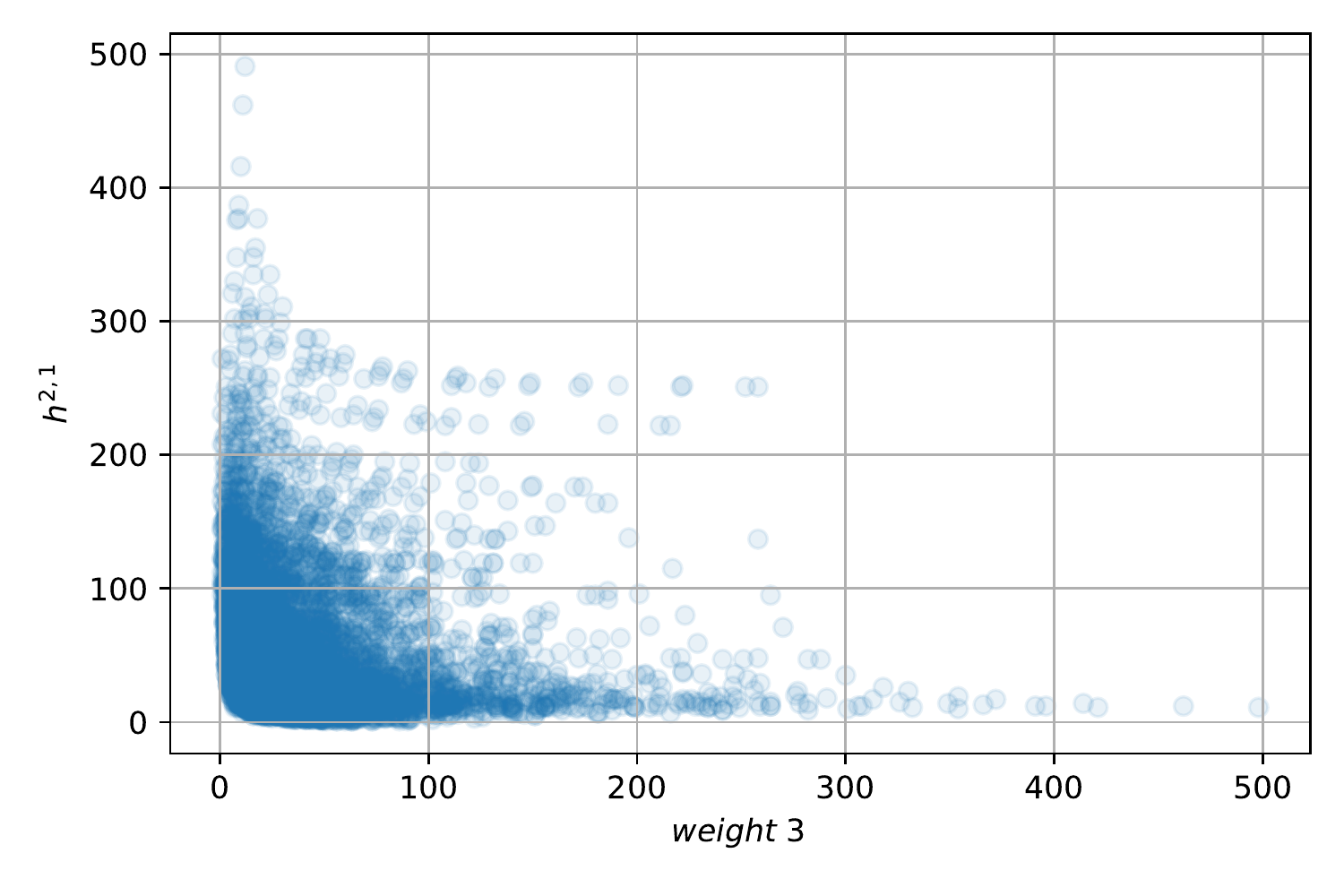}
    	\caption{}\label{h21w3}
    \end{subfigure}\\
\end{figure}
\begin{figure}[H]\ContinuedFloat
	\begin{subfigure}{0.48\textwidth}
		\centering
		\includegraphics[width=\textwidth]{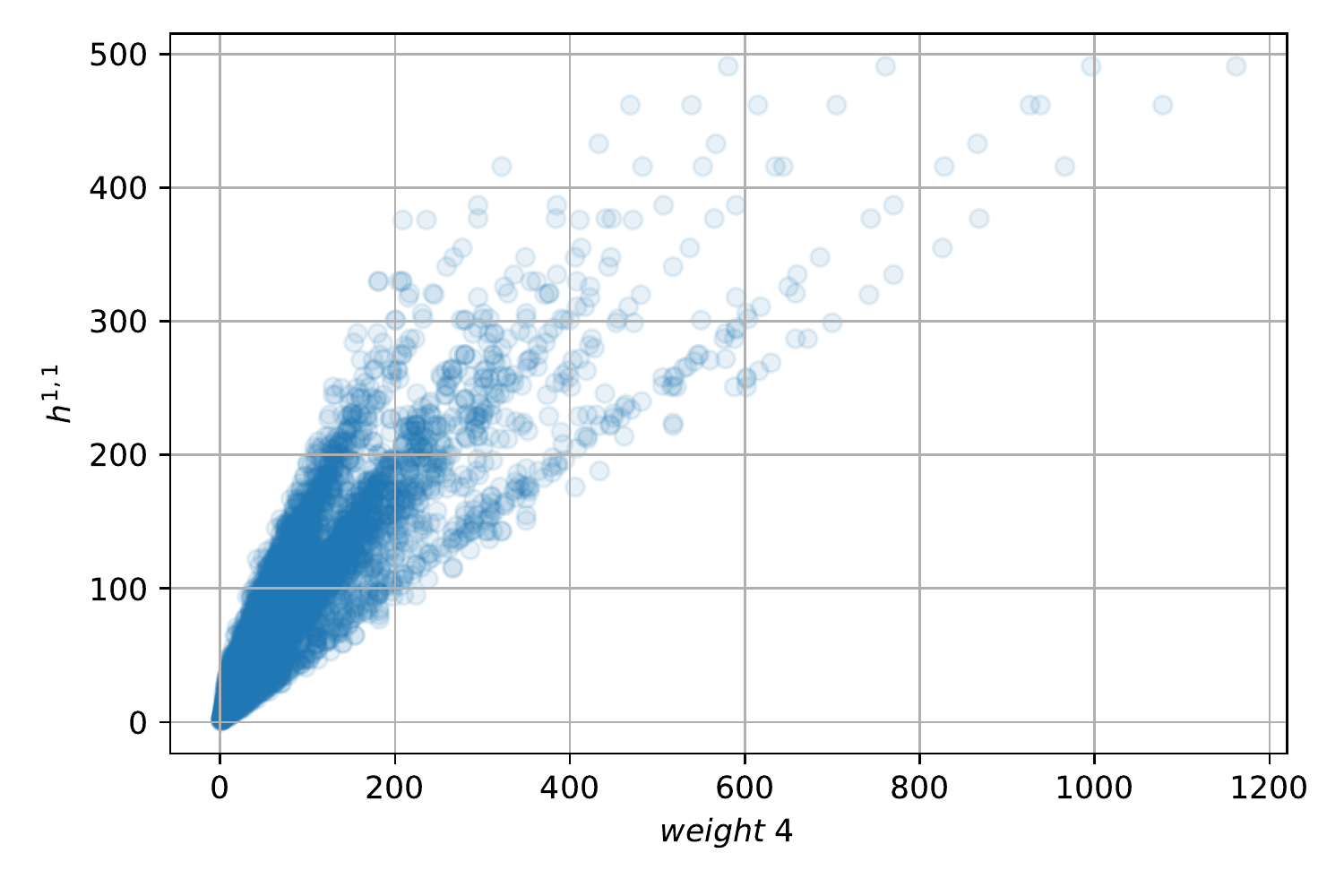}
		\caption{}\label{h11w4}
	\end{subfigure} 
    \begin{subfigure}{0.48\textwidth}
    	\centering
    	\includegraphics[width=\textwidth]{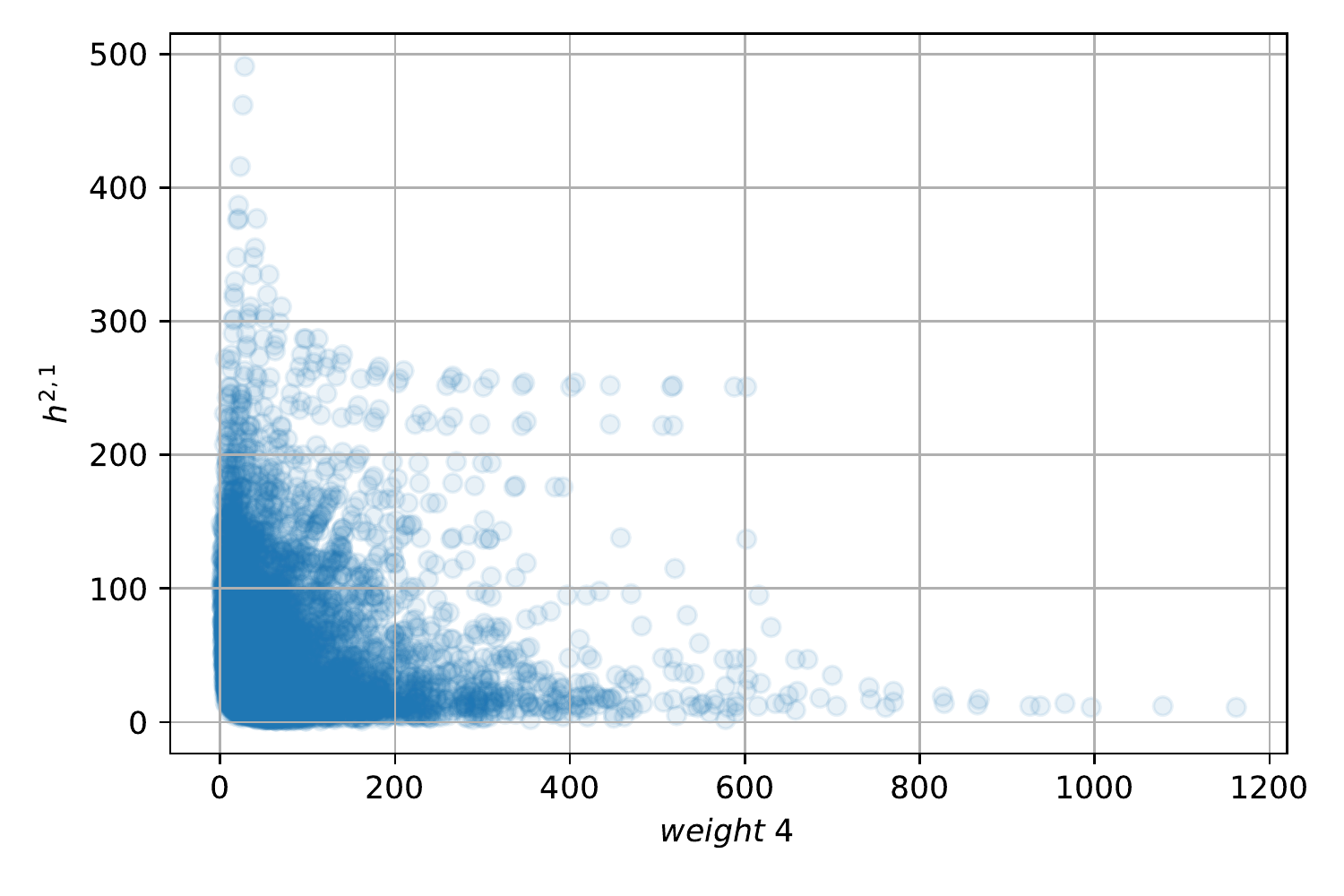}
    	\caption{}\label{h21w4}
    \end{subfigure}
\caption{Plots of the non-trivial Hodge numbers $\{h^{1,1},h^{2,1}\}$ against each of the first 4 weights in the CY 5-vectors. Behaviour is similar to  that with the final weight, showing a linear relationship to $h^{1,1}$ and a relationship preserving the mirror symmetry structure for $h^{2,1}$.}\label{ExtraHodgePlots}
\end{figure}

\subsection{Additional Misclassification Analysis}\label{additionalmisclass}
Distributions of correctly and incorrectly classified CY 5-vectors for each of the other architectures (Support Vector Machine and Neural Network), trained on 50 CY and 50 non-CY 5-vectors, are given in figure \ref{extramisclassplots}.
Note the architectures had the same hyperparameters as in previous investigation of section \S\ref{mlCY}.

The behaviour is similar to that for the Logistic Regressor, where training with Random 5-vectors improves determination for high $h^{2,1}$, whilst training with Transverse 5-vectors improves determination for high $h^{1,1}$.

\begin{figure}[H]
	\centering
	\begin{subfigure}{0.48\textwidth}
		\centering
		\includegraphics[width=\textwidth]{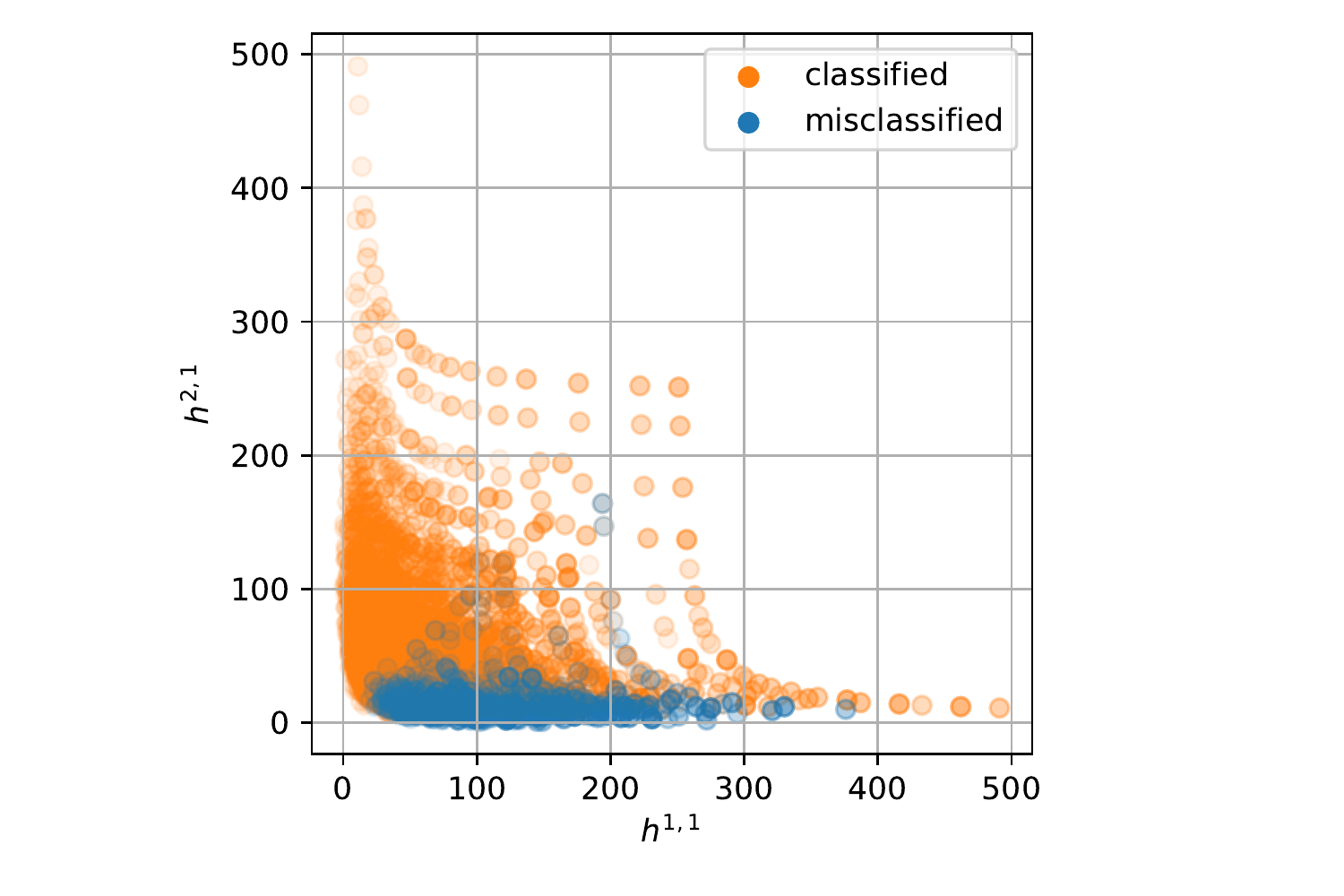}
		\vspace{-0.8cm}
		\caption{SVM trained with Random}\label{svmR}
	\end{subfigure} 
    \begin{subfigure}{0.48\textwidth}
    	\centering
    	\includegraphics[width=\textwidth]{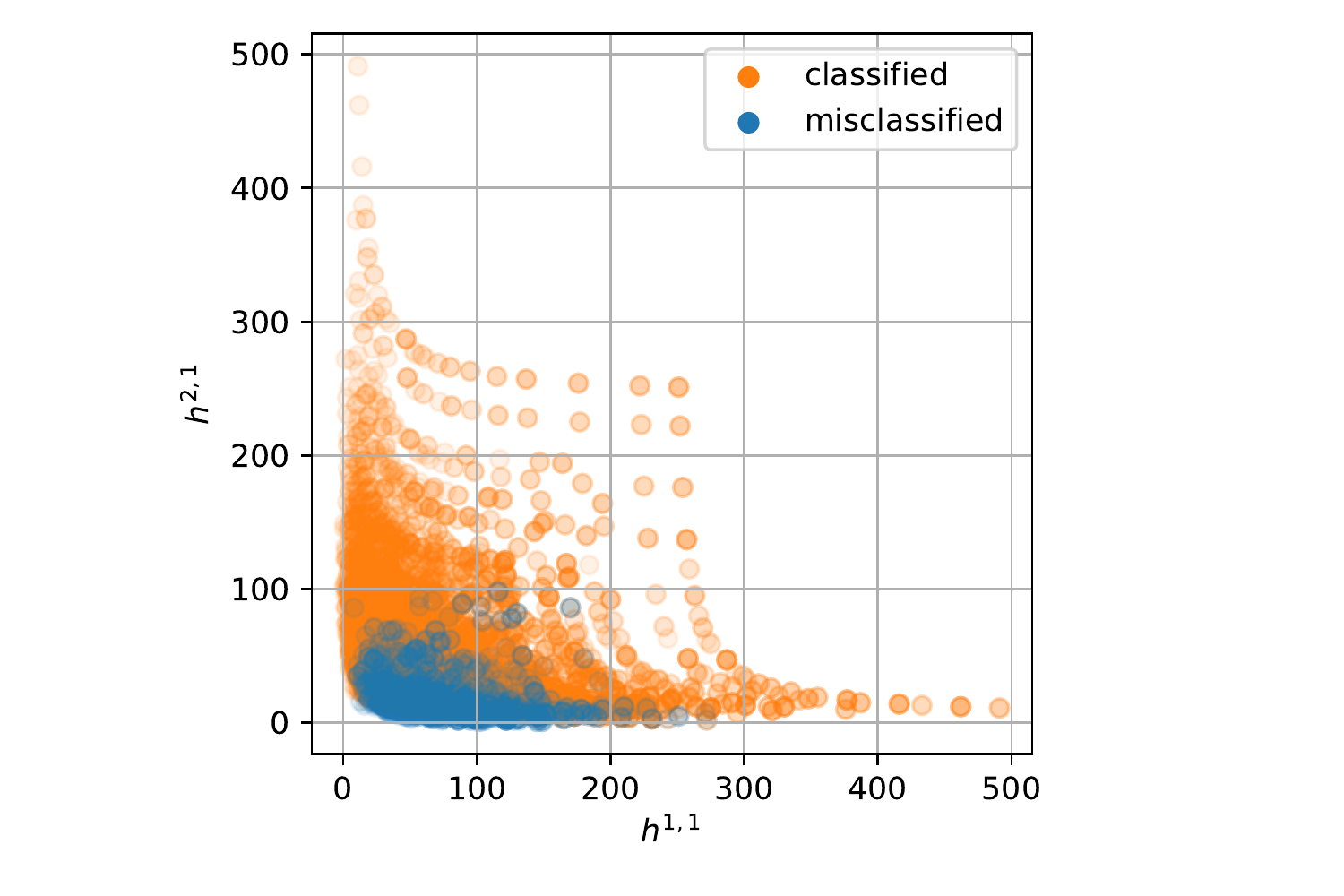}
		\vspace{-0.8cm}
    	\caption{NN trained with Random}\label{nnR}
    \end{subfigure} 
\end{figure}
\begin{figure}[H]\ContinuedFloat
    \begin{subfigure}{0.48\textwidth}
    	\centering
    	\includegraphics[width=\textwidth]{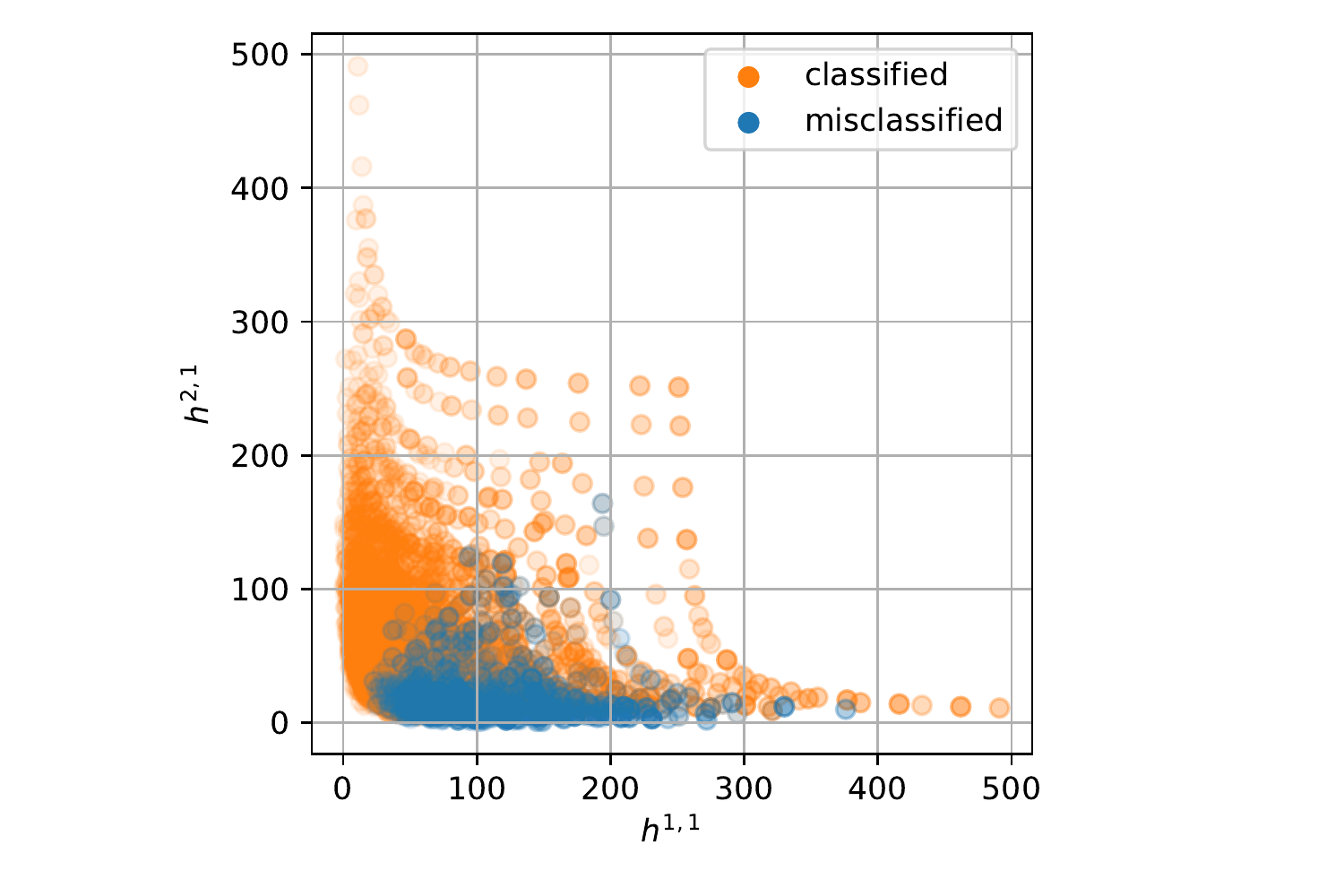}
		\vspace{-0.8cm}
    	\caption{SVM trained with Coprime}\label{svmC}
    \end{subfigure}
	\begin{subfigure}{0.48\textwidth}
		\centering
		\includegraphics[width=\textwidth]{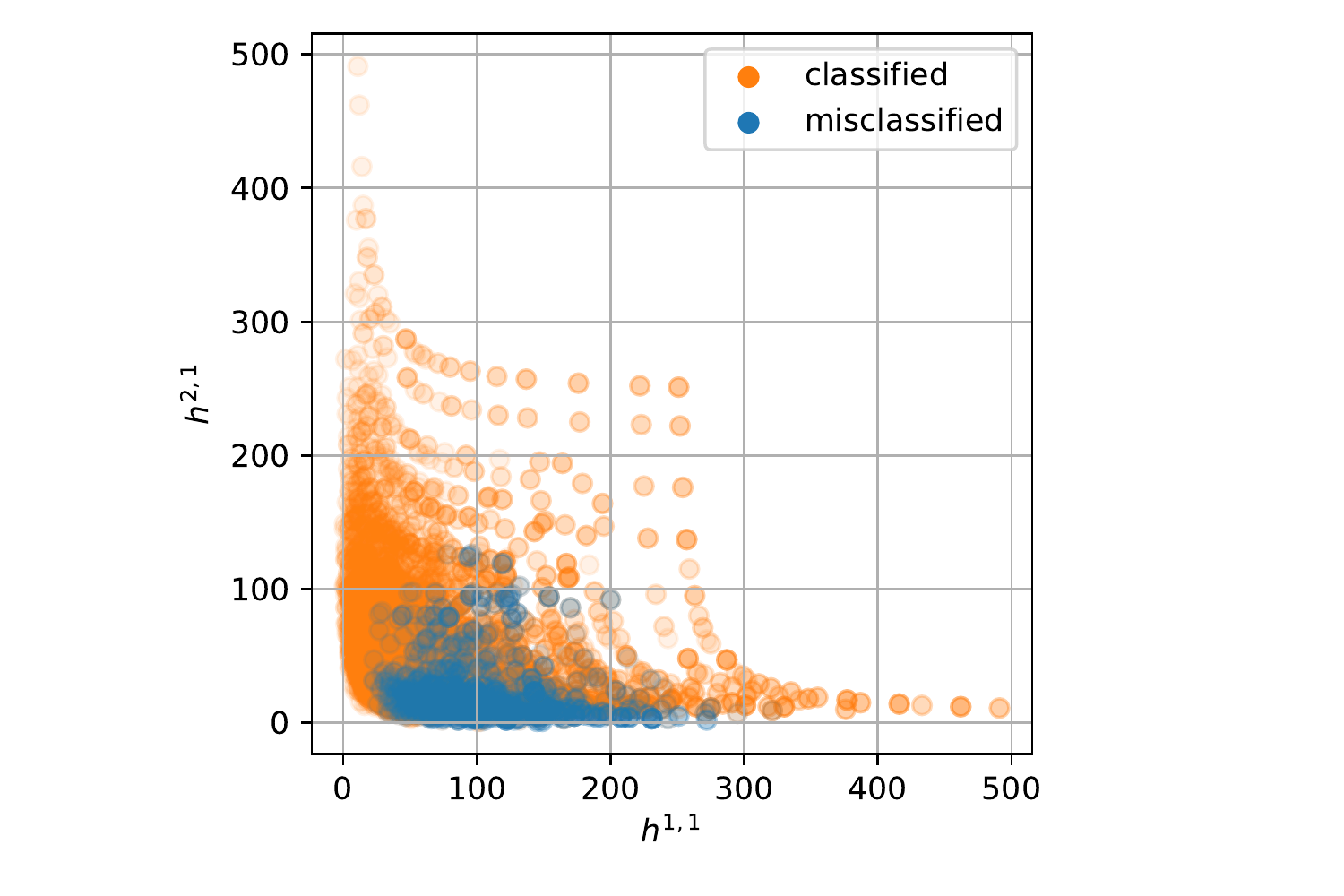}
		\vspace{-0.8cm}
		\caption{NN trained with Coprime}\label{nnC}
	\end{subfigure} \\
	\begin{subfigure}{0.48\textwidth}
		\centering
		\includegraphics[width=\textwidth]{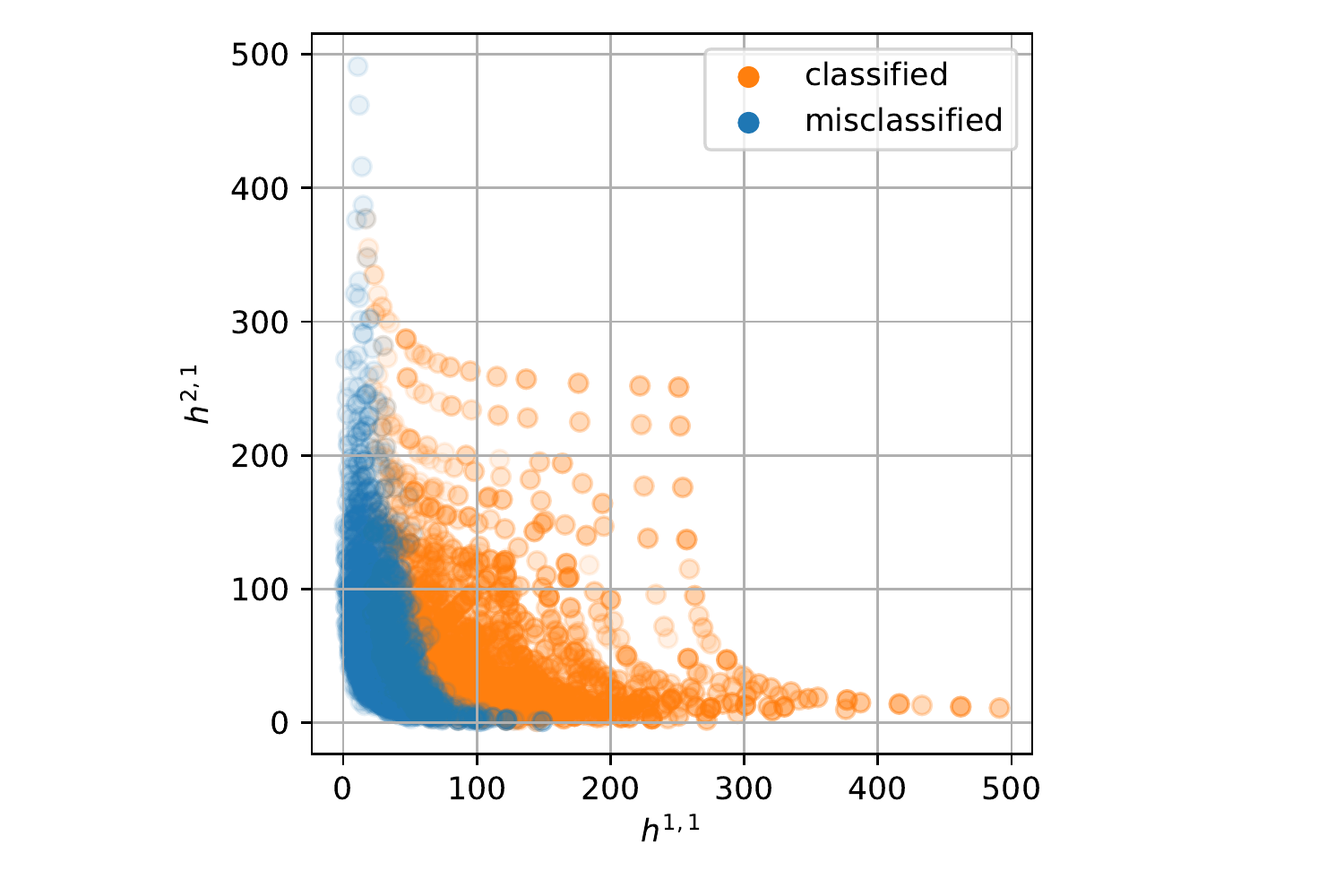}
		\vspace{-0.8cm}
		\caption{SVM trained with Transverse}\label{svmT}
	\end{subfigure} 
    \begin{subfigure}{0.48\textwidth}
    	\centering
    	\includegraphics[width=\textwidth]{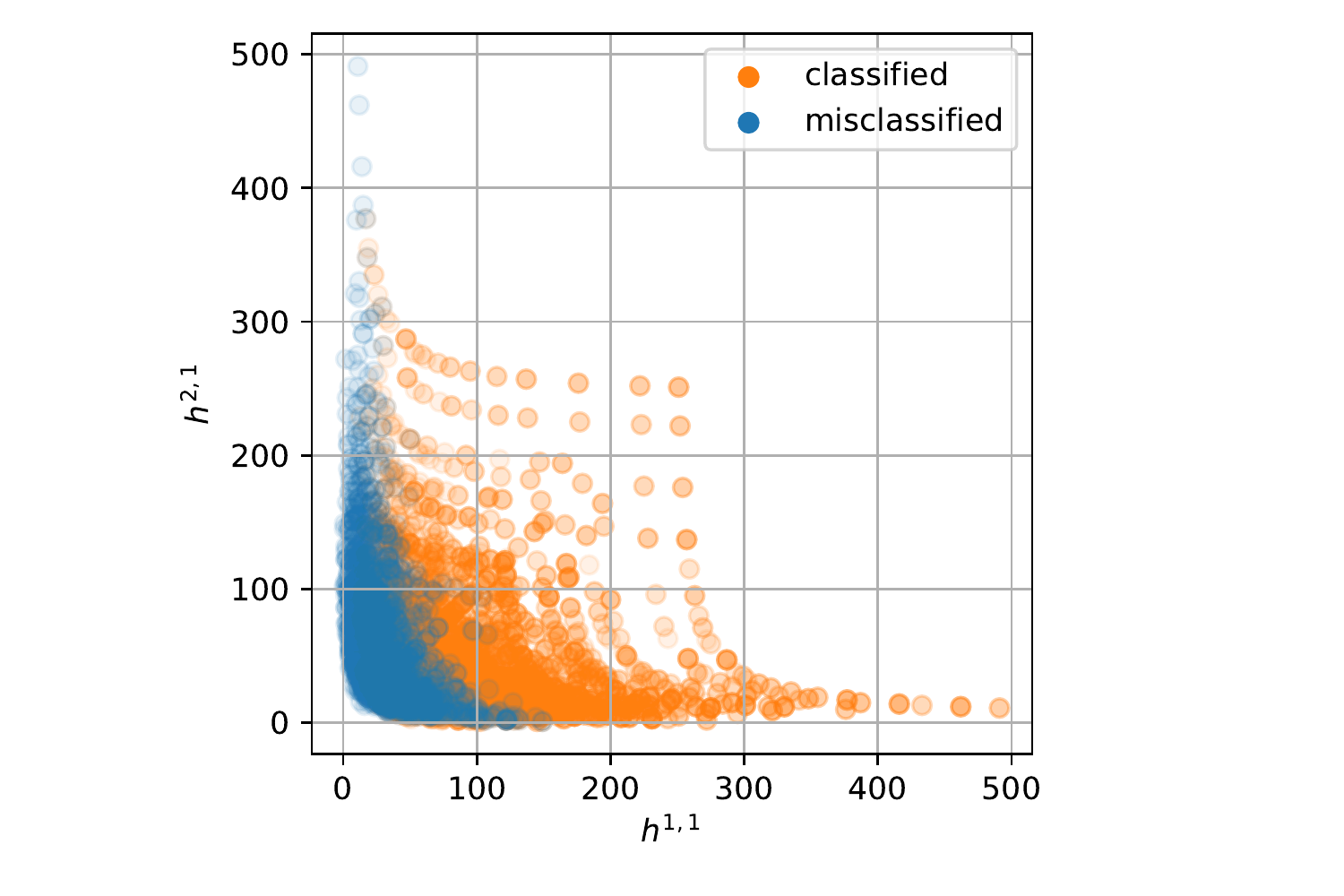}
		\vspace{-0.8cm}
    	\caption{NN trained with Transverse}\label{nnT}
    \end{subfigure}
\caption{Classified and misclassified CY 5-vectors plotted with respect to Hodge numbers, where prediction was performed by either of the architectures: Support Vector Machine (SVM), or Neural Network (NN); trained with each of the non-CY datasets respectively.}\label{extramisclassplots}
\end{figure}

\addcontentsline{toc}{section}{References}
\bibliographystyle{utphys}
\bibliography{references}

\end{document}